\documentstyle[graphicx]{mn}
\def\tabspace{\noalign{\vspace*{0.5mm}}}
\def\gx{{GX~339$-$4}}
\newcommand\rxte{\textsl{RXTE}}
\newcommand\pca{\textsl{PCA}}
\newcommand\hexte{\textsl{HEXTE}}
\newcommand\asca{\textsl{ASCA}}
\newcommand\sis{\textsl{SIS}}
\newcommand\gis{\textsl{GIS}}

\newcommand\aproxgt{\mathrel{%
      \rlap{\raise 0.511ex \hbox{$>$}}{\lower 0.511ex \hbox{$\sim$}}}}
\newcommand\aproxlt{\mathrel{%
      \rlap{\raise 0.511ex \hbox{$<$}}{\lower 0.511ex \hbox{$\sim$}}}}
\def\errtwo#1#2#3{$#1^{+#2}_{-#3}$}
\def\errtwa#1#2#3{$#1^{+#2}_{-#3}\times10^{-3}$}

\begin{document}
\title[Coronal Temporal Correlations in \gx]{Coronal--Temporal Correlations
  in \gx: Hysteresis, Possible Reflection Changes, and Implications for
  ADAFs} 
\author[M.A. Nowak, J. Wilms, \& J.B. Dove]{
  M.A.~Nowak$^{1,2}$, J. Wilms$^3$, J.B. Dove$^4$ \\
  $^1$ JILA, University of  Colorado, Campus Box 440, Boulder, CO
  80309-0440, U.S.A.; \\
  $^2$ current address, MIT-CXC, NE80-6077,  77 Massachusetts Ave.,
  Cambridge, MA ~02139, U.S.A. \\
  $^3$ Institut f\"ur Astronomie und Astrophysik,
  Abt.~Astronomie, Waldh\"auser Str. 64, D-72076 T\"ubingen, Germany; \\
  $^4$ Dept. of Physics, Metropolitan State College of
  Denver, C.B. 69, P.O. Box 173362, Denver, CO 80217-3362, U.S.A. \\
}

\maketitle

\begin{abstract} 
  We present spectral fits and timing analysis of Rossi X-ray Timing
  Explorer observations of \gx. These observations were carried out over a
  span of more than two years and encompassed both the soft/high and
  hard/low states. Two observations were simultaneous with Advanced
  Satellite for Cosmology and Astrophysics observations.  Hysteresis in the
  soft state/hard state transition is observed.  The hard state exhibits a
  possible anti-correlation between coronal compactness (i.e., spectral
  hardness) and the covering fraction of cold, reflecting material.  The
  correlation between `reflection fraction' and soft X-ray flux, however,
  appears to be more universal.  Furthermore, low flux, hard state
  observations--- taken over a decline into quiescence--- show that the Fe
  line, independent of `reflection fraction', remains broad and at a
  roughly constant equivalent width, counter to expectations from Advection
  Dominated Accretion Flow models.  All power spectral densities (PSD) of
  the hard state X-ray lightcurves are describable as the sum of just a few
  broad, quasi-periodic features with frequencies that roughly scale as
  coronal compactness, $\ell_{\rm c}$, to the $-3/2$ power.  This is
  interpretable in a simple, toy model of an \emph{efficient} spherical
  corona as variations of $\ell_{\rm c} \propto R_{\rm t}$, where $R_{\rm
    t}$ is the `transition radius' between the corona and an outer thin
  disc.  Similar to observations of Cyg X-1, time lags between soft and
  hard variability anti-correlate with coronal compactness, and peak
  shortly after the transition from the soft to hard state.  A stronger
  correlation is seen between the time lags and the `reflection fraction'.
  These latter facts might suggest that the time lags are associated with
  the known, spatially very extended, synchrotron emitting outflow.
\end{abstract}

\section{Introduction}\label{sec:intro}

\subsection{Coronal Models}\label{sec:coronal}

\begin{table*}
\caption{Observing log of the observations of GX~339$-$4. \label{tab:log_table}}
\begin{center} 
\begin{tabular}{lcrcccccc}
\hline
Obs. & Date & Exp.$^a$ & 3--9\,keV$^{b}$ & 9--20\,keV$^{b}$ & 20--200\,keV$^c$
     & Optical & Radio \\   
& (y.m.d) & (s)   & & $10^{-9}\,\rm
     erg\,cm^{-2}\,s^{-1}$  &  &  ${\rm M_B}$, ${\rm M_V}$, ${\rm M_I}$ & mJy  \\
\hline
P20181\_01 & 1997.02.03 &   5000 & 1.06 & 1.02 & 4.95 &  &  
      $^d$\errtwo{9.1}{0.2}{0.2}, \errtwo{7.0}{0.7}{0.7} &\\
P20181\_02 & 1997.02.10 &  10500 & 0.94 & 0.91 & 4.58 &  &  
      $^d$\errtwo{8.2}{0.2}{0.2}, \errtwo{6.3}{0.7}{0.7}  &\\
P20181\_03 & 1997.02.17 &   8200 & 0.90 & 0.87 & 4.53 &  &  
      $^d$\errtwo{8.7}{0.2}{0.2}, \errtwo{6.1}{0.7}{0.7} &\\
P20056\_01 & 1997.04.05 &   2200 & 0.91 & 0.87 & 4.29 &  &  &\\
P20056\_02 & 1997.04.10 &   2100 & 1.09 & 1.03 & 4.77 &  &  &\\
P20056\_03 & 1997.04.11 &   1800 & 1.09 & 1.02 & 4.77 &  &  &\\
P20056\_05 & 1997.04.15 &   1800 & 1.16 & 1.06 & 5.05 &  &  &\\
P20056\_06 & 1997.04.17 &   1300 & 1.19 & 1.11 & 5.18 &  &  &\\
P20056\_07 & 1997.04.19 &   2100 & 1.17 & 1.07 & 4.75 &  &  &\\
P20056\_08 & 1997.04.22 &   2000 & 1.10 & 1.06 & 4.42 &  &  &\\
P20181\_04 & 1997.05.29 &   9900 & 0.60 & 0.59 & 3.02 &  &  &\\
P20181\_05 & 1997.07.07 &  10000 & 0.24 & 0.25 & 1.48 &  &  &\\
P20181\_06 & 1997.08.23 &  11400 & 0.72 & 0.71 & 1.84 &  &  &\\
P20181\_07 & 1997.09.20 &  10500 & 0.95 & 0.90 & 4.49 &  &  &\\
P20181\_08 & 1997.10.28 &  10500 & 0.62 & 0.47 & 3.21 &  &  &\\
P40108\_01 & 1999.01.12 &  12300 & 0.63 & 0.14 & 0.20 &  & $^e<0.2$, $<0.2$ &\\
P40108\_02 & 1999.01.22 &  13100 & 0.53 & 0.12 & 0.20 &  & $^e<0.2$, $<0.2$ &\\
P40108\_03 & 1999.02.12 &   6800 & 0.47 & 0.42 & 1.84 & $^f$17.9, 17.2, 15.9 &
      $^e$\errtwo{4.6}{0.1}{0.1}, \errtwo{6.4}{0.1}{0.1} &\\
P40108\_04 & 1999.03.03 &  16600 & $^g$0.47 & 0.45 & 2.40 &  $^g$17.6,
      16.9, 15.6 &  $^e$\errtwo{5.7}{0.1}{0.1}, \errtwo{6.1}{0.1}{0.1} &\\
P40108\_05 & 1999.04.02 &   9200 & $^g$0.49 & 0.48 & 2.75 &  &
     $^e$\errtwo{5.1}{0.1}{0.1}, \errtwo{4.8}{0.1}{0.1} &\\
P40108\_06 & 1999.04.22 &  13100 & 0.23 & 0.23 & 1.31 & &
     $^e$\errtwo{3.1}{0.0}{0.0}, \errtwo{2.2}{0.1}{0.1} &\\
P40108\_07 & 1999.05.14 &   9800 & 0.08 & 0.07 & 0.38 &  &
     $^e$\errtwo{1.4}{0.0}{0.0}, \errtwo{1.3}{0.1}{0.1} &\\
\hline
\end{tabular}
\end{center}
\small $^a$Exposure times rounded to the closest 100\,s;
       $^b$\pca\ Flux; 
       $^c$\hexte\ Flux, normalized to \pca;
       $^d$8.6\,GHz, 0.8\,GHz data, from Wilms et al. \shortcite{wilms:99aa}; 
       $^e$8.6\,GHz, 4.8\,GHz data, from Corbel et al. \shortcite{corbel:00a};
       $^{f,g}$data from YALO telescope, 5-6 days after X-ray observation
       ($f$) and within $\pm1$\,day of X-ray observations ($g$), courtesy
           C. Bailyn and R. Jain;
       $^h$\gis\ Flux 0.39 and 0.38 $\times 10^{-9}~{\rm
           erg~cm^{-2}~s^{-1}}$, respectively
\end{table*}


Most galactic black hole candidates (BHCs) show transitions between
spectrally soft states and spectrally hard states.  Hard spectral states
appear to occur at bolometric luminosities $\aproxlt 5\%$ of the Eddington
limit, $L_{\rm Edd}$, while soft spectral states occur at higher
luminosities \cite{nowak:95a}. A possible interpretation of this luminosity
sequence is that state transitions are related to changes in the mass flow,
$\dot{M}$, through the accretion disc.  The nature of the transition
between states, and, equally importantly, the geometry of the accretion
flow within any given state, is still uncertain and a matter of vigorous
debate (see Done \nocite{done:01a} 2001 for a review).

Most current models favor Comptonization as the major physical process to
produce the observed hard state X-ray spectrum (although see Markoff et al.
\nocite{markoff:01a} 2001). In these models, soft photons from an accretion
disc are Compton-upscattered in a hot ($kT_{\rm e}\sim 150\,\rm keV$),
thermal electron plasma, the accretion disc corona (ADC). Early models of
this type usually assumed that the ADC sandwiched a classical Shakura \&
Sunyaev \shortcite{shakura:73a} accretion disc
\cite{sunyaev:79a,haardt:93a}. It has generally been assumed that this
corona is produced via magnetohydrodynamical instabilities
\cite{balbus:91a,stone:96a}.  It has been shown, however, that such a
sandwich geometry does not reproduce the observed X-ray spectra. The
temperatures required to produce the hard power law cannot be reached since
reprocessing of $\sim 50$\% of the hard Comptonized radiation results in a
very large Compton cooling rate of the corona \cite{dove:97a}.

In recent years, a large set of different ADC geometries have been proposed
in order to circumvent the ``heating problem''. All of these models have
small covering factors for the ADC or attempt to otherwise reduce the
fraction of reprocessed hard radiation returning to the ADC as soft
radiation.  Models discussed include: patchy, static, coronae above the
accretion disc \cite{stern:95b,gierlinski:97a,poutanen:99a}; in- or
outflowing coronae where the amount of reprocessed radiation is decreased
due to relativistic beaming \cite{shrader:99a,beloborodov:99a}; heavily
ionized reflector models \cite{nayakshin:98a,nayakshin:01a,done:01a}; and
models in which the ADC and the accretion disc are physically separated. In
the latter models, the ADC is typically represented by a very hot,
geometrically thick accretion disc in the central regions around the black
hole, surrounded by a cold, geometrically thin and optically thick
accretion disc. Models of this type are the ``sphere+disc'' models of Dove
et al. \shortcite{dove:97b}--- where it is assumed that the matter flowing
through the corona is efficiently converting its potential energy into
radiation--- and the Advection Dominated Accretion Flows (ADAFs; e.g., Esin
et al. \nocite{esin:97c} 1997)--- where most of the potential energy of the
accreting matter is advected into the black hole.  These latter models also
postulate that a large fraction of the seed photons for Comptonization come
from synchrotron radiation, due to the motion of the hot electrons in the
magnetic field of the advective flow.

A further complication of ADC models is the interface between the cold
accretion disc and the ADC. In ``slab-like'' ADC models, the high
ionization parameter at the ADC-disc interface leads to the formation of an
ionized transition layer which can strongly affect the reflectivity of the
accretion disc \cite{ross:93a,nayakshin:98a,nayakshin:01a,done:01a}. The
``sphere+disc'' models, notably, fail to reproduce the observed iron
fluorescence features, possibly due to the fact that any overlap between
the accretion disc and the ADC is usually not modelled (Dove et al.
\nocite{dove:98a} 1998, although see Zdziarski et al.
\nocite{zdziarski:98a} 1998). Finally, due to the closeness of the ADC to
the central black hole, relativistic smearing of spectral features needs to
be accounted for (Done \& \.Zycki \nocite{done:99a} 1999), although many
models have ignored such effects.

Observationally, the situation is equally challenging. For ``sphere+disc''
models, radiatively efficient spherical models have suggested coronal radii
as small as $\sim 30\,GM/c^2$ \cite{wilms:99aa}, while ADAF models
typically prefer radii $\aproxgt 200\,GM/c^2$ \cite{esin:97c}. Attempts to
measure the coronal size mostly have been unsuccessful
\cite{nowak:99a,nowak:99b}.  On the other hand, attempts to model the
observed broad band X-ray and $\gamma$-ray spectra with the ADC models have
generally been quite successful
\cite{dove:98a,gierlinski:97a,gierlinski:99a,zdziarski:98a,shrader:99a}.
Many fundamentally different Comptonization models appear to give equally
good fits to the data, thus making a decision between these models from a
purely observational point very difficult.  One potential discriminant
among models has emerged with the suggestion of Zdziarski, Lubi\'nski, \& Smith
\shortcite{zdziarski:99a} that there is an \emph{anti-correlation} between
the spectral hardness of hard state BHC spectra and the covering fraction
of cold, reflecting material (the so-called `$\Gamma$-$\Omega/2\pi$
correlation', where $\Gamma$ is the photon number flux power law spectral
index, i.e., $F_\gamma \propto E^{-\Gamma}$, and $\Omega/2\pi$ is the
covering fraction of the cold reflector).  Different power-law+reflection
models which fit the same data equally well, have implied very different
underlying fit parameters. (Contrast the fits of \gx\ data presented by
Wilms et al. \nocite{wilms:99aa} 1999 to those presented by Revnivtsev,
Gilfanov, \& Churazov \nocite{revnivtsev:99a} 2001.)  Finally, we note that
the ``Comptonization paradigm'' itself has recently been challenged, with
Markoff et al.  \shortcite{markoff:01a} postulating that synchrotron
radiation from the radio outflow/jet, now observed in many BHC systems
during the hard state, might well contribute a large fraction of the
observed X-ray flux.

\subsection{Observations of \gx}\label{sec:observations}
The galactic black hole \gx\ is well-suited to perform a new study of the
relative merits of different accretion disc models, as it is the only
persistent source that has been observed in all spectral states of BHCs
\cite{ilovaisky:86a,grebenev:91a,miyamoto:91a,mendez:97a,wilms:99aa}.
Furthermore, the emission from \gx\ is likely completely dominated by the
accretion flow, all the way from radio through gamma-ray wavelengths, and
it shows strong correlations among these energy bands
\cite{hanni:98a,wilms:99aa,corbel:00a,fender:01a}.  In general, radio
emission at $\aproxlt 10$\,mJy levels is present, is positively
correlated with the X-rays in the hard state, and is quenched during the
soft state \cite{hanni:98a,fender:99b,corbel:00a}. Brightness temperature
arguments suggest that, during the hard state, the radio emitting outflow
must extend to $> 10^{12.5}$\,cm \cite{wilms:99aa}.  For the hard state,
the optical light from this system is likely dominated by the accretion
flow onto the compact object as well, with a large fraction of the hard
state optical flux possibly being synchrotron emission from the extended
outflow \cite{imamura:90a,steiman:90a}.

In this paper we apply a variety of spectral models and timing analyses to
22~observations of \gx\ obtained with the \textsl{Rossi X-ray Timing
  Explorer} (\rxte), some of which were also simultaneous with the
\textsl{Advanced Satellite for Cosmology and Astrophysics} (\asca),
optical, and/or radio observations. The observation log, with fluxes in
selected energy bands, is presented in Table~\ref{tab:log_table}.  Assuming
isotropic emission, a $4\,M_\odot$ compact object mass, and a distance of
4\,kpc, the 3--200\,keV luminosities of the hard state observations listed
in Table~\ref{tab:log_table} correspond to $2\times10^{-3}$--$0.03~L_{\rm
  Edd}$.

The observations were chosen for analysis as follows.  Eight of these
observations (P20181) were originally extensively analyzed by us
\cite{wilms:99aa,nowak:99c}. We had found that those data, with the
exception of the faintest observation, showed few differences among their
variability properties \cite{nowak:99c}, and that their spectra could be
well-fit by either a sphere+disc coronal model, or by a power-law plus
ionized reflection model \cite{wilms:99aa}.  For the latter spectral model,
we did not find a hardness-reflection fraction anti-correlation, but
instead we found a hardness-ionization parameter anti-correlation.  This is
in contrast to the results of Revnivtsev, Gilfanov, \& Churazov
\shortcite{revnivtsev:99a}, who, \emph{analyzing the same data} with a
power law+reflection model (with gaussian smearing applied to mimic
relativistic smearing), did find the suggested anti-correlation.  We have
chosen to reanalyze these observations, as well as seven others (P20056)
that represented the brightest and softest hard state observations
presented by Revnivtsev, Gilfanov, \& Churazov \shortcite{revnivtsev:99a}.

We present seven other observations (P40108) that are original to this
work.  All of these observations were performed simultaneously with radio
observations (see Corbel et al. \nocite{corbel:00a} 2000), two were
performed simultaneously with \textsl{Advanced Satellite for Astrophysics
  and Cosmology} (\asca) observations, and three were performed
near-simultaneously with optical monitoring.  An additional five
observations exist in this series (again, with a number of simultaneous
radio, optical, and \asca\ observations); however, these latter
observations, in contrast to the first seven, are too faint to perform
useful timing analyses.  These fainter observations will be discussed in a
future work (Corbel et al. 2001; in preparation).

\subsection{Outline}\label{sec:outline}

In \S\ref{sec:fits} we discuss the spectral analysis in detail,
concentrating on the results obtained with three different Comptonization
models. Before discussing the analyses in detail, however, we begin with a
comparison of the relative strengths and weaknesses of each of these
models.  In \S\ref{sec:timing} we describe the timing behaviour of \gx, and
correlate this behaviour with the results of our spectral analyses.  Twenty
out of twenty two of our observations occurred in the hard state. In
\S\ref{sec:soft} we describe our analysis of the two soft state
observations.  We assess in \S\ref{sec:crab} the robustness of our fits by
presenting ratios of the \gx\ data to observations of the Crab nebula and
pulsar. We explore the reality of the hardness-reflection fraction
anti-correlation in \S\ref{sec:real}.  In \S\ref{sec:simple}, we present a
new, simple coronal model that reproduces some of the coronal-temporal
correlations that we find in \S\ref{sec:timing}. We discuss the
implications of our results for ADAF models in \S\ref{sec:adaf}, and we
briefly discuss alternative models to thermal Comptonization in
\S\ref{sec:alternative}.  We summarize our results in \S\ref{sec:summary}.

Our results rely on many observations with several satellites. In order not
to deter the reader from the results of the analysis with the
technicalities of the data extraction, we present our data extraction
strategies separately in Appendix~\ref{sec:anal}.

\section{Spectral Fits}\label{sec:fits}
\subsection{The Blind Touching the Elephant: The Relative Merits of Various
  Coronal Models}\label{sec:blind}

Here we discuss the Comptonization codes of Poutanen \& Svensson
\shortcite{poutanen:96b}, Coppi \shortcite{coppi:99a}, and Dove, Wilms \&
Begelman \shortcite{dove:97a}. Although each has been used separately to
fit various BHC spectra, this work is the first to attempt to
systematically compare several of them.  All three will be used in fits of
the \gx\ data.  Each model has unique attributes and considers one or more
facets of the Comptonization problem more `correctly' or `completely' than
the others; however, none are truly fully self-consistent models.

The code of Poutanen \& Svensson \shortcite{poutanen:96b} ({\tt compps} in
the X-ray spectral fitting package, XSPEC; Arnaud \nocite{arnaud:96a} 1996)
is essentially a `one-zone' model in that it considers injection of (either
blackbody or multi-temperature disc blackbody; Mitsuda \nocite{mitsuda:84a}
1994) seed photons with a prescribed flux and a fixed geometry into a
corona of uniform optical depth and temperature and/or uniform distribution
of non-thermal electrons. The corona also has a fixed geometry, although a
variety of geometries are considered. A fraction of the escaping photons
can be reflected off of cold, partially ionized material; however, this
fraction is not self-consistently determined, nor are any of the reflected
photons further reprocessed by the corona.  Smearing of the reflection due
to relativistic disc motion and gravitational effects, however, can be
applied. The Fe edge is modelled in the reflection, but the Fe line is not.
Furthermore, the temperature of the corona and its (photon-photon collision
produced) electron-positron pair optical depth are not self-consistently
calculated, and must be checked \emph{a posteriori}.

The model of Coppi \shortcite{coppi:99a}, available from that author
as the XSPEC model {\tt eqpair}, does calculate a self-consistent
coronal temperature and pair optical depth by using the coronal
compactness (proportional to energy released in the corona divided by
its radius) as the fundamental fit parameter. Furthermore, it allows
one to specify both a thermal and non-thermal compactness for the
corona, as well as a compactness for the (blackbody) seed photons. A
spherical corona with seed photons distributed according to the
diffusion equation is modelled; therefore, \emph{all} the seed photons
pass through the corona.  Reflection is implemented in a similar
manner to the {\tt compps} model, but smearing of the reflection is
not considered.  We will use this model extensively in the following
sections, and we will present coronal compactness, $\ell_{\rm c}$,
instead of coronal temperature, $kT_{\rm e}$, coronal $y$-parameter,
or power-law photon index, $\Gamma$, as a fundamental fit parameter.

The coronal model of Dove, Wilms, \& Begelman \shortcite{dove:97a},
available from those authors as the XSPEC model {\tt kotelp}, utilizes
Monte Carlo spectra calculations that are then stored in an interpolation
grid.  For a given seed photon spectrum and geometry and a given coronal
geometry and initial optical depth, the non-uniform coronal temperature
distribution, pair optical depth, energy balance between corona and disc,
and reflection fraction--- with subsequent
re-Comptonizations/re-reflections of this component--- are all
self-consistently calculated.  In this work we shall use the `sphere+disc'
geometry described by Dove et al. \shortcite{dove:98a}, i.e. a
multi-temperature disc blackbody seed photon distribution with $kT(R)
\propto R^{-3/4}$ and maximum temperature of $kT=150$\,eV emanating from an
outer thin disc surrounding a central spherical corona.  (Other geometries
and seed photon distributions are possible.)  Again, the fundamental
parameter is the coronal compactness, $\ell_{\rm c}$, relative to the disc
compactness (fixed to 1), not the average coronal temperature.  For
$\ell_{\rm c} \aproxlt 1$, the \emph{intrinsic} energy generation of the
disc dominates the system flux.  For $\ell_{\rm c} \aproxgt 1$, the seed
photon spectrum is dominated by hard X-rays reprocessed by the disc
\cite{dove:97b}.  The Fe line and edge are both calculated as part of the
reflection component; however, in current implementations no relativistic
smearing and only solar abundances are considered.

As is evident from the above capsule summaries, the various models do
represent to some extent the ``blind man touching the elephant'', each
feeling out a different portion of the puzzle.  Even though the {\tt
  compps} model represents a `one zone' corona, it is still the most
flexible in terms of geometry of both the corona and the properties of the
reflector.  (Given sufficient computer resources to run grids of models
{\tt kotelp} can be equally flexible.)  The {\tt compps} model, however,
lacks the self-consistent energy balance provided by {\tt eqpair} or {\tt
  kotelp}.  The {\tt eqpair} model has self-consistent energy and pair
balance, but lacks the flexible geometry of the other two models.  The {\tt
  kotelp} model self-consistently determines the coronal energy balance,
the seed photon flux (i.e., intrinsic disc flux vs. reprocessed flux), and
the reflected component, but is the least flexible of the models in terms
of currently available fit parameters.

\subsection{Considerations for Model Fitting}\label{sec:consider}

A complication in utilizing the above models is that degeneracies are
common, especially as regards the issue of reflection. Models of the Fe
line/edge region are strongly affected by assumptions concerning the Fe
abundance, the shape of the seed photon and incident hard spectrum, the
reflection fraction, the reflector's ionization parameter, the coronal
geometry and outflow properties (e.g., Beloborodov \nocite{beloborodov:99a}
1999), and the reflector inclination.  All but the first of this list could
easily vary from observation to observation. (The reflector inclination
could vary due to disc warping effects; Pringle \nocite{pringle:96a} 1996,
Maloney \& Begelman \nocite{maloney:97a} 1997.) Indeed, we have found
adequate fits \emph{for the same observation} with reflection fractions
ranging from 0 to 1, Fe line equivalent widths ranging from 50--400\,eV,
coronal temperatures ranging from 20--200\,keV and coronal optical depths
ranging from 0.1--3.

Another consideration perhaps unfamiliar to those who do not practice X-ray
astronomy and perhaps too readily accepted by those who do, is that model
fits often have reduced $\chi^2$ substantially below unity.  This is in
part due to the adoption of systematic errors (see the Appendix), and
indicates that we are fitting some systematic features of the satellite
response matrices.  Fits to the (presumed power-law) Crab nebula plus
pulsar clearly indicate the presence of systematic errors in the \pca\ 
response. Furthermore, without systematic errors, the low energy \pca\ data
would dominate the fits, while the high energy \hexte\ data, useful for
measuring coronal temperatures and optical depths, would add very little
statistical weight.

Obtaining a reduced $\chi^2$ substantially below one indicates that an
\emph{individual} observation is over-parameterized.  In the fits described
below, we try to minimize any over-parameterization by freezing numerous
parameters, especially in the line region, and we have tended to adopt
fairly small systematic errors in high signal-to-noise regions of the
spectrum (e.g., below 4\,keV).  The reduced $\chi^2$ for the fits we
present are always $>0.6$ (cf. Revnivtsev, Gilfanov, \& Churazov
\nocite{revnivtsev:99a} 2001).  We would consider this too small if we were
fitting a single observation; however, we are concerned with the
\emph{relative} trends of the fit parameters. Here we consider 20 separate,
but uniformly measured and uniformly fit hard state observations.  For
example, small differences in the line region, even when some fraction of
this region represents systematic efects, can be very statistically
significant given the large effective area of the \pca.  \emph{Differences}
in fit parameters, therefore, can be meaningful. Even then we endeavor not
to over-interpret the parameter trends (see, for example, the discussion of
Fig.~\ref{fig:eq_lkt_setc} ).  To give the reader some indication of the
role played by systematic errors, in Table~\ref{tab:kot} we present a set
of fits for observation P20181\_01 (italicized), where we adopt no
systematic errors.  The error bars for parameters associated with narrow
features (e.g., line widths) decrease somewhat.  Overall, however, the
``bestfit'' parameters are not substantially changed, although the $\chi^2$
has greatly increased.

Given the above concerns about model degeneracies and systematic errors, we
explicitly examine the ratios of our data to those from the Crab nebula and
pulsar.  We discuss the implications of these comparisons for the
interpretation of our fits in \S\ref{sec:crab}.  We further discuss the
relative parameter trends in the sections below.

\subsection{PCA Fits: {\tt eqpair+gauss}}\label{sec:pca_eqpair}

As discussed by Wilms et al. \shortcite{wilms:99aa}, there typically has
been a difference between the best-fit photon indices for \pca\ and \hexte\ 
observations of the Crab pulsar and nebula of $\Delta \Gamma \aproxgt 0.1$.
Recent \pca\ response matrices have slightly decreased this discrepancy;
however, differences still remain.  Partly for this reason, numerous
authors have chosen not to include the \hexte\ data in their fits of GBHC
spectra (e.g., Revnivtsev, Gilfanov, \& Churazov \nocite{revnivtsev:99a}
2001).  Here we briefly explore whether this is justified.

For our fits to \pca\ data, we have applied the {\tt eqpair} Comptonization
model with an additional gaussian line profile with a peak energy fixed at
6.4\,keV. We further have fixed the blackbody seed photon compactness to be
1, the reflection to be from cold, neutral material with solar abundances,
the angle of the reflecting medium with respect to our line of sight to be
$45^\circ$, and the compactness of the non-thermal electrons to be 0.  The
fit parameters are therefore the compactness of the thermal electrons, the
coronal optical depth, the seed photon blackbody temperature, the overall
normalization, the line width, and the line amplitude. Folding in the
systematic uncertainties described in the Appendix, these models yield
extremely good fits to all the hard state observations, with typical
reduced $\chi^2 \sim 0.3$.  The fitted lines have widths of $\sigma\approx
1$\,keV and equivalent widths ranging from 100--300\,eV.  The reflection
fractions range from $\Omega/2\pi \approx 0.3$--0.6.  The seed photon
temperature is typically $\approx 600$\,eV, the coronal optical depth is
typically $\approx 2$, and the coronal compactness is typically $\ell_{\rm
  c} \approx 1$--7, which here implies coronal temperatures of tens of keV.

\begin{figure}
\centerline{
\includegraphics[width=0.35\textwidth,angle=270]{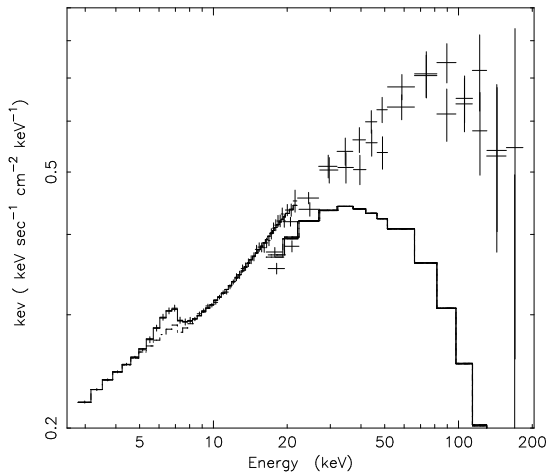}
}
\caption{\small Unfolded energy spectrum of \gx\ (P40108\_03) 
  with the {\tt eqpair} Comptonization model fitted \emph{to the \pca\ data
    only}. The \hexte\ data, assuming the model fit to the \pca\ data, is
  also shown assuming a relative normalization constant of 0.9 (see
  Appendix) for both the \hexte\ A and B clusters as compared to the \pca.
  (Here error bars are 1-$\sigma$.) \protect{\label{fig:eq_unfold}}}
\end{figure}

Fig.~\ref{fig:eq_unfold}, which presents the unfolded spectrum for
observation P40108\_03, shows that these fits fail utterly to reproduce the
\hexte\ data, even accounting for likely systematic differences between
\pca\ and \hexte.  The `unfolded spectrum' shows the best fitting model,
here plotted as $E^2\times F_\gamma(E)$, multiplied by the ratio of the
data in detector space to the spectral model folded through the
experimental response function of the detectors.  It is important to note
in Fig.~\ref{fig:eq_unfold} that the apparent visual significance of narrow
spectral features should not be taken too strongly.  An `unfolded spectrum'
has a tendency to exaggerate the \emph{assumed} spectral features,
especially those with widths comparable to or narrower than the
instrumental resolution.  In this instance, however, the unfolded spectrum
does highlight the complete inadequacy of the extremely well-fitting \pca\ 
model in describing the broad-band \hexte\ data.

\subsection{PCA+HEXTE Fits: {\tt kotelp} Models}\label{sec:kotelp}

\begin{figure*}
\centerline{
\includegraphics[width=0.44\textwidth]{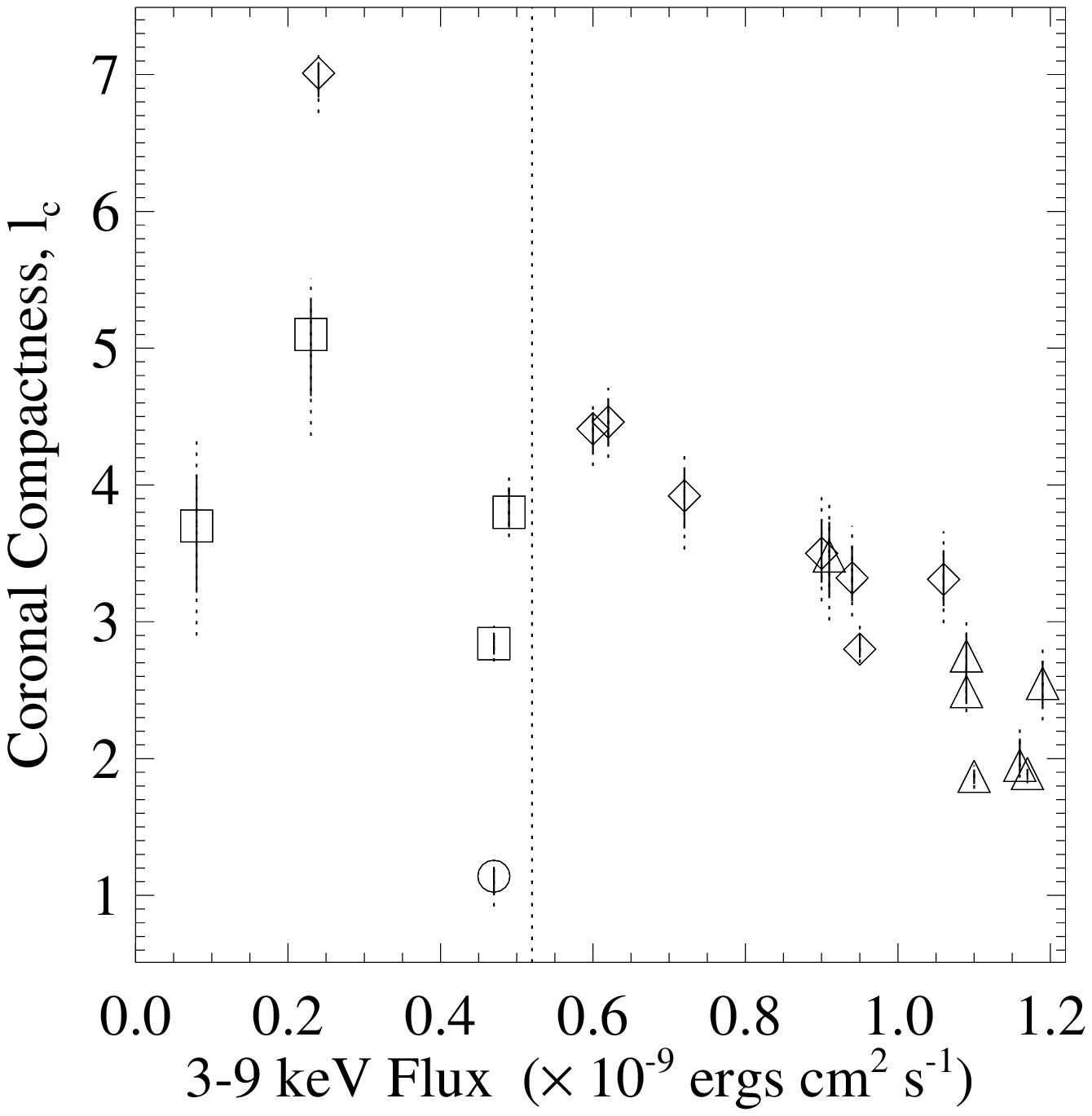}}
\centerline{
\includegraphics[width=0.44\textwidth]{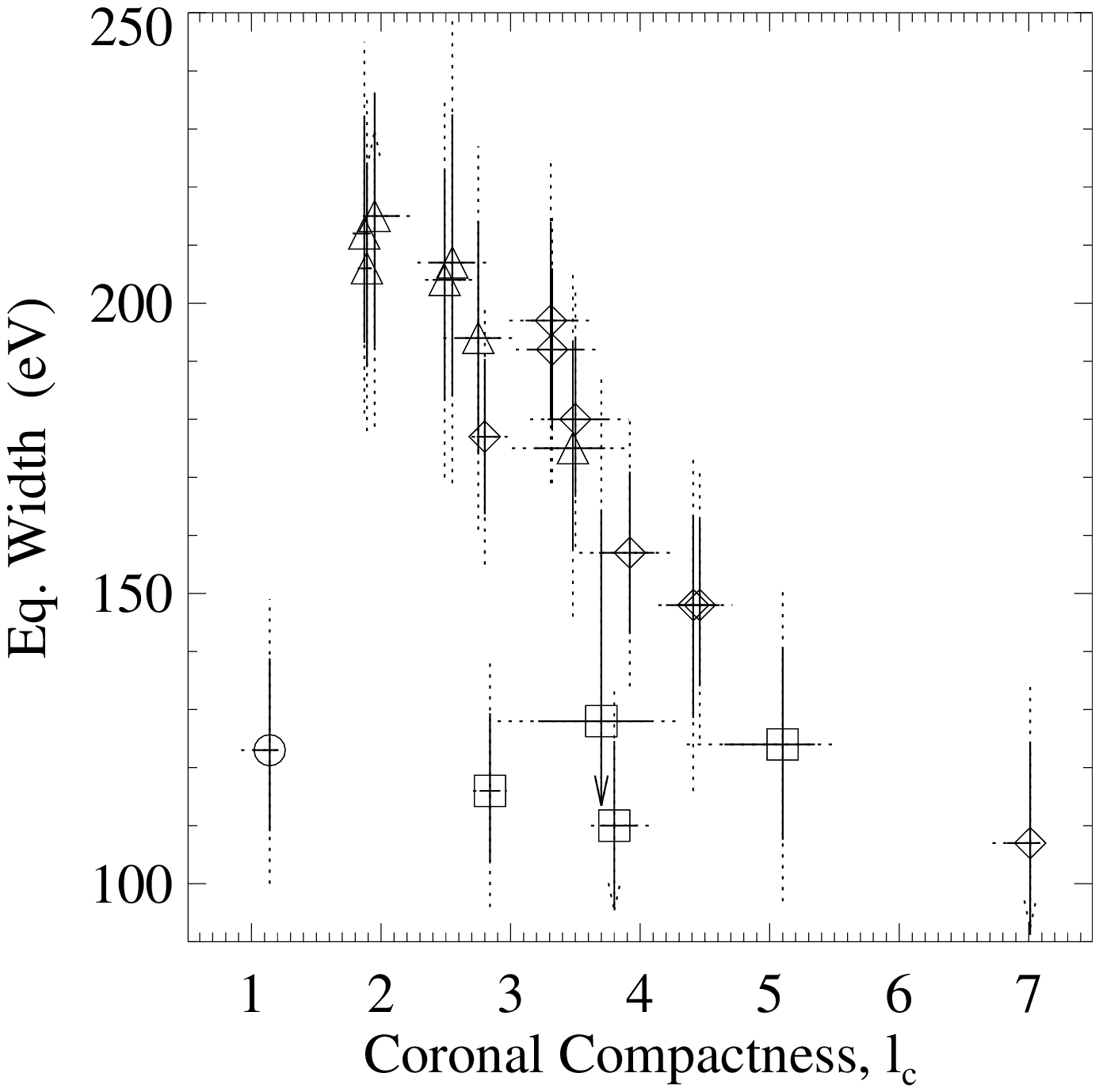}
}
\caption{\small Results of the {\tt kotelp} model
  fit to joint \pca\ and \hexte\ data of \gx.  {\it Top:} Coronal
  compactness vs.  \pca\ flux in the 3--9\,keV band.  {\it Bottom:}
  Equivalent width of the Fe line vs. coronal compactness.  Throughout this
  work, unless otherwise noted, diamonds represent the 1997 hard state
  observations previously discussed by Wilms et al.  (1999).  Triangles
  represent 1997 hard state observations discussed by Revnivtsev, Gilfanov,
  \& Churazov (2001).  Squares represent 1999 hard state observations.
  Circles represent the first 1999 hard state observed after \gx\ returned
  from a protracted soft state.  The dotted vertical line represents the
  lowest 3--9\,keV \pca\ flux observed for the soft state observations
  discussed here.  The dotted error bars are the 90\% confidence limit
  (i.e., $\Delta \chi^2 = 2.71$ for one interesting parameter), and the
  solid error bars are equal to the 90\% confidence level error bars
  divided by $\sqrt{2.71}$.  \protect{\label{fig:kot_gss}}}
\end{figure*}

We fit the simultaneous \pca\ and \hexte\ data with the {\tt kotelp} model
with a `sphere+disc' geometry as described by Dove et al.
\shortcite{dove:98a}.  Here we choose to average the spectrum over
inclination angles $\mu \equiv \cos \theta = 0.4$--0.6.  These models do
not adequately fit the data, and typically produce reduced $\chi^2 > 3$,
primarily due to large, broad residuals in the Fe line/edge region, as one
might expect for a relativistically broadened line.  (We previously had
noted this for {\tt kotelp} fits to \rxte\ data of Cyg~X-1; Dove et al.
\nocite{dove:98a} 1998.)

In order to mimic such a line for the {\tt kotelp} model, we performed a
series of fits where we added a narrow, fixed energy and negative amplitude
gaussian function to remove the $\approx 30$\,eV equivalent width (EW) line
produced by the code.  We then added a broad gaussian function with peak
energy fixed at 6.4\,keV, and refit the models.  Fitted parameters are
presented in Table~\ref{tab:kot} in the Appendix.  These models produced
markedly improved fits, with reduced $\chi^2$ ranging from 0.7--2.5, with
an average reduced $\chi^2$ of 1.6.  The three faintest observations fit
best. Most observations show negative residuals, as might be expected from
an unmodelled edge, near 10\,keV.  A large fraction of $\chi^2$, however,
comes from the 3--4\,keV region, where 1999 observations show positive
residuals, i.e., a soft excess, and 1997 observations show negative
residuals.  

We note that if we instead fit near edge-on {\tt kotelp} models, the P20056
fits improve by $\Delta \chi^2 \approx 10$, since the contribution of the
seed photon spectrum is reduced compared to the Comptonized spectrum, and
hence the negative 3--4\,keV residuals are reduced.  The P20181
observations fit approximately the same, but the 1999 fits become much
worse as the 3--4\,keV excess increases.  If we instead choose a nearly
face-on {\tt kotelp} model, as the seed photon spectrum is now emphasized
compared to the Comptonized spectrum, the positive 3--4\,keV residuals are
then reduced. Such fits to the 1999 observations improve by $\Delta \chi^2
\approx 30$.  The fits to the 1997 observations, however, become much
worse. We discuss these points further in \S\ref{sec:eqpair},
\S\ref{sec:asca}, and \S\ref{sec:crab}.

Fitted compactnesses range from 1--7, and fitted optical depths range from
1--1.6.  These parameters translate to average coronal temperatures of
90--170\,keV.  (Note, however, that face-on {\tt kotelp} models produce
lower compactnesses and higher optical depths, resulting in a corona with a
lower temperature; see \S\ref{sec:asca}.)  Equivalent widths of the added
broad line range from 110--220\,eV, and widths of the lines range from
$\sigma \approx 0.4$--0.9.

Two trends stand out from these fits.  As Fig.~\ref{fig:kot_gss} shows,
there is a strong anti-correlation between compactness and 3--9\,keV \pca\ 
flux for the 1997 hard state observations.  In 1999, \gx\ is seen to have
returned from the soft state with an initially very soft spectrum that
hardened as the source faded, before beginning to soften slightly at very
low flux levels.  As previously noted with {\it Ginga} data
\cite{miyamoto:95a}, there is clear evidence for hysteretic behaviour. In
1997, \gx\ achieved a higher 3--9\,keV flux, \emph{without entering the
  soft state}, than the lowest 3--9\,keV soft state flux measured shortly
prior to the return to the hard state in 1999.  The `overlap region'
between the hard and soft states appears to be in the 3--9\,keV \pca\ flux
range of $\approx 0.5$--$1.2\times10^{-9}~{\rm ergs ~cm^2~s^{-1}}$.  Of
course, the issue of the ordering of the bolometric luminosities is less
clear, especially since we have not measured the flux below 3\,keV, which
likely accounts for the largest fraction of the soft state flux.  The
observations represented by triangles in these figures are among the
highest flux levels ever detected for the hard state of \gx, whereas other
`high state' and `very high state' observations have exhibited
substantially greater 3--9\,keV fluxes \cite{belloni:99a,miyamoto:91a}.

The second correlation that stands out, also shown in
Fig.~\ref{fig:kot_gss}, is the anti-correlation between the equivalent
width of the broad line and the coronal compactness.  In previous fits to a
fraction of the 1997 hard state observations (diamonds; Wilms et al.
\nocite{wilms:99aa} 1999), utilizing \pca\ data only, such a correlation
was \emph{not} detected.  Aside from not including \hexte\ data, those
previous fits utilized an earlier version of the \pca\ response matrix with
larger systematic uncertainties.  Here to some extent the Fe line
equivalent width might be serving as a reasonable proxy for reflection
fraction.  If this is the case, however, then whereas a correlation is seen
prior to the 1997 transition to the soft state, the correlation is
apparently absent after the 1999 return to the hard state.

We considered two other combinations of models with the basic {\tt kotelp}
model in order to assess the robustness of the above correlations.  Here we
shall discuss only the highlights of the results.  For both of these
models, the line produced by the {\tt kotelp} code was again subtracted as
above.  In the first case, we added the XSPEC {\tt laor} model
\cite{laor:91a}, which gives the line profile for a relativistically
broadened line from a disc rotating about a Kerr black hole\footnote{To
  some extent, the broad red tail of the {\tt laor} line can offset
  differences between the \hexte\ and systematically softer \pca\ 
  detectors. A hard power law extrapolated backward from the \hexte\ energy
  band to the \pca\ energy band could leave a soft excess, even if in
  reality no such excess existed.}.  These models improved the reduced
$\chi^2$ of the fits to 0.7--2.1, with an average reduced $\chi^2$ of 1.2.
(The observation immediately following the return to the hard state,
P40108\_03, produced the worst fit, again due to a soft excess at $\aproxlt
4$\,keV.)  The line energy, which was allowed to be a free parameter,
ranged from 6.5--6.8\,keV, and the line equivalent widths ranged from
150--350\,eV.  There essentially was no change from any of the trends and
correlations discussed above.

For the second model, in addition to using the {\tt laor} model to describe
the line region, we also added an ionization edge at $\approx 8$\,keV.
This further improved the fits to a reduced $\chi^2$ ranging from 0.7--1.4,
with an average of 0.9.  (As before, observation P40108\_03 was the most
poorly fit due to a soft excess at energies $\aproxlt 4$\,keV.)  The line
energies were as above and their equivalent widths ranged from
100--270\,eV. The additional ionization edge had an energy that ranged from
8.0--9.5\,keV and that also was positively correlated with the {\tt laor}
line energy.  The edge optical depth ranged from $\tau=0.02$--0.07.
Although the line equivalent width correlations discussed above were
slightly weakened, the trends shown in Fig.~\ref{fig:kot_gss} again remain
basically unaltered.  Taking the above results as a potential signature of
reflection changes, we turned to fits with the {\tt eqpair} model.

\subsection{PCA+HEXTE Fits: {\tt eqpair} Models}\label{sec:eqpair}

\begin{figure*}
\centerline{
\includegraphics[width=0.35\textwidth]{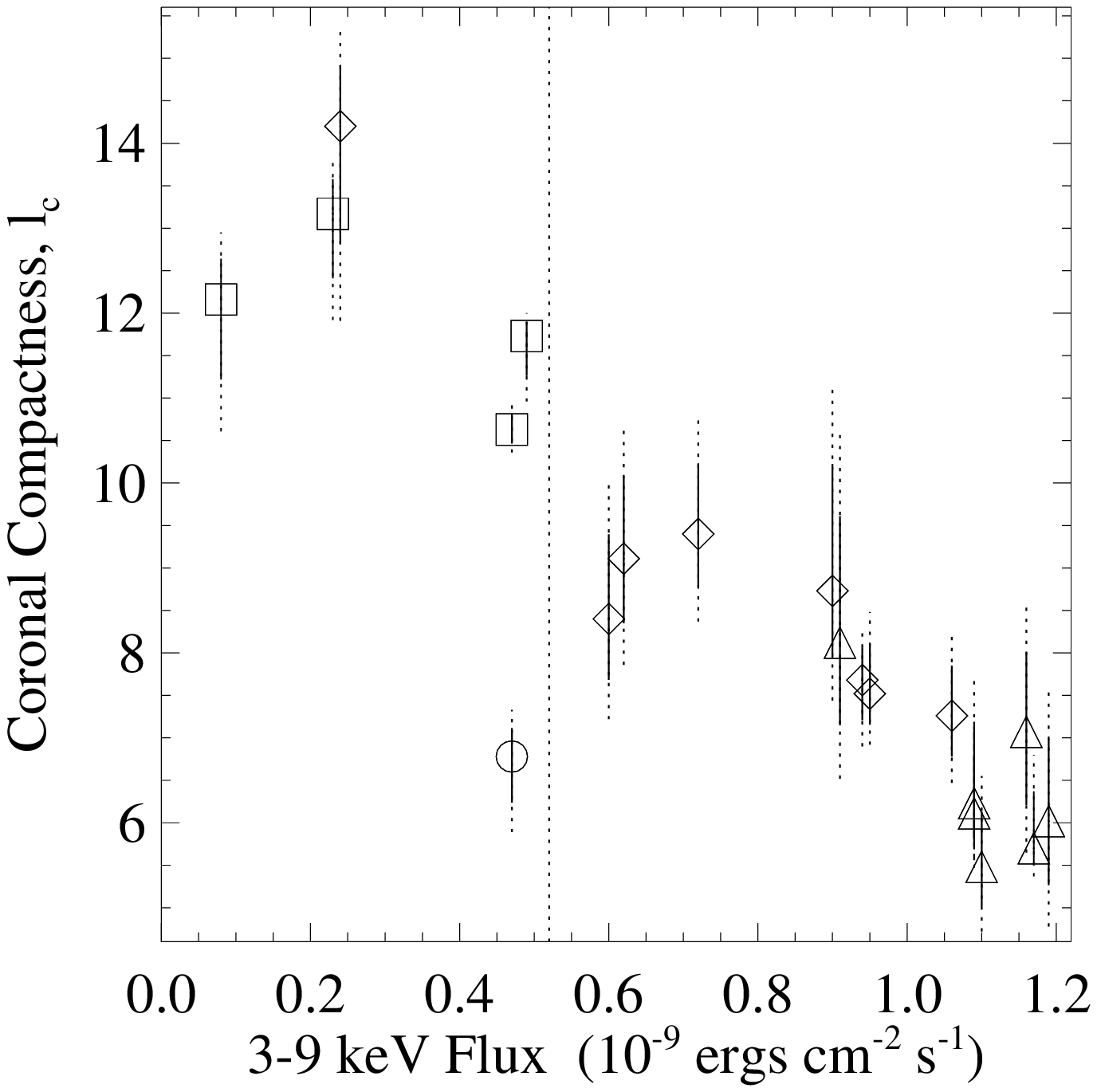}
\includegraphics[width=0.35\textwidth]{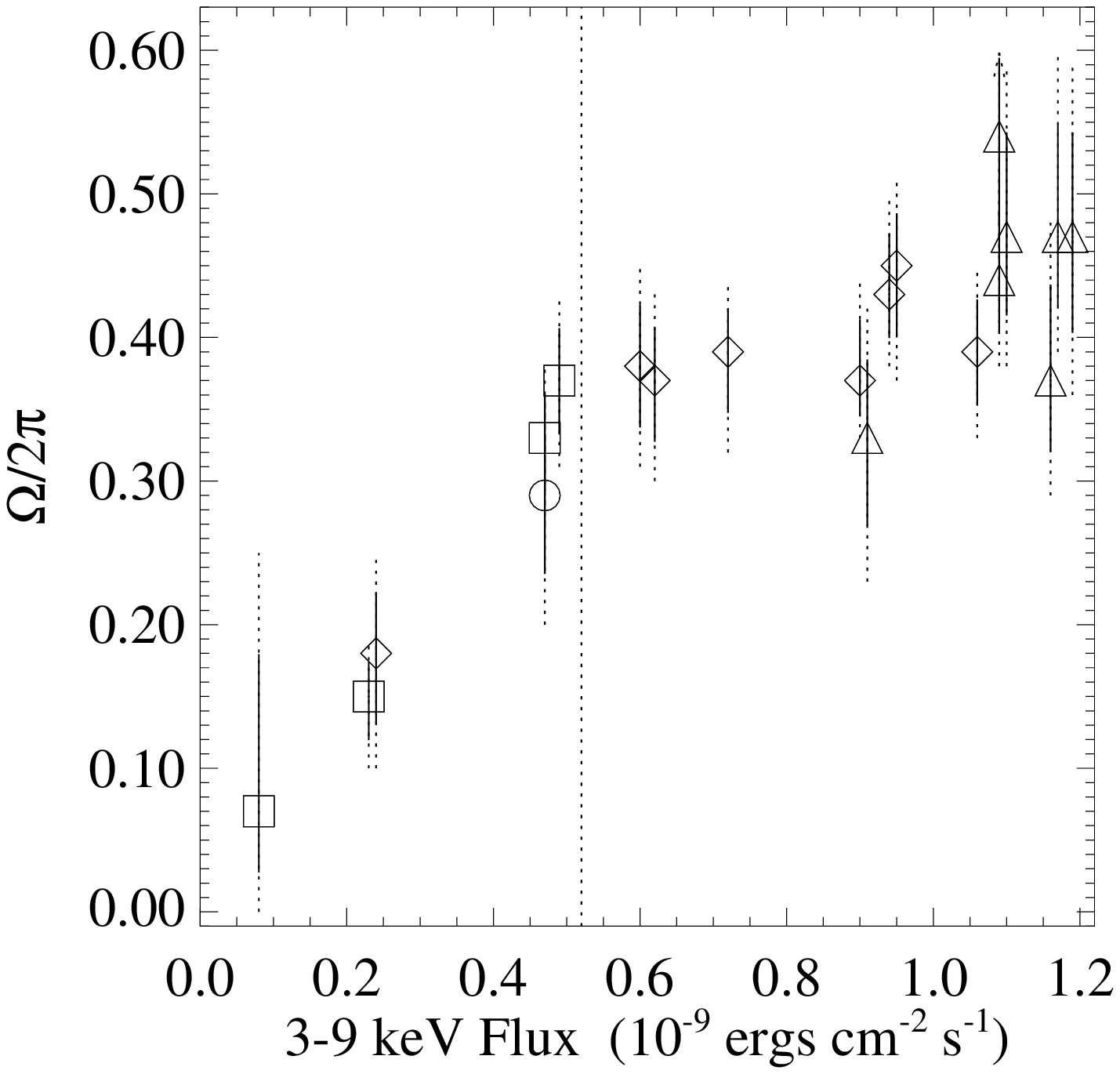}
\includegraphics[width=0.35\textwidth]{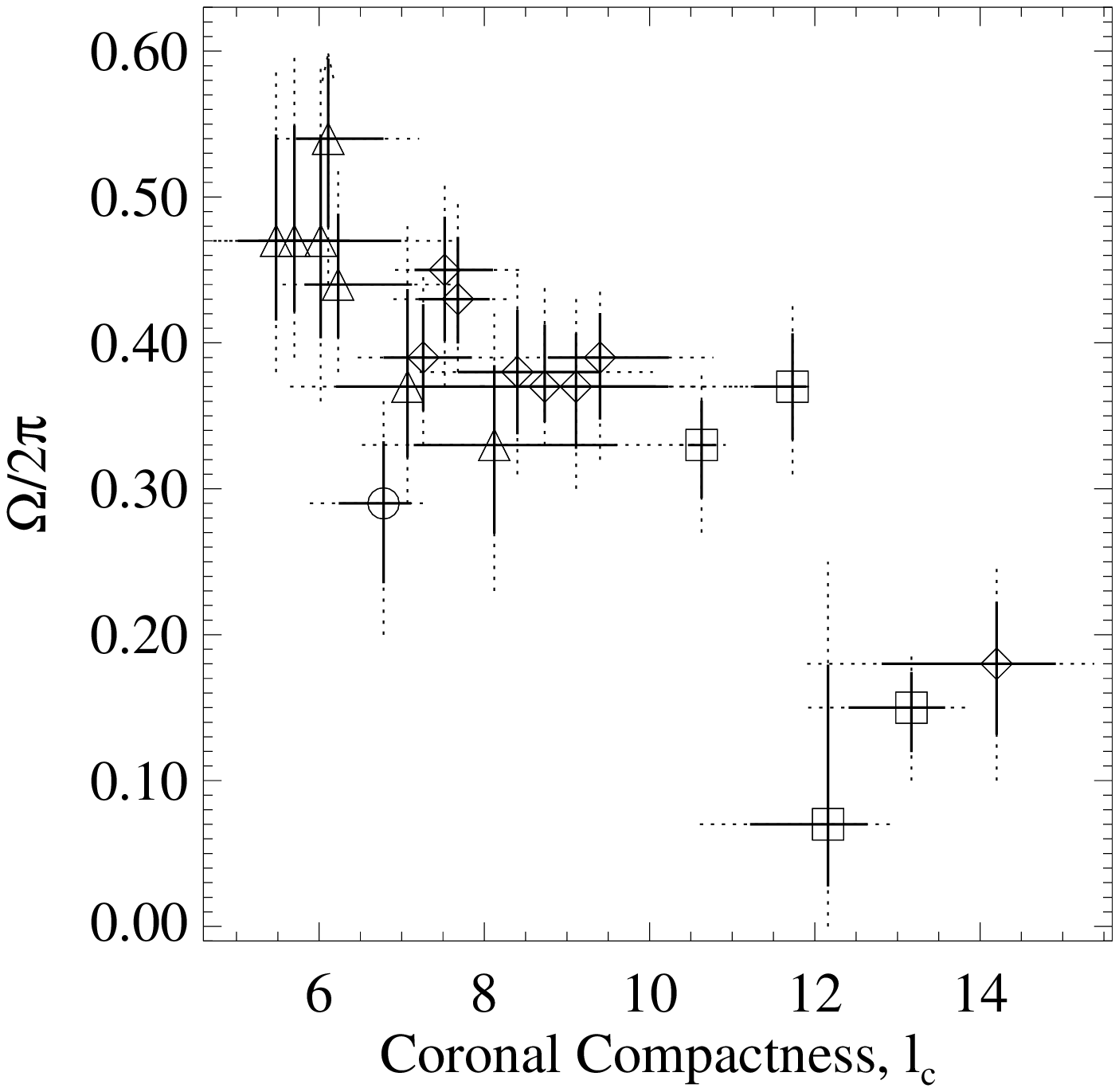}
}
\caption{\small Results of the {\tt eqpair} model
  fit to joint \pca\ and \hexte\ data of \gx.  {\it
    Left:} Coronal compactness vs.  \pca\ flux in the 3--9\,keV band.  {\it
    Middle:} Reflection fraction vs. \pca\ flux in the 3--9\,keV band.
  {\it Right:} Reflection fraction vs. coronal compactness.
  \protect{\label{fig:eq_lkt_seta}}}
\end{figure*}

\begin{figure*}
\centerline{
\includegraphics[width=0.35\textwidth]{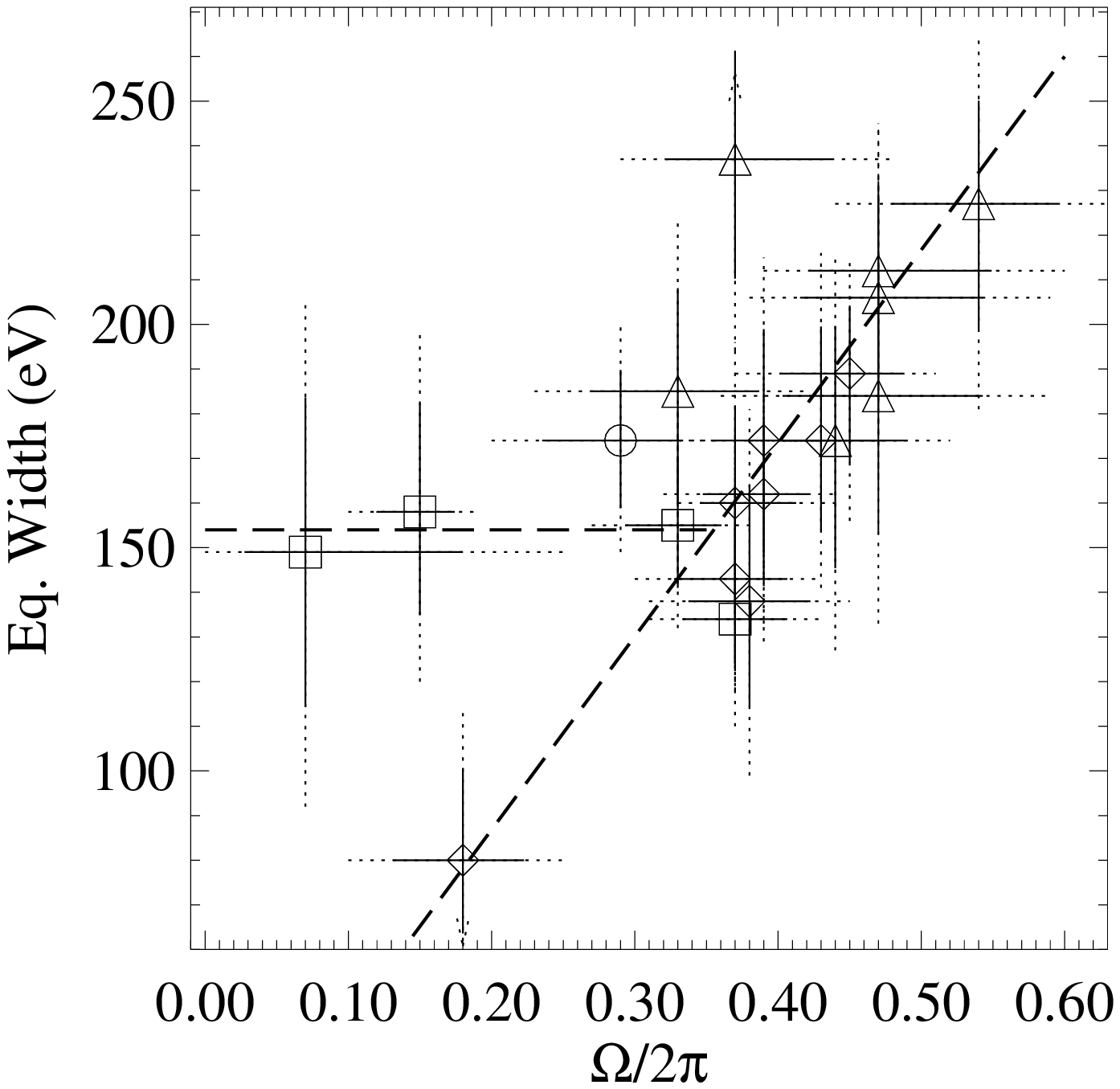}
\includegraphics[width=0.35\textwidth]{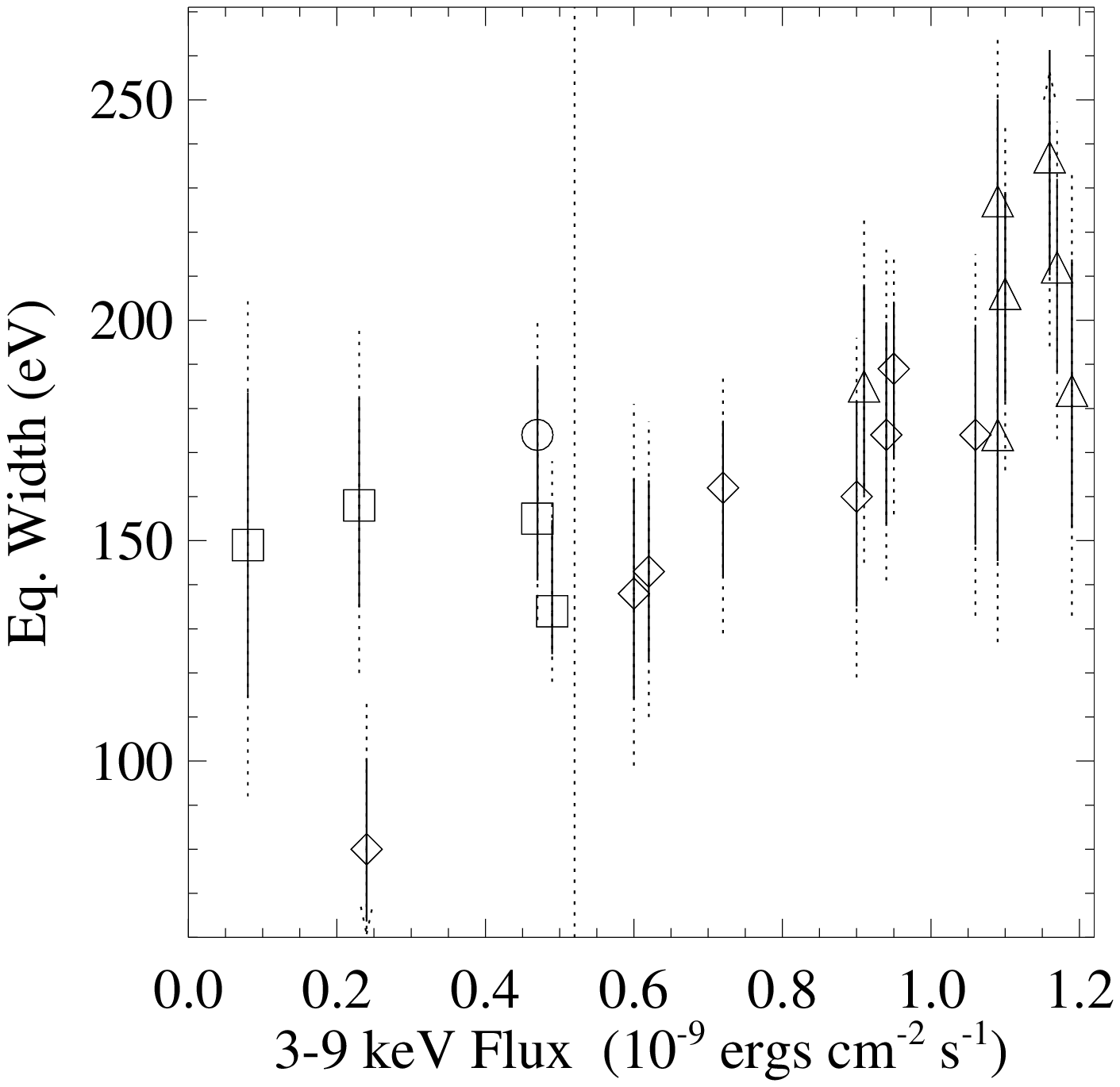}
\includegraphics[width=0.35\textwidth]{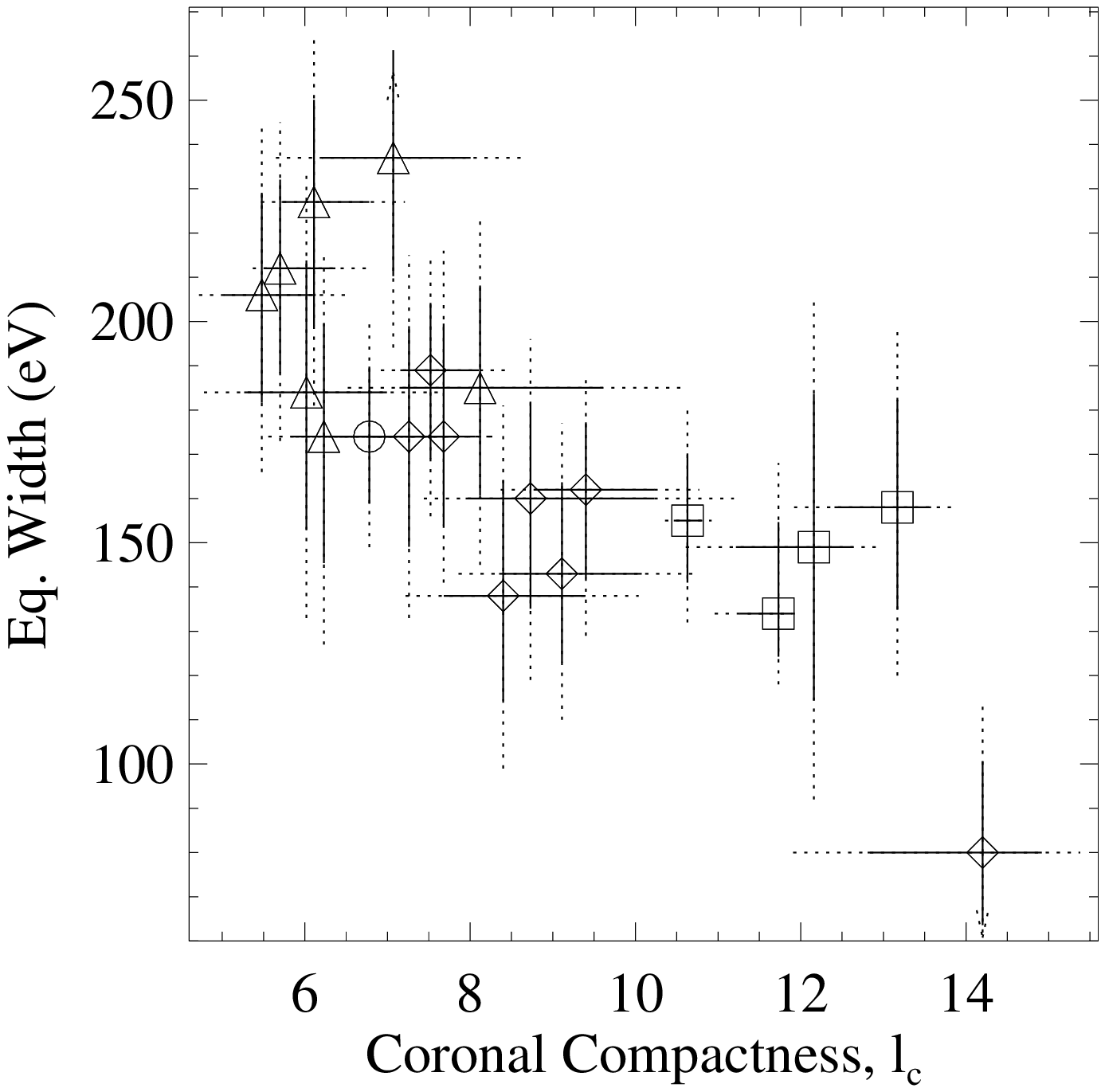}
}
\caption{\small Results of the {\tt eqpair} model
  fit to joint \pca\ and \hexte\ data of \gx.  {\it
    Left:} Equivalent width of the Fe line vs.  reflection fraction,
  $\Omega/2\pi$.  Dashed lines show: the expected trend if the line
  equivalent width is proportional to reflection fraction (with
  0 intercept), and the mean line equivalent width for the P40108
  observations. {\it Middle:} Equivalent width of the Fe line vs. \pca\ 
  flux in the 3--9\,keV band.  {\it Right:} Equivalent width of the Fe line
  vs. coronal compactness.  \protect{\label{fig:eq_lkt_setb}}}
\end{figure*}

\begin{figure*}
\centerline{
\includegraphics[width=0.44\textwidth]{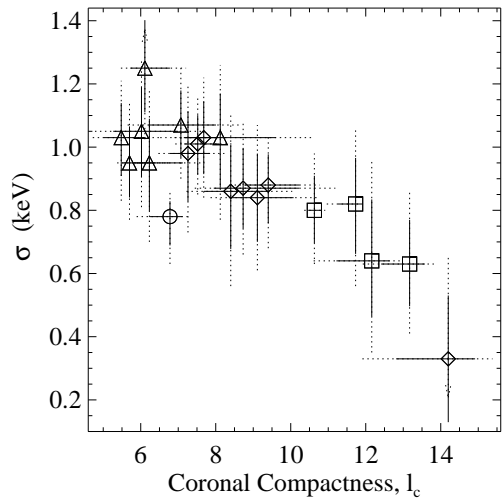}
}
\caption{\small Width of the Fe line vs. the coronal 
  compactness, for the {\tt eqpair} model fit to
  joint \pca\ and \hexte\ data of \gx.  \protect{\label{fig:eq_lkt_setc}}}
\end{figure*}

For the joint fits to the \pca\ and \hexte\ data discussed here, we used
the exact same {\tt eqpair} model as in \S\ref{sec:pca_eqpair}, except that
we instead fixed the Fe abundance to be 4 times solar, which is more
consistent with the large line equivalent widths but low reflection
fractions.  (Lower or higher abundances, however, produced only slightly
worse fits.)  We found acceptable fits to the joint data with parameters
very similar to the \pca\ only fits (\S\ref{sec:pca_eqpair}); however, a
completely different set of parameters described the joint data slightly
better.  These models, with a reduced $\chi^2 \approx$ 0.6--1.0, had seed
photon temperatures in the range of 30--100\,eV, coronal compactnesses in
the range of 5--14 (yielding coronal temperatures of $\sim 200$\,keV, given
the typically fitted optical depth of 0.1--1), and reflection fractions of
0.1--0.5.  Line widths were broad, and line equivalent widths were $\approx
80$--240\,eV.  Since these are considered to be the more ``usual
parameters'' (see Poutanen, Krolik, \& Ryde \nocite{poutanen:97b} 1997), we
present these results in Figs.~\ref{fig:eq_lkt_seta}--\ref{fig:eq_lkt_setc}
and give the parameters in Table~\ref{tab:kot}.

As shown in Fig.~\ref{fig:eq_lkt_seta}, the same basic pattern of
compactness vs. 3--9\,keV flux was seen, i.e., higher fluxes yielded softer
spectra, and \gx\ returned in 1999 from the soft state with an initially
soft spectrum.  The variation, however, is less pronounced, especially on
the return from the soft state in 1999.  This is because for the {\tt
  eqpair} model the seed photon temperature is not a fixed quantity.  The
relatively soft spectra for the first three 1999 hard state observations
(P40108\_03--\_05) are modelled with a combination of both low compactness
and low seed photon temperature ($\approx 30$\,eV).  We note also that
these three spectra indicate coronae that are pair-dominated, as the fitted
initial electron optical depth was $\tau_{\rm es} = 0.01$ (we set this value as
a lower limit to the fits), but the total optical depths were $\tau \approx
0.3$.  The error bars on the initial electron optical depths, however, were
large, and non-pair-dominated plasmas are allowed.  The soft positive
residuals seen for observation P40108\_03 with the {\tt kotelp} model fits
were absent with the {\tt eqpair} fit.

Reflection fraction and Fe line correlations are readily noticed in
Figs.~\ref{fig:eq_lkt_seta} and \ref{fig:eq_lkt_setb}.  The most striking
pattern is the reflection fraction, $\Omega/2\pi$, vs. the 3--9\,keV \pca\ 
flux.  There is a positive correlation of fitted reflection fraction with
flux, except in the overlap/hysteretic region between the hard and soft
states.  There the reflection fraction is quasi-independent of flux, except
at the very highest flux levels (triangles).  Correlating the reflection
fraction with compactness is less clear.  Ignoring the P20056 observations
(triangles), and the 3 lowest flux observations (P20181\_05,
P40108\_06--\_07), no correlation would be seen between reflection fraction
and compactness, as discussed by Wilms et al. \shortcite{wilms:99aa}.  Only
by including the highest flux (triangles) and lowest flux observations does
a possible anti-correlation between compactness and reflection fraction
appear.  Furthermore, the faintest three observations, taken by themselves,
indicate exactly the opposite trend.

The Fe line equivalent width, as for the reflection fraction, is
anti-correlated with compactness if one considers all the observations, but
does not show this anti-correlation if one ignores the brightest
(triangles) and faintest (P20181\_05) observation (see Wilms et al.
\nocite{wilms:99aa} 1999).  This is also seen by correlating the Fe line
equivalent width with the reflection fraction.  Prior to the 1997
transition to the soft state, reflection fraction and Fe line equivalent
width are correlated, after the 1999 return to the hard state, there is a
lower limit to the Fe line equivalent width of $\approx 150$\,eV.  This is
comparable to what we found for the {\tt kotelp} fits discussed in
\S\ref{sec:kotelp}, except that there the 1999 post-transition equivalent
widths were consistent with the lowest equivalent width measured prior to
the 1997 soft state transition.

There also is an apparent anti-correlation between physical line width,
$\sigma$, and coronal compactness, as shown in Fig.~\ref{fig:eq_lkt_setc}.
This is similar to the anti-correlation between spectral hardness and
`smearing width' claimed by Revnivtsev, Gilfanov, \& Churazov
\shortcite{revnivtsev:99a}.  The low reduced $\chi^2$ of our fits, however,
indicate that we may already be over-parameterizing the data.  Furthermore,
this anti-correlation is greatly weakened if we fit the brightest
observations (triangles) with fixed line energies of 6.7\,keV instead. The
line widths then decrease by $\Delta\sigma \approx 0.2$\,keV.

\subsection{ASCA Observations}\label{sec:asca}

Two of the \rxte\ observations discussed in this work, P40108\_04 and
P40108\_05, were obtained simultaneously with \asca.  Compared to
\rxte, \asca\ had a narrower band pass (effectively 1--10\,keV for
observations obtained as late as 1999), an approximately 4 times
smaller effective area, but a much higher spectral resolution
($\aproxlt 0.2$\,keV for these observations) and a lower background
due to the nature of its instruments and its arcminute spatial
resolution.  Furthermore, as it is an independently calibrated
instrument (see the mission web pages at {\tt
  http://heasarc.gsfc.nasa.gov/docs/asca/ascagof.html} for a
discussion of \asca\ calibration issues), it provides a useful check
of the \rxte\ results.

Analyzed independently of \rxte, these two \asca\ observations can be
fitted with models very similar to those employed by Wilms et al.
\shortcite{wilms:99aa} to study three archival \asca\ observations of \gx.
Specifically, these observations are well fit by a multi-temperature disc
blackbody spectrum with peak temperature of $\approx 0.15$\,keV, a broken
power law with a photon index $\sim 1.8$ below 4\,keV and $\sim1.6$ above,
and a gaussian line at 6.4\,keV\footnote{Note that, consistent with the
  analysis of \asca\ observations of Cyg~X-1 presented by Ebisawa et al.
  \shortcite{ebisawa:97a}, Wilms et al. \shortcite{wilms:99aa} searched for
  fits with \emph{narrow} ($\sigma \sim 0.1$\,keV) Fe lines.  The fitted
  lines had equivalent widths $\approx 40$\,eV, and the fits yielded
  reduced $\chi^2 \approx 1$ for the faintest observations, but 1.4 for the
  brightest. As compared to the \asca\ observations discussed here, the
  observations discussed by Wilms et al. \shortcite{wilms:99aa} had
  3--9\,keV fluxes that were approximately 0.3, 0.5, and 1.8 times as
  great.}  These observations prefer Fe lines with equivalent widths of
$\approx 220$\,eV and that are broad ($\sigma \approx 0.8$\,keV),
although the exact parameter values for the line do depend upon the fitted
continuum model.

\begin{figure*}
\centerline{
\includegraphics[width=0.44\textwidth,angle=270]{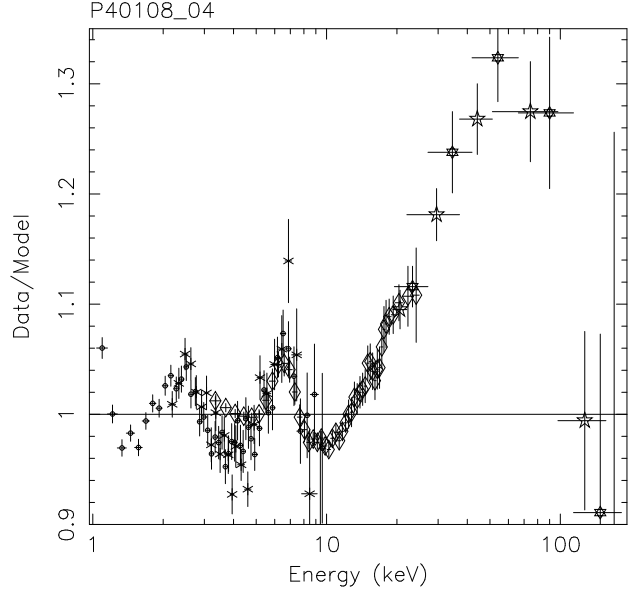}
}
\caption{Residuals for an absorbed power law model fit simultaneously to 
  the \sis, \gis, and \pca\ data (P40108\_04).  Circles and crosses are the
  \gis\ and \sis\ data, respectively.  Diamonds are the \pca\ data. Stars
  are the \hexte\ data (not included in the fit), with normalization
  constants chosen to match the \pca\ residuals. (Error bars here are
  1-$\sigma$ errors.)  \protect{\label{fig:ratios}}}
\end{figure*}

One might hope that including the \rxte\ data would remove enough of the
ambiguities in the continuum model so as to further constrain the line
model.  In Fig.~\ref{fig:ratios} we plot the residuals (detector space data
divided by the model folded through the response matrices of the detectors)
for an absorbed power law ($N_{\rm H}$ fixed to $6 \times 10^{21}~{\rm
  cm^{-2}}$) fit to 1--22\,keV data (1--10\,keV in \gis, 2--10\,keV in
\sis, and 3--22\,keV in \pca). Although both \asca\ and \pca\ agree on the
presence of a strong, broad line-like feature near 6.4\,keV, there are
clear systematic differences among the \gis, \sis, \pca, and \hexte\ 
detectors.  As we note in the Appendix, \pca\ and \hexte\ yield different
power law indices for observations of the Crab nebula and pulsar, with
\hexte\ giving a systematically harder index.  The \asca\ detectors were
calibrated to yield Crab indices comparably hard to that found by \hexte.
Furthermore, both the \sis\ and \gis\ detectors show a break near 4\,keV
that is absent in the \pca\ data, as well as additional spectral structure
near 1 and 2\,keV.

\begin{figure*}
\centerline{
\includegraphics[width=0.44\textwidth]{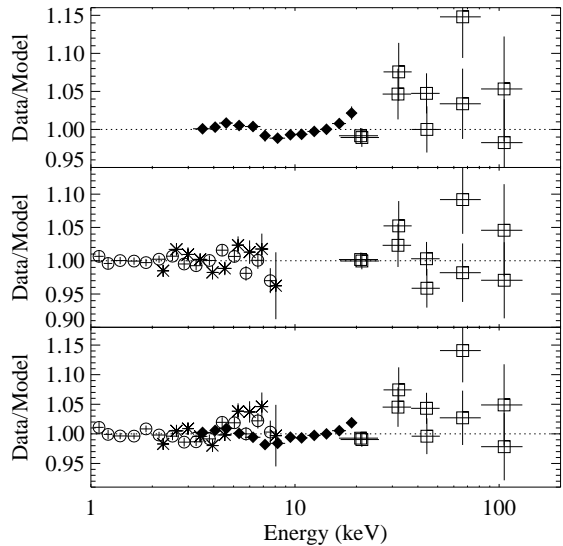}
}
\caption{Residuals for corona plus dust scattering halo models (see text)
  fit to: \pca\ and \hexte\ data (top); \gis, \sis, and \hexte\ data
  (middle); and \gis, \sis, \pca, and \hexte\ data (bottom). (Error bars
  here are 1-$\sigma$ errors.)  \protect{\label{fig:dust}}}
\end{figure*}

Ignoring the \pca\ data, data from the \gis, \sis, and \hexte\ detectors
can be reasonably well-fit with the same disc blackbody, broken power law,
plus gaussian line models mentioned above, but with an additional caveat.
The power law above 4\,keV (\emph{uniform} from \asca\ through \hexte\ 
energy bands) is exponentially cutoff in the \hexte\ bands at energies only
$\aproxgt 80$\,keV (i.e., the {\tt highecut} model in XSPEC; see also Wilms
et al. \nocite{wilms:99aa} 1999).  The reduced $\chi^2$ for such models are
$\approx 1.3$, with the greatest deviations coming from the 8--10\,keV and
17--25\,keV regions, where the model over-predicts the flux.  An even
better fit to just the \asca+\hexte\ data, however, can be obtained with a
near face-on {\tt kotelp} model with the addition of a dust scattering
halo, as is known to exist in this system from the work of Predhel et al.
\shortcite{predhel:91a}, and references therein.  We use the {\tt dust}
model from XSPEC, except here we \emph{add} a component by fitting in
XSPEC: absorption $\times$ (model + [model $-$ dust$\times$model]). This
accounts for scattering of X-rays \emph{into} our line of sight.  The extra
halo component has two parameters: $f_{\rm halo}$, which is essentially the
amplitude of the scattered component, and $S_{\rm halo}$, which is related
to size of the halo and effectively determines the energy ranges scattered
by the halo.  Residuals for these fits (allowing the $N_{\rm H}$ column to
be a free parameter) are presented in Fig.~\ref{fig:dust}, and the
parameters are listed in Table~\ref{tab:asca}.

These fits to the \asca+\hexte\ data are extremely good, with reduced
$\chi^2 \approx 1$ for P40108\_04.  (The reduced $\chi^2$ for P40108\_05 is
larger predominantly due to some disagreement between the \sis\ and \gis\ 
detectors in the 8--10\,keV range.)  The fitted $N_{\rm H}$ column (chosen
to be the same for both the point source and dust spectrum) is in good
agreement with prior measurements as well as considerations from optical
extinction and 21\,cm measurements (see the discussion and references of
Zdziarski et al. \nocite{zdziarski:98a} 1998).  The fractional contribution
of the dust halo is $\approx 10\%$ at 1\,keV, in agreement with the results
of Predehl et al.  \shortcite{predhel:91a}.  The break in the \asca\ 
spectra near 4\,keV arises from a combination of the red tail of the Fe
line, the curvature of the Comptonized spectrum towards low energy, the
prominent blue tail of the seed photons due to the face-on disc, and the
dust halo.  We note that these face-on {\tt kotelp} models, as compared to
the models of \S\ref{sec:kotelp}, have slightly increased coronal
compactnesses, but greatly increased coronal optical depths. This yields
average coronal temperatures of $\approx 50$\,keV, as opposed to the
$\approx150$\,keV for the models of \S\ref{sec:kotelp}.

If we add the \pca\ data to the fits, the neutral hydrogen column and dust
parameters do not greatly change, but the coronal compactness increases and
the optical depth drops.  The overall fit in terms of reduced $\chi^2$ is
not as good as for the \asca+\hexte\ fits alone, as is clear from
Fig.~\ref{fig:dust}.  The detectors, as for Fig.~\ref{fig:ratios}, again
show clear systematic differences.  However, the level of residuals shown
in Fig.~\ref{fig:dust}, i.e., $\aproxlt 1$--5\% for the \gis, \sis, and
\pca\ detectors, are as low one can reasonably expect to achieve given the
differing systematic uncertainties of these detectors.

\section{Timing Analysis}\label{sec:timing}

\begin{figure*}
\centerline{
\includegraphics[width=0.33\textwidth]{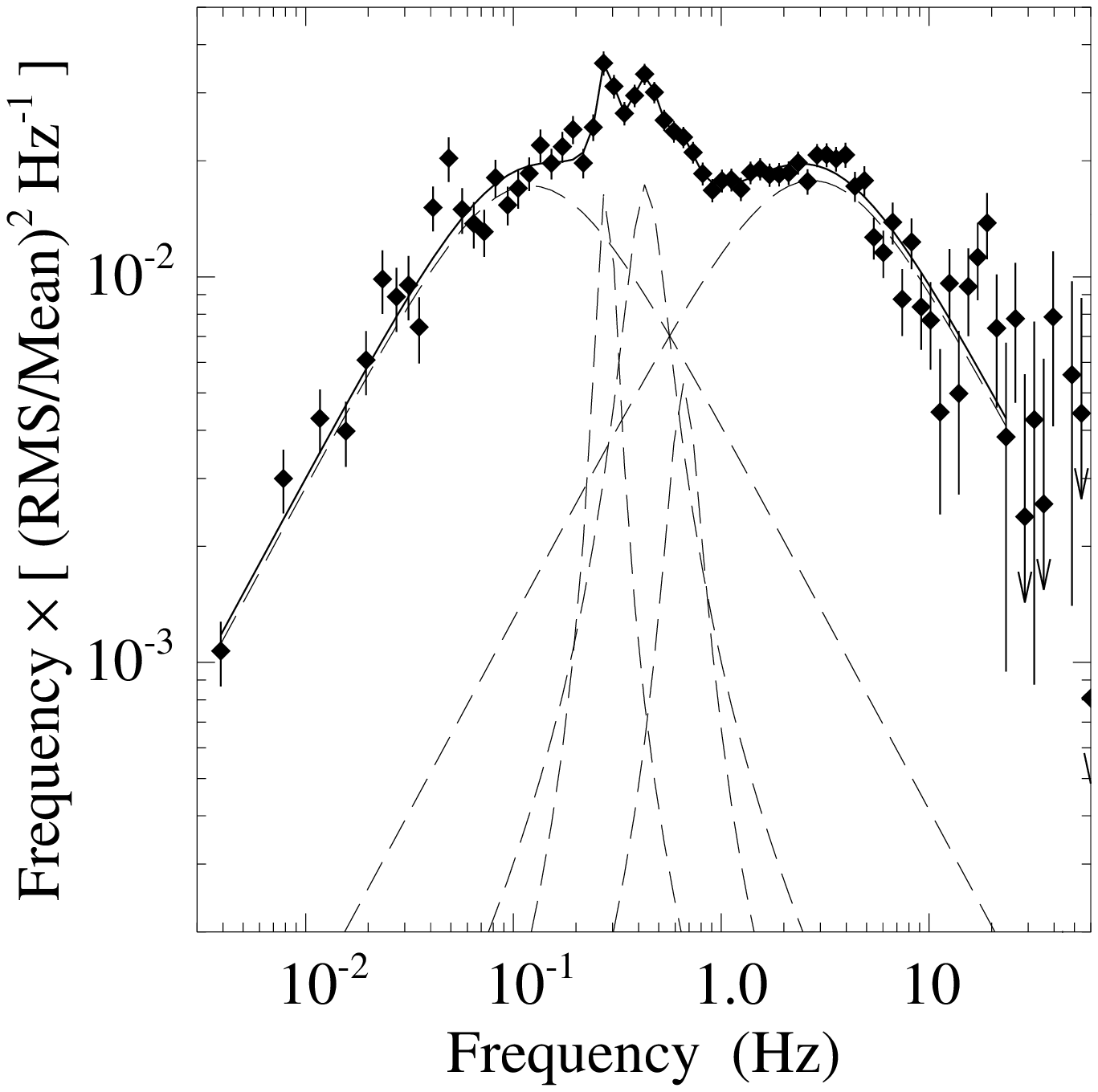}
\includegraphics[width=0.33\textwidth]{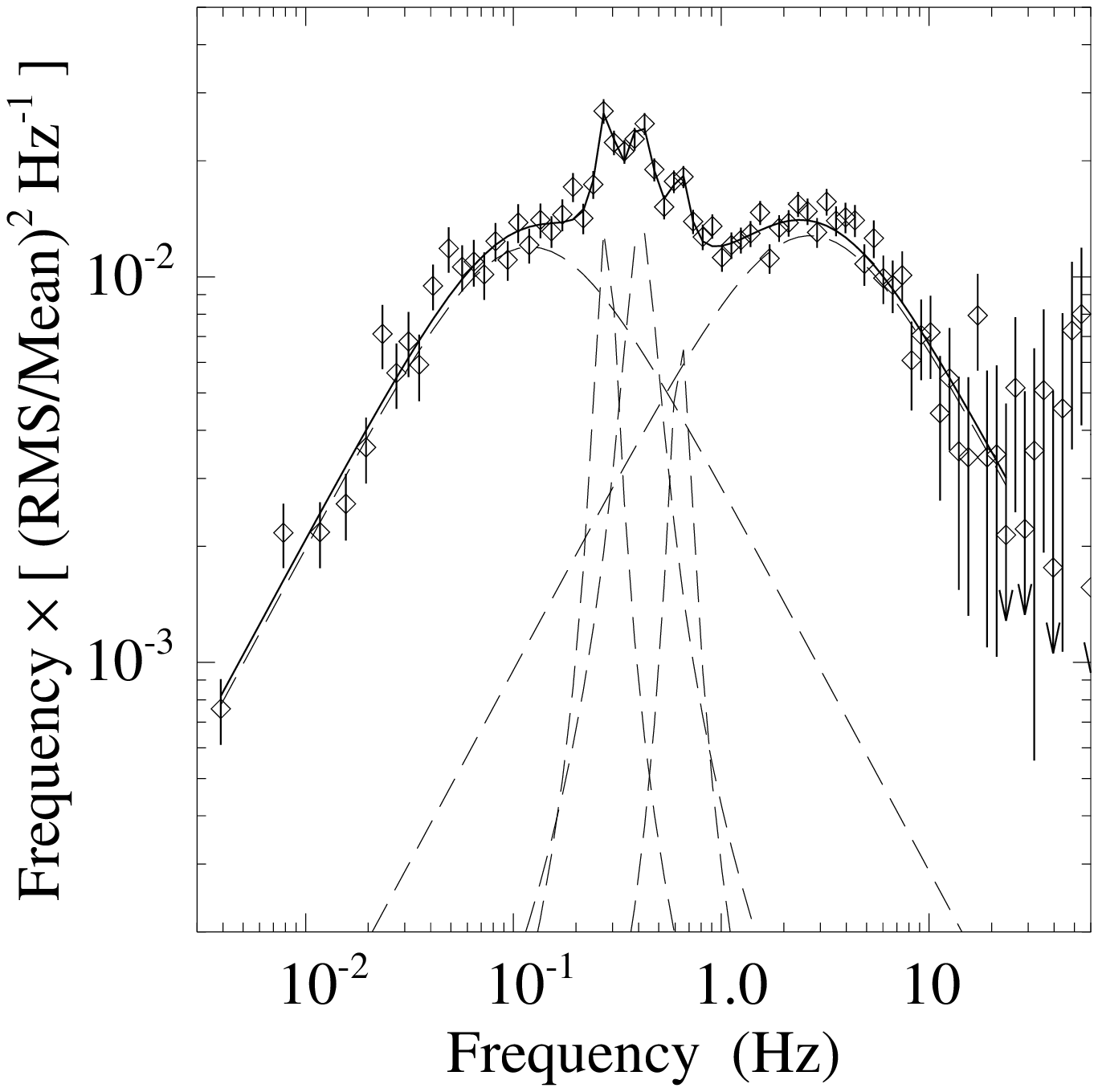}
\includegraphics[width=0.33\textwidth]{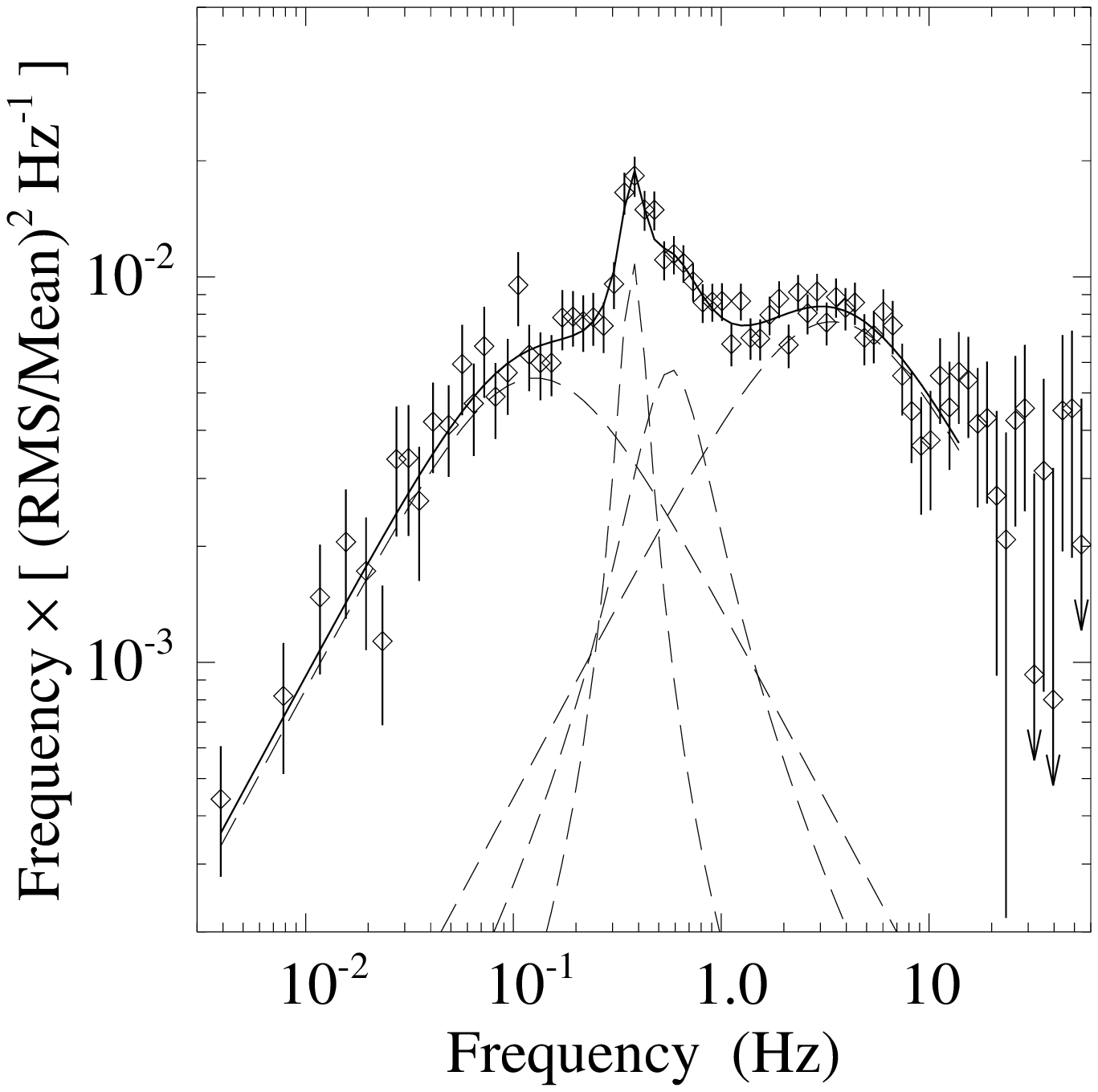}
}
\caption{\small Fits to several PSD of \gx.  Solid line shows the best-fit
  model, while the dashed lines show the individual components of the
  model.  {\it Left:} Observation P40108\_05 in the 0--5\,keV band. {\it
    Middle:} Observation P40108\_05 in the 10--20\,keV band. {\it Right:}
  Observation P20056\_05 in the 10--20\,keV band. (Error bars here are
  1-$\sigma$ errors.)  \protect{\label{fig:psds}}}
\end{figure*}

As discussed in \S\ref{sec:observations}, these observations were selected
partly because with them we were able to perform timing as well as spectral
analyses.  Timing analyses of the P20181 observations previously have been
presented by Nowak, Wilms, \& Dove \shortcite{nowak:99c} and Nowak
\shortcite{nowak:00a}.  Fourier power spectral densities (PSD) only of the
P20181 and P20056 observations previously have been presented by
Revnivtsev, Gilfanov, \& Churazov \shortcite{revnivtsev:99a}. The PSD fits
presented below most closely follow the discussion by Nowak
\shortcite{nowak:00a}.

Following the work of Wijnands \& van der Klis \shortcite{wijnands:99a} and
Psaltis, Belloni, \& van der Klis \shortcite{psaltis:99a}, Nowak
\shortcite{nowak:00a} suggested that the best representation of the hard
state PSD of GBHC in general, and \gx\ in specific, was the sum of only a
few, broad, quasi-periodic features.  Nowak \shortcite{nowak:00a} showed
that a composite PSD of the seven brightest P20181 observations (which
exhibit relatively uniform variability properties; see Nowak, Wilms, \&
Dove \nocite{nowak:99c} 1999, and the discussion below), could be described
by a zero frequency centered Lorentzian function with a cutoff frequency at
0.1\,Hz, plus additional quasi-periodic oscillations (QPO) at frequencies
0.34\,Hz (and its harmonic), 2.5\,Hz, and 18\,Hz.  Following the suggestion
of Psaltis, Belloni, \& van der Klis \shortcite{psaltis:99a}, the latter
three frequencies might be the analogues of the `horizontal branch', `lower
kiloHertz', and `upper kiloHertz' QPO features seen in neutron star Z-sources.
(For a description of the Z-source features, see van der Klis
\nocite{vanderklis:96a} 1996.)

\begin{figure*}
\centerline{
\includegraphics[width=0.44\textwidth]{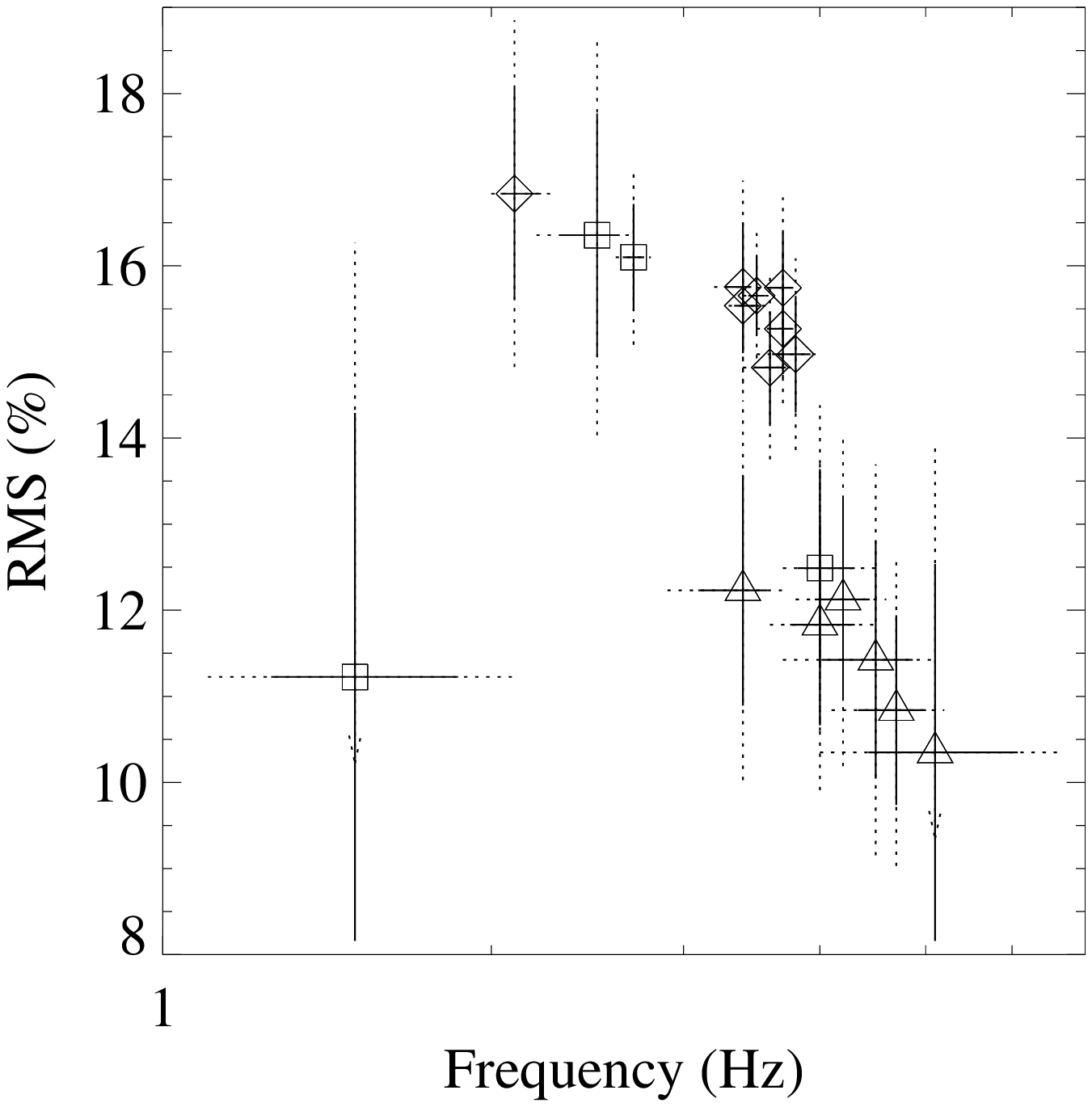}}
\centerline{
\includegraphics[width=0.44\textwidth]{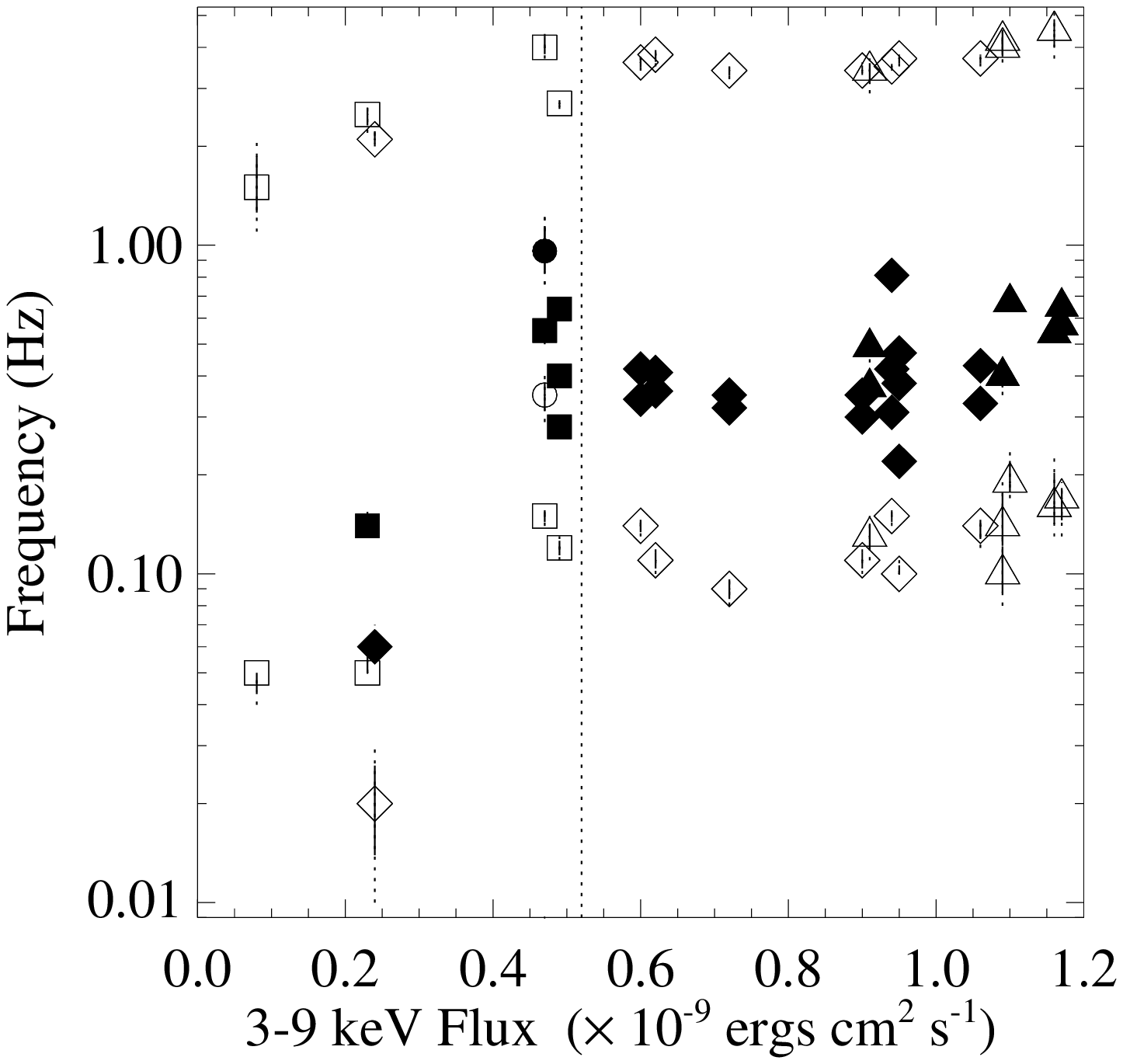}}
\centerline{
\includegraphics[width=0.44\textwidth]{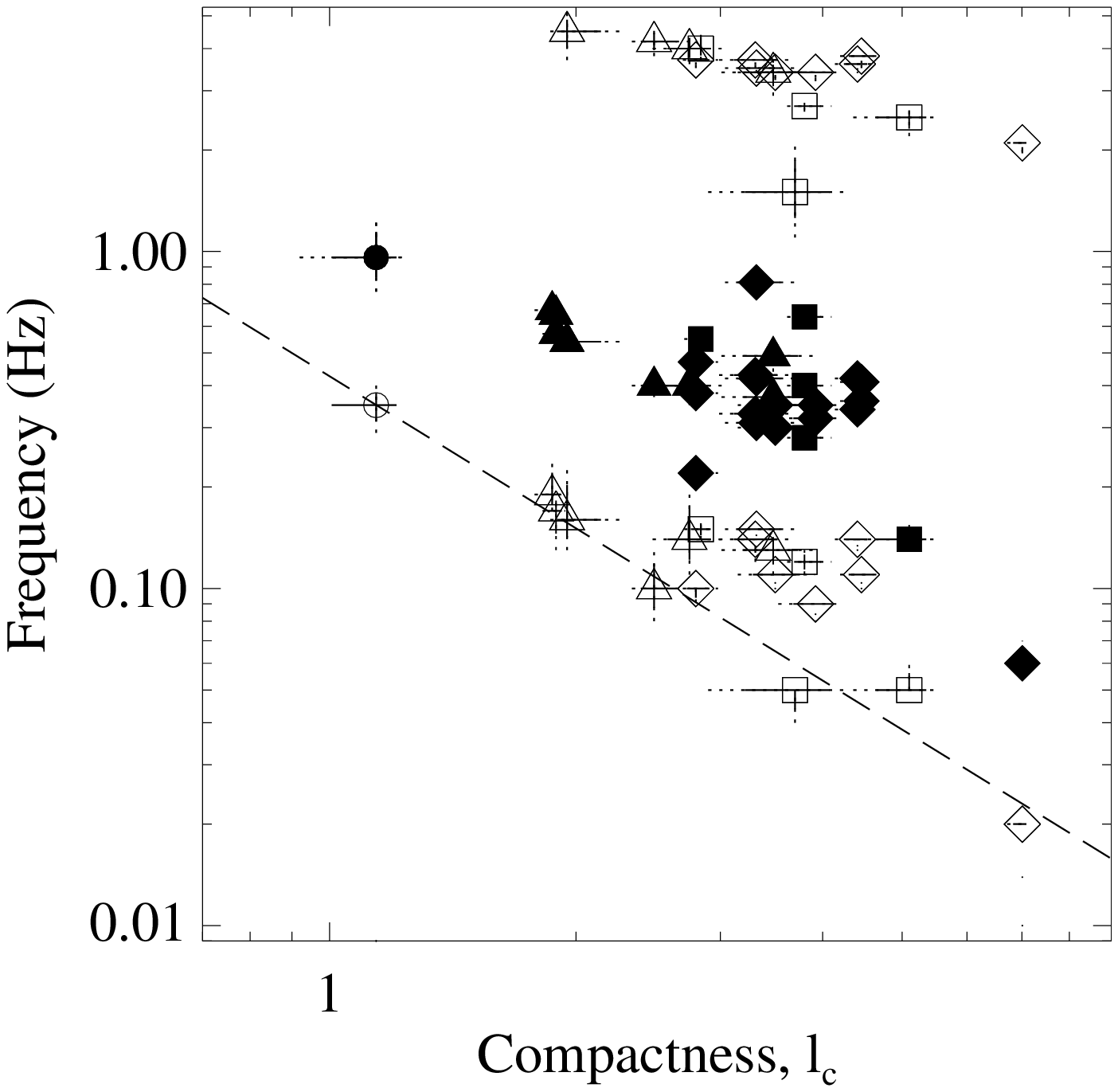}
}
\caption{\small   {\it Top:} rms variability vs. frequency for the 
  highest frequency fit component to the PSD (10--20\,keV band).  {\it
    Middle:} Characteristic PSD frequencies (10--20\,keV band) vs. \pca\ 
  flux measured from 3--9\,keV.  Clear symbols are the roll-over frequency
  for the zero frequency centered Lorentzian functions (zfc-Lor$_1$ and
  zfc-Lor$_2$) in Tables~\ref{tab:qpo1} and \ref{tab:qpo2}.  Solid symbols
  are the peak frequencies of QPO$_1$--QPO$_3$ in these tables.  {\it
    Bottom:} Same frequencies as above, accept here plotted vs.
  compactness from the {\tt kotelp+gauss} fits of Table~\ref{tab:kot}.
  Dashed line shows $f\propto \ell_c^{-3/2}$.
  \protect{\label{fig:qpovsstuff}}}
\end{figure*}

Here we considered each observation individually; therefore, we were unable
to fit any features with peak frequencies $\aproxgt 10$\,Hz. Furthermore,
we chose to fit a zero frequency centered Lorentzian function,
\begin{equation}
P(f)=\frac{A}{1+(f/f_{\rm c})^2}
\end{equation}
to the lowest ($\aproxlt 0.1$\,Hz) and highest
($\aproxgt 1$\,Hz) frequency features.  For the middle frequencies
($f\approx 0.3$\,Hz), we fit a QPO function of the form
\begin{equation}
P(f) = \pi^{-1} \frac{R^2 Q f_0}{f_0^2 + Q^2(f-f_0)^2} ~~,
\end{equation}
where $f_0$ is the resonant frequency and $Q$ is the quality factor, with
$Q\approx \Delta f/f_0$ being the fractional width of the QPO.  (For $Q \ll
1$, the maximum of $P(f)$ occurs at $f \gg f_0$. Partly for this reason, we
instead chose to fit a zero frequency centered Lorentzian function at
high frequencies, where the signal-to-noise was low.  A number of
observations, most notably P20056\_05, were better fit, however, with a
high frequency QPO feature. We fit multiple QPO features in the 0.3\,Hz
range, so long as the $\chi^2$ of the fits improved by more than 15 for
each additional feature.  Most PSDs required two such features, although
several required three (Fig.~\ref{fig:psds}).

Observation P40108\_03 required only two features: the zero frequency
centered Lorentzian function and a second feature which we identified as a
QPO. As for all the other PSDs (with the exception of P40108\_07; see
below), this `QPO' represented a distinct maximum in the plot of frequency
times the PSD power (e.g., Fig.~\ref{fig:psds}).  P40108\_07 also showed
only two features, but they both were broad, they were widely separated in
frequency, and they were of comparable amplitude in plots of frequency
times PSD power. We therefore identified both of these features with the
zero frequency centered Lorentzian functions.

Fig.~\ref{fig:psds} shows several examples of the PSD fits, and parameters
for the high energy band fits are presented in Tables~\ref{tab:qpo1} and
\ref{tab:qpo2}.  In general, the root mean square (rms) variability was on
the order of 30\%, with the soft energy band showing slightly higher rms
variability than the high energy band.  Decreased variability at high
energy has been argued by Revnivtsev, Gilfanov, \& Churazov
\shortcite{revnivtsev:99a}, via ``frequency resolved spectroscopy'', as
indicative of the ``reflection spectrum'' having little variability.
However, the PSD retains no phase information (which, in reality, could be
a complicated mixture of components; see Nowak \nocite{nowak:00a} 2000). If
a highly variable ``power law'' is $180^\circ$ out of phase with a highly
variable ``reflection component'' (as suggested for the
``$\Gamma$-$\Omega/2\pi$'' correlation; Zdziarski, Lubi\'nski, \& Smith
\nocite{zdziarski:98a} 1999), low variability would result at high energy.

There is some trend for the PSD fit components to show lower rms
variability with higher frequency.  This trend is most pronounced for the
highest frequency fit component, as we show in Fig.~\ref{fig:qpovsstuff}.
We also show in Fig.~\ref{fig:qpovsstuff} the fitted frequencies plotted
against the measured 3--9\,keV flux.  Whereas there is a large degree of
variation among the low flux frequencies, there is relatively less
frequency variation in the `hysteretic region' above the lowest measured
soft state 3--9\,keV flux.  The P20056 observations at very high flux show
a slight upward trend in frequency.  Observations P20181\_05, P40108\_06,
and P40108\_07, all at fairly low fluxes, all exhibit frequencies markedly
lower than those exhibited by observations in the `hysteretic region'.
Observation P40108\_03, the first observation after the return to the hard
state in 1999, although exhibiting a relatively lower variability
amplitude, shows a sharp spike upward in characteristic PSD frequencies.

Note that although there are clear systematic trends, there is not a very
large variation in the characteristic frequency of a given PSD feature.
For example, the cutoff frequencies of the Lorentzian functions fit at high
frequencies vary by less than a factor of 3 overall.  di Matteo \& Psaltis
\shortcite{dimatteo:99a} have argued that the limited range of frequency
variations exhibited by GBHC argue for a comparably small (less than a
factor of 2) variation in the (cylindrical) radial size scale of any
central corona or in the size scale of any ``transition radius'' between an
advection and non-advection dominated flow.  Assuming that the highest
observed frequency is generated in the outer, non-advection dominated disc
(e.g., Psaltis \& Norman \nocite{psaltis:00a} 2000), 1.8\,Hz implies a
transition radius $\aproxlt 270~GM/c^2$, for a $4\,M_\odot$ central object.
However, a feature with a frequency as high as 20\,Hz, such as discussed by
Nowak \shortcite{nowak:00a} (which would not be detectable within these
individual observations) would imply a radius $\aproxlt 60~GM/c^2$.  The
rise of the characteristic PSD frequencies with increasing flux, followed
by a plateau at high fluxes, is very reminiscent of the behaviour seen in
the neutron star source 4U~1820$-$30 \cite{zhangw:98a,kaaret:99a}. Such a
plateau has been argued as evidence for the accretion disc extending inward
all the way to its marginally stable orbit \cite{zhangw:98a,kaaret:99a}.

In Fig.~\ref{fig:qpovsstuff} we also show these frequencies plotted vs. the
coronal compactness from the {\tt kotelp+gauss} fits of
Table~\ref{tab:kot}.  There is a clear trend for harder spectra to show
systematically lower frequencies for all of the characteristic PSD
features.  Such a trend was noted by di Matteo \& Psaltis
\shortcite{dimatteo:99a} and Revnivtsev, Gilfanov, \& Churazov
\shortcite{revnivtsev:99a}, although those authors parameterized hardness
by the photon index, $\Gamma$, and they did not include \hexte\ data in
their fits.  Observations in the `hysteretic region', which have relatively
similar frequencies, deviate from this trend slightly. The overall
behaviour, however, is roughly consistent with the PSD frequencies being
proportional to $\ell_c^{-3/2}$ (although the highest frequency exhibits a
slightly flatter trend).  We comment upon this further in
\S\ref{sec:simple} below.

\begin{figure*}
\centerline{
\includegraphics[width=0.33\textwidth]{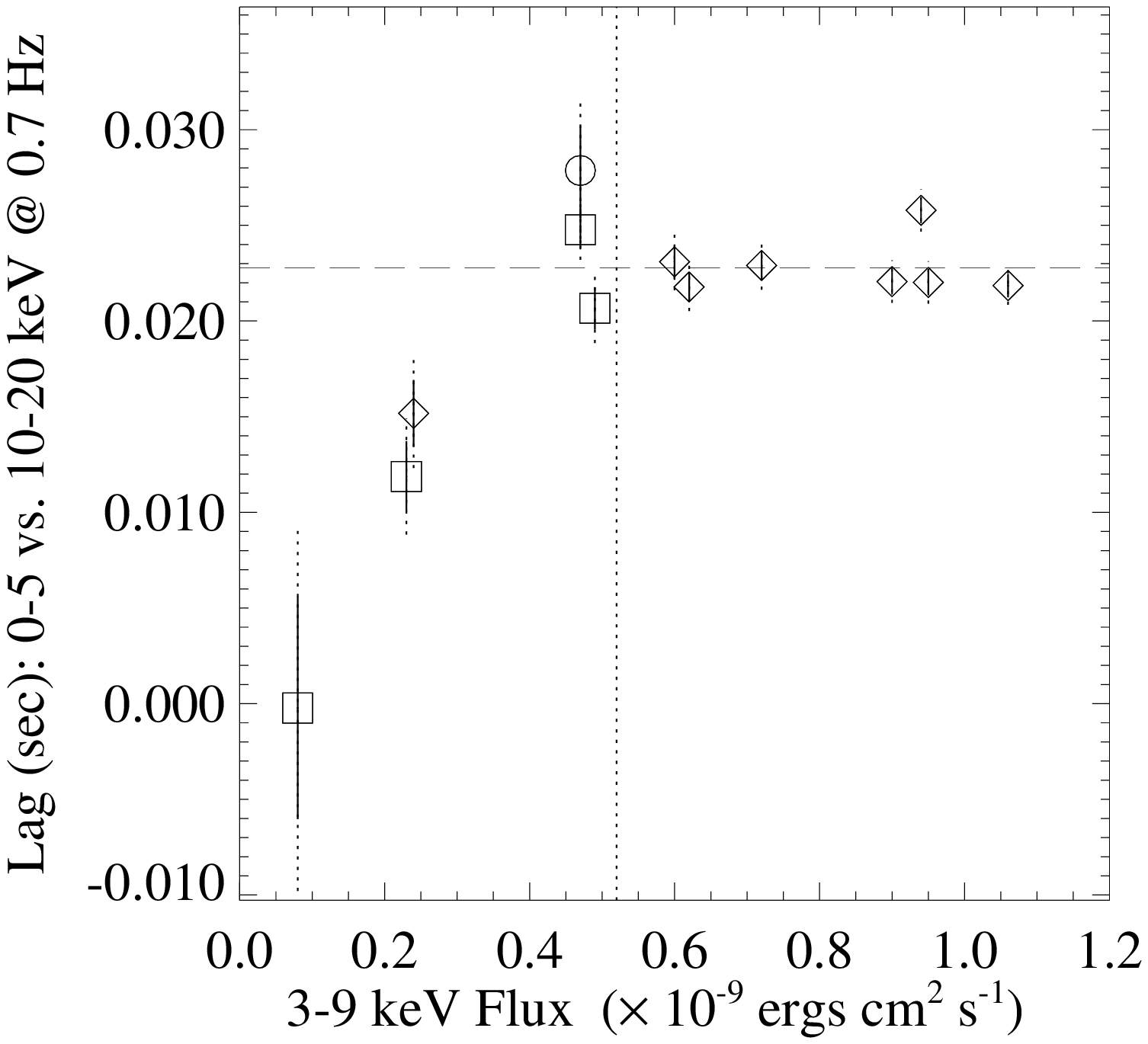}
\includegraphics[width=0.33\textwidth]{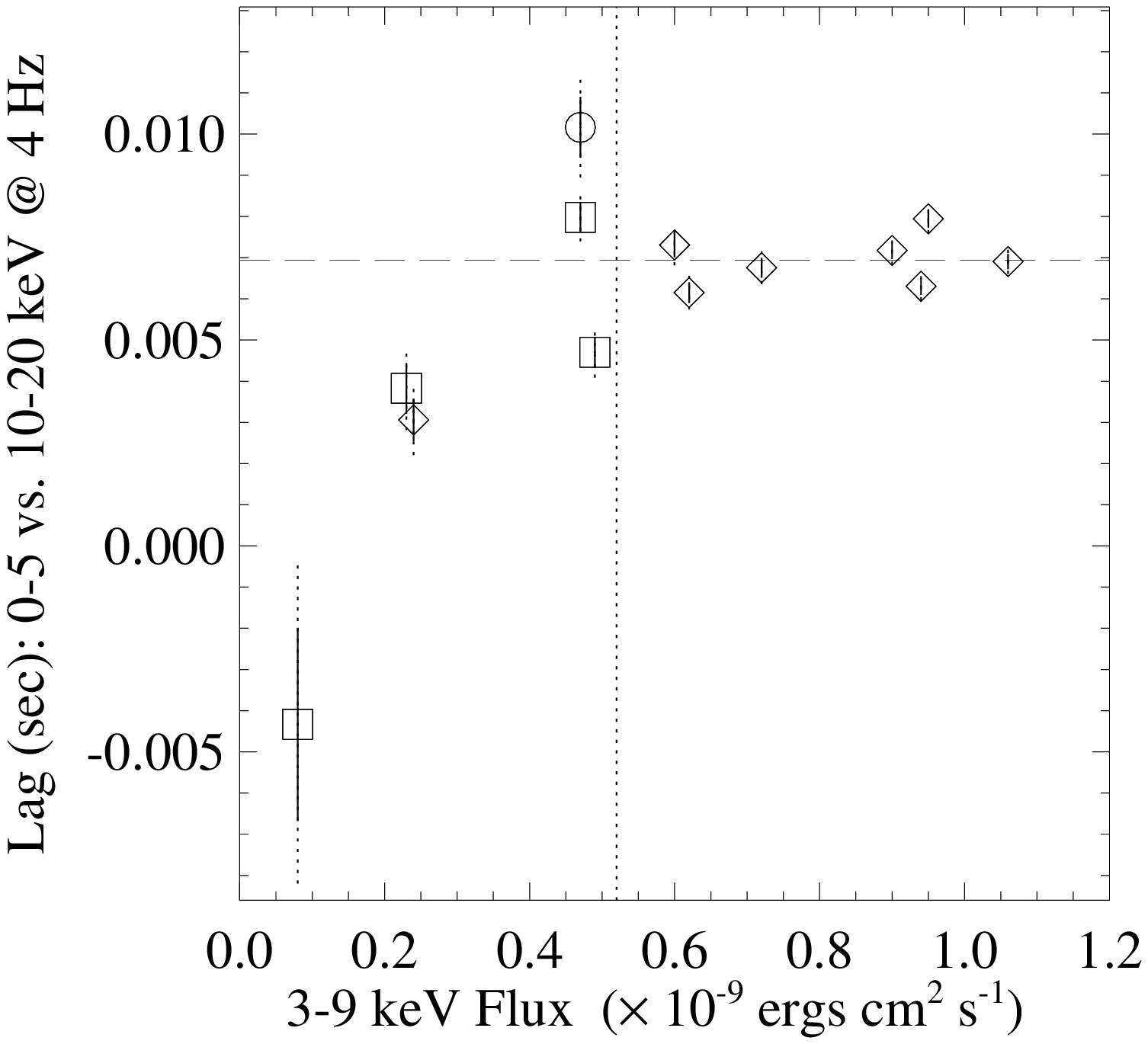}
\includegraphics[width=0.33\textwidth]{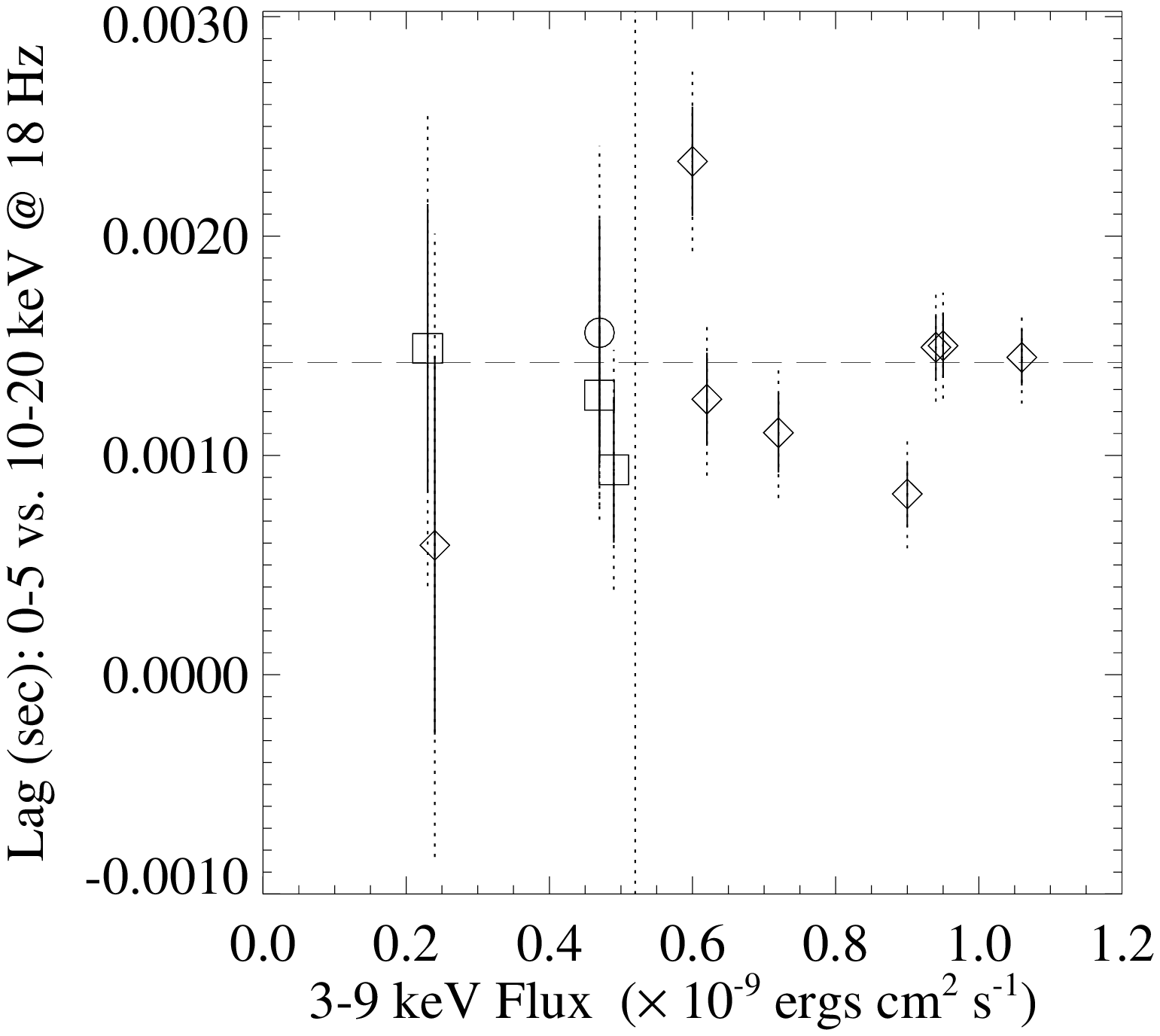}
}
\caption{\small Time lags between variability in the 0--5\,keV band and in
  the 10--20\,keV band vs. \pca\ flux measured from 3--9\,keV.  Positive
  values indicate the hard band variability lagging behind the soft band
  variability.  Dashed line corresponds to the mean time lag of all P20181
  observations, excluding P20181\_05.  {\it Left:} Average frequency
  0.7\,Hz.  {\it Middle:} Average frequency 3.7\,Hz.  {\it Right:} Average
  frequency 18\,Hz.  \protect{\label{fig:lagsvsfl}}}
\end{figure*}

\begin{figure*}
\centerline{
\includegraphics[width=0.33\textwidth]{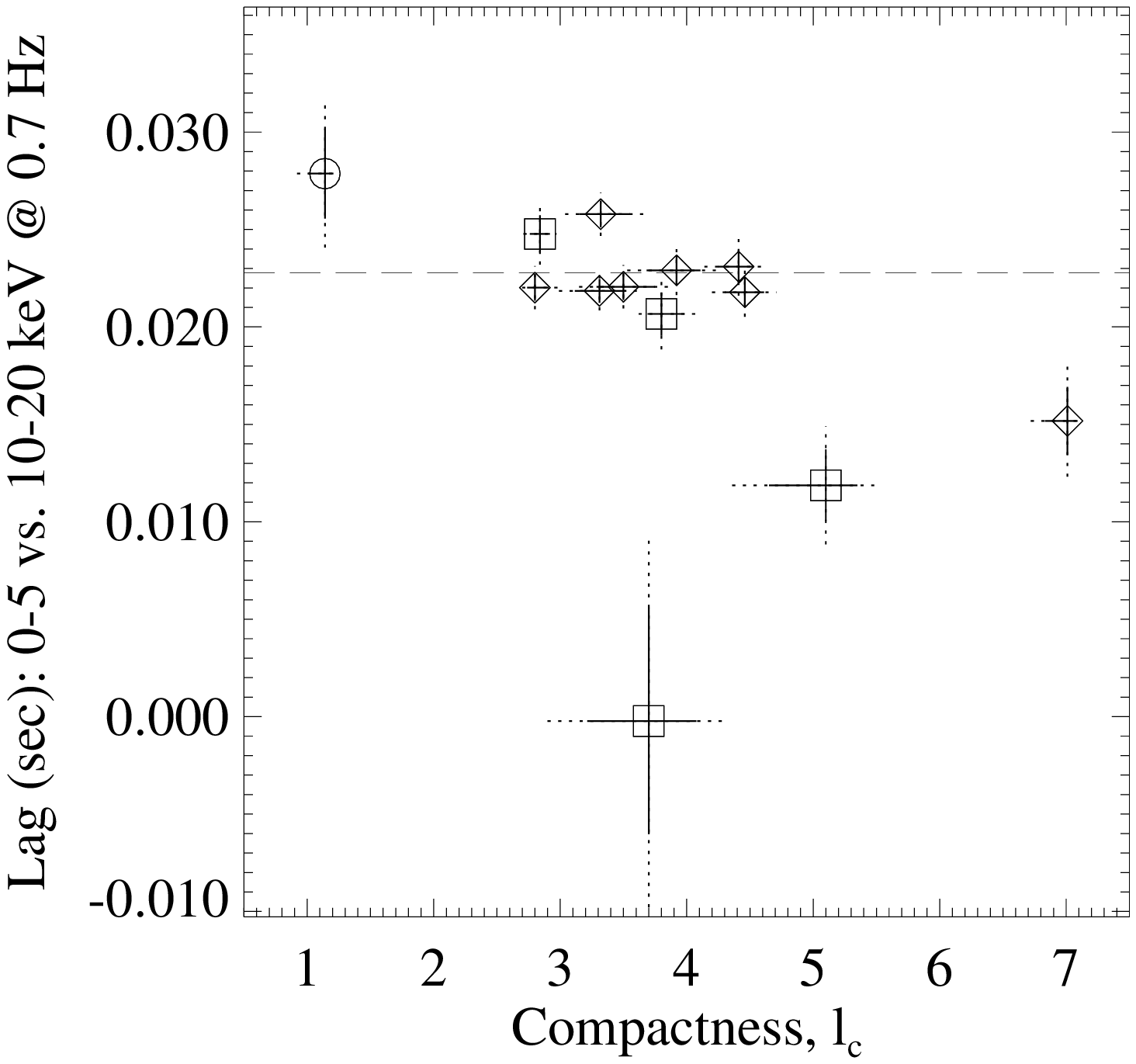}
\includegraphics[width=0.33\textwidth]{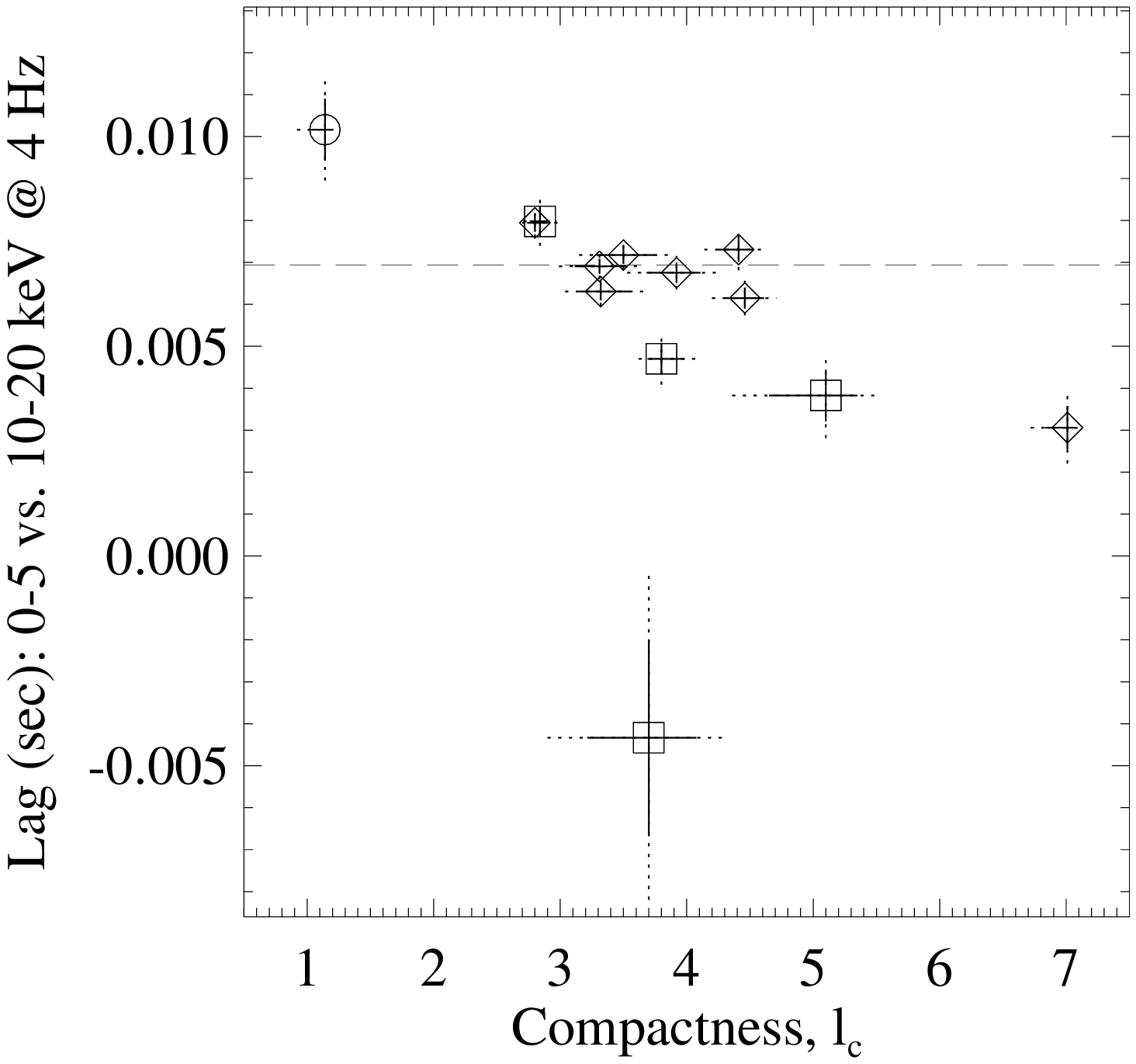}
\includegraphics[width=0.33\textwidth]{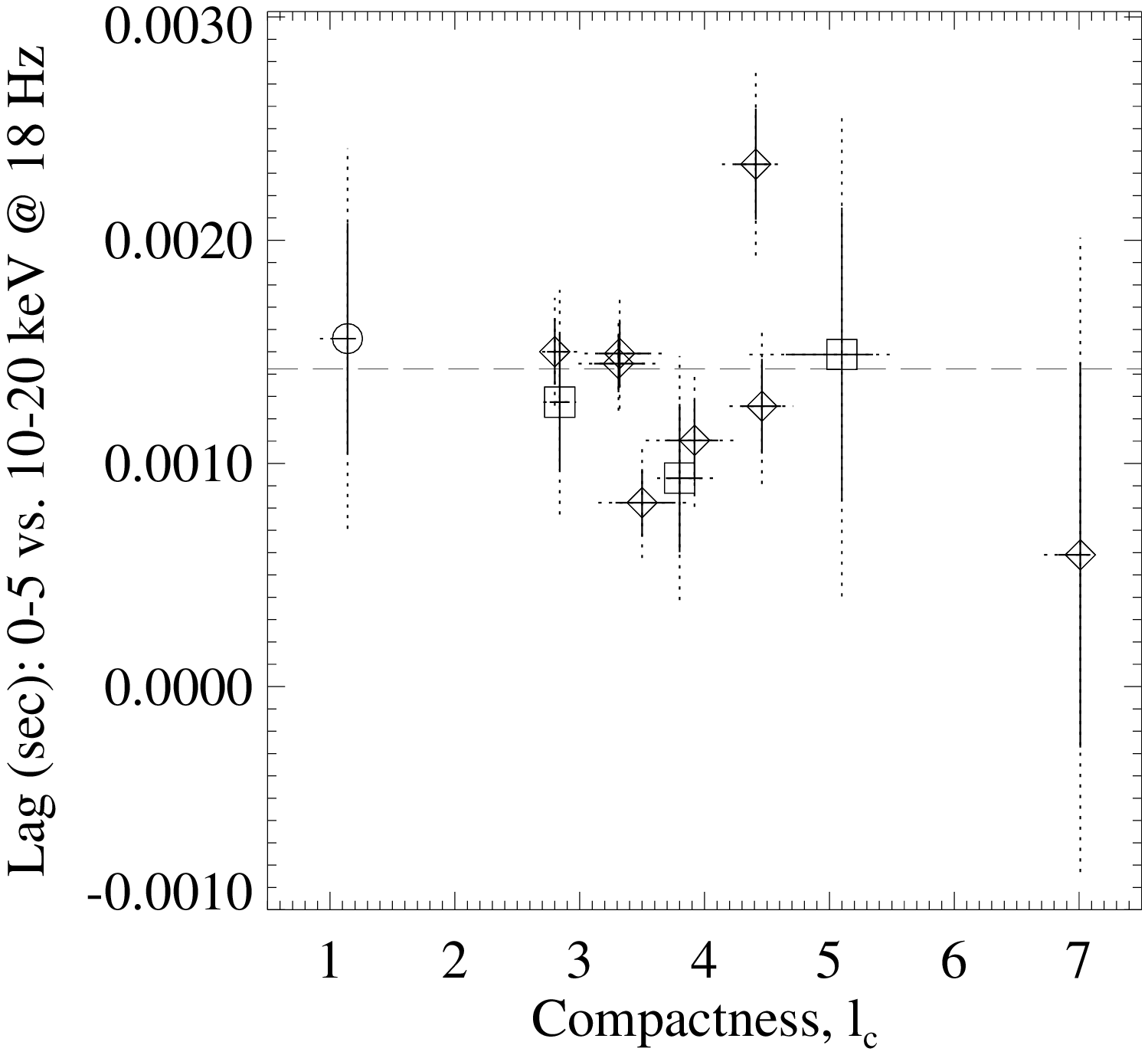}
}
\caption{\small Same as in Fig.~\ref{fig:lagsvsfl}, except here plotted
  vs. compactness from the {\tt kotelp+gauss} fits of Table~\ref{tab:kot}.
  {\it Left:} Average frequency 0.7\,Hz.  {\it Middle:} Average frequency
  3.7\,Hz.  {\it Right:} Average frequency 18\,Hz.
  \protect{\label{fig:lagsvslc}}}
\end{figure*}

\begin{figure*}
\centerline{
\includegraphics[width=0.33\textwidth]{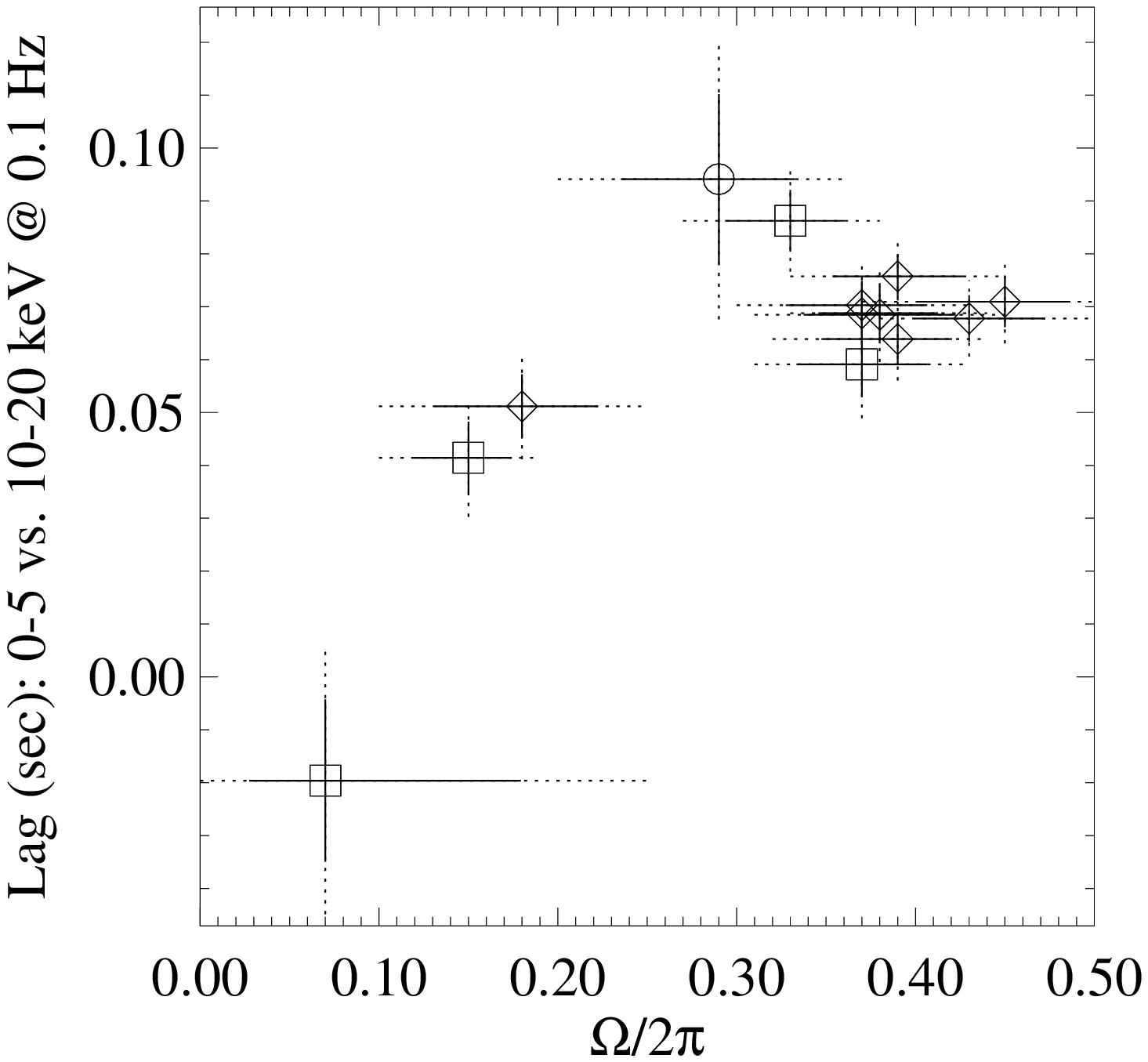}
\includegraphics[width=0.33\textwidth]{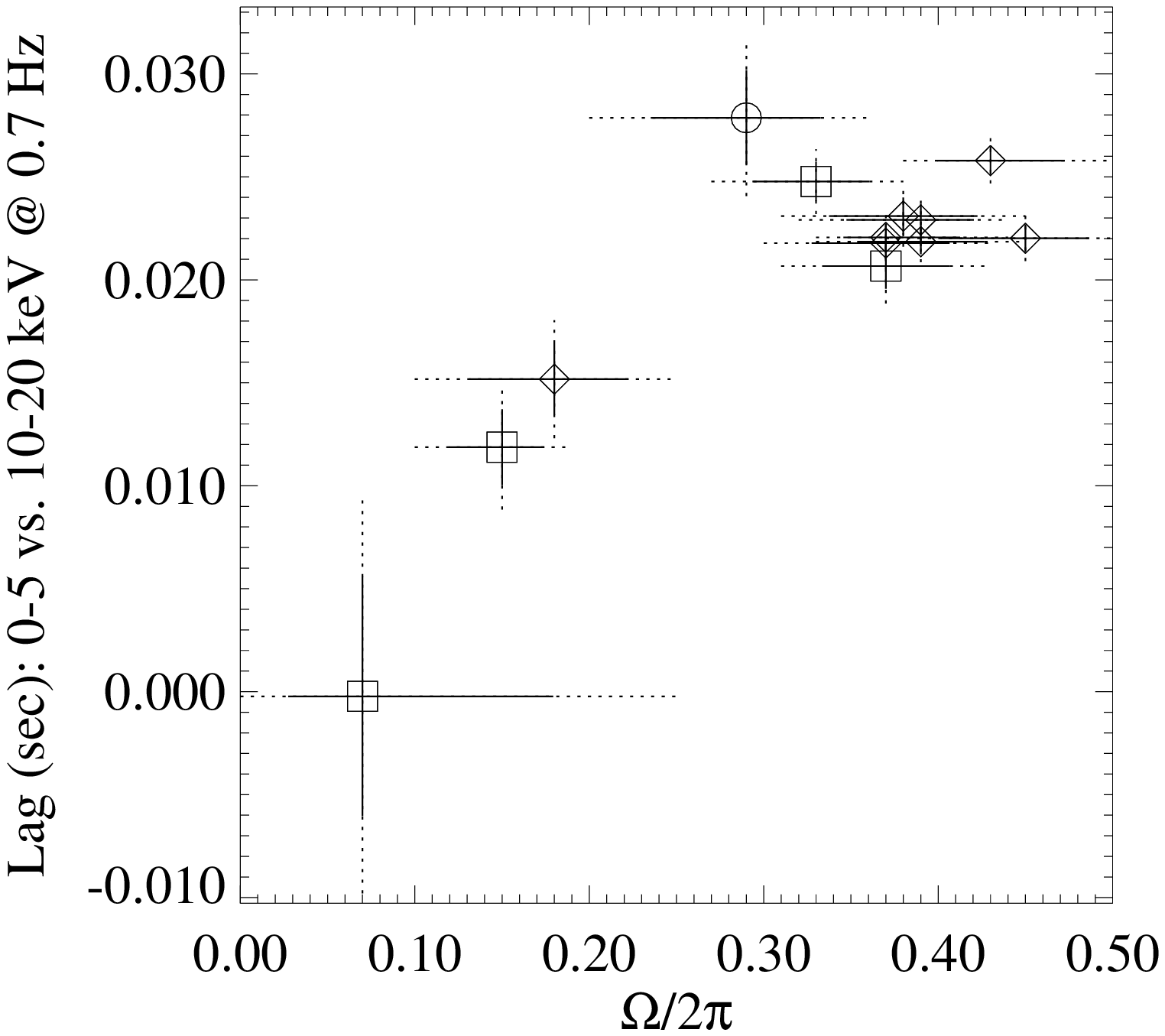}
\includegraphics[width=0.33\textwidth]{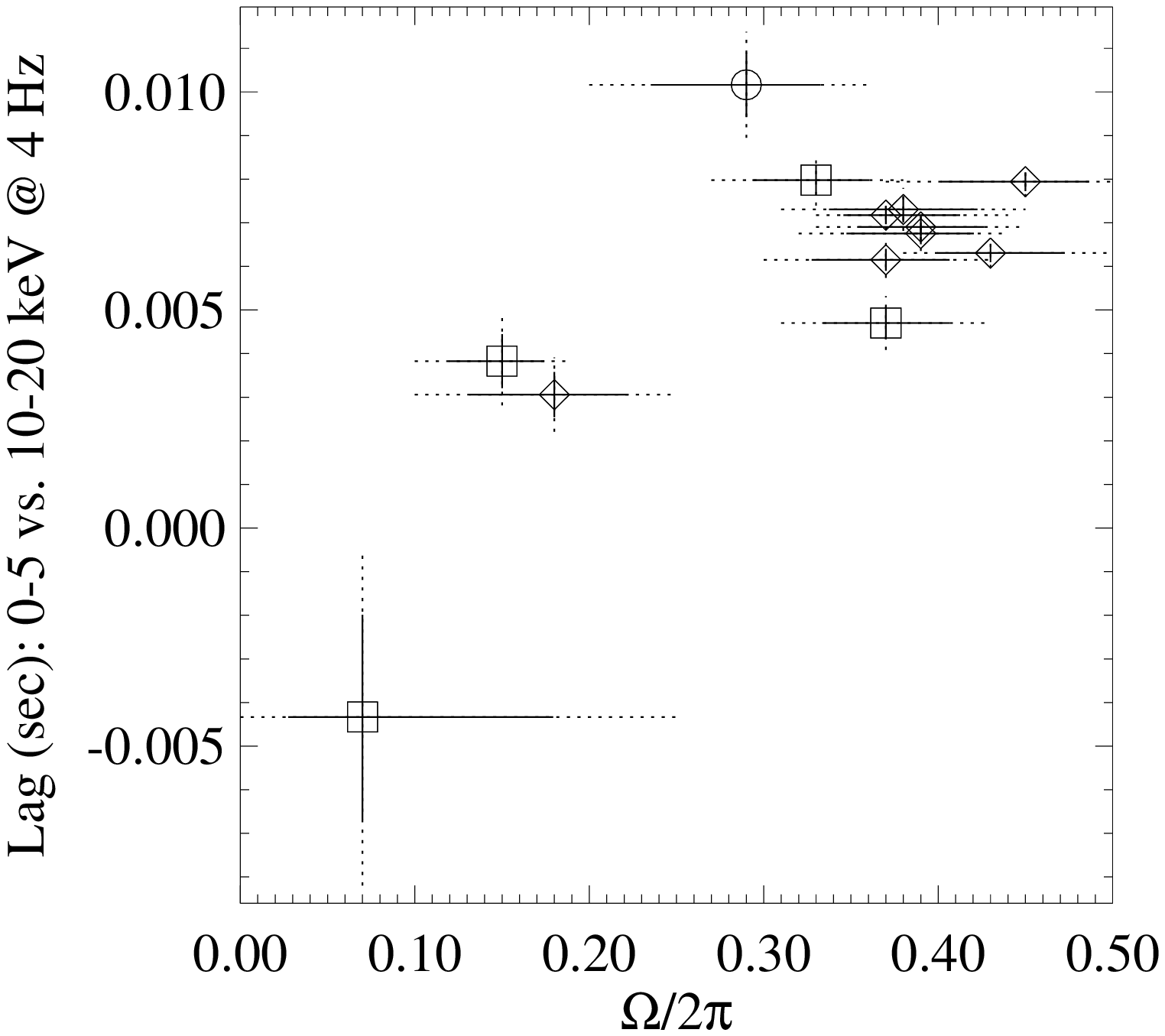}
}
\caption{\small Same as in Fig.~\ref{fig:lagsvsfl}, except here plotted
  vs. reflection fraction from the {\tt eqpair+gauss} fits of
  Table~\ref{tab:kot}.  {\it Left:} Average frequency 0.1\,Hz.  {\it
    Middle:} Average frequency 0.7\,Hz.  {\it Right:} Average frequency
  3.7\,Hz.  \protect{\label{fig:lagsvsref}}}
\end{figure*}

We have also calculated the average time lag between the variability in the
0--5\,keV and the 10--20\,keV energy bands.  In general, the broad band
variability of GBHC in their hard state shows the hard variability lagging
behind the soft variability, with the lag increasing both for lower Fourier
frequencies and for larger energy separations (Miyamoto \& Kitamoto
\nocite{miyamoto:89a} 1989; Miyamoto et al. \nocite{miyamoto:92a} 1992;
Nowak et al. \nocite{nowak:99a} 1999; Nowak, Wilms, \& Dove
\nocite{nowak:99c} 1999; and references therein).  As seen in
Figs.~\ref{fig:lagsvsfl}, \ref{fig:lagsvslc}, and \ref{fig:lagsvsref}, the
variability time lags span a large dynamic range from $\approx
10^{-3}$\,sec at high Fourier frequencies, to $\approx 0.1$\,sec at low
frequency.  Various models have attributed this broad range of observed
time delays to photon diffusion in a very extended corona
\cite{kazanas:97a}, wave propagation in a disc \cite{nowak:99b}, or
temporally correlated flares covering an accretion disc
\cite{poutanen:99a}.  In order to both increase the signal to noise and to
describe the overall characteristic behaviour of the time delays, we have
averaged the calculated time lags over a range of a factor of 5 in Fourier
frequency.  The resulting average time lags are plotted vs. 3--9\,keV flux
in Fig.~\ref{fig:lagsvsfl}.

At the highest frequency, we see that the time delays are roughly
consistent with being at a common minimum value of $\approx 10^{-3}$\,Hz.
We have previously argued that such a minimum value of the time delay sets
a lower limit of roughly $30~GM/c^2$ for the size scale of any corona
\cite{nowak:99b}.  For lower Fourier frequencies, we see that in the
`hysteretic region' time lags are relatively uniform from observation to
observation. For observation P40108\_03, coincident with the 1999 soft to
hard state transition, there is an increase in the characteristic time lag.
A very similar, but more dramatic, time lag increase also has been seen in
the soft/hard state transitions (or `failed' state transitions) of Cyg~X-1
\cite{pottschmidt:00a}.  At lower fluxes, the time lag is seen to drop.

Nowak, Wilms, \& Dove \shortcite{nowak:99c} have previously argued that the
simultaneous decrease in the characteristic PSD frequencies and in the time
lags seen for observation P20181\_05 (i.e., the lowest flux diamond points
in the figures) argues against the flare models of Poutanen \& Fabian
\shortcite{poutanen:99a}, where one expects the PSD time scales (i.e.,
inverse frequencies) and the time lags to be correlated.  The trend
exhibited by observation P20181\_05 for the PSD time scales and time lags
to be \emph{anti-correlated} is seen to continue towards lower flux.
Interestingly, the lowest flux observation shows that the power spectra
averaged about 0.1\,Hz and 4\,Hz exhibit a lag of the soft variability
behind the hard variability.  We note, however, that as discussed by Nowak
\shortcite{nowak:00a}, the net overall time lags could in fact be made up
of several independent processes--- which we would be averaging over in these
figures--- some of which have soft variability lagging hard variability, and
some of which have hard variability lagging soft variability.

In Fig.~\ref{fig:lagsvslc} we show the time lags plotted against the
coronal compactness from the {\tt kotelp+gauss} fits of
Table~\ref{tab:kot}.  Again, the plateau of the observations in the
hysteretic region are apparent.  The overall trend is for less
compact/softer spectra to show longer time lags than more compact/harder
spectra, albeit the faintest two observations deviate from this trend. This
deviation for the faintest two observations is similar to that seen for the
anti-correlation between `reflection fraction' and compactness (from the
{\tt eqpair} fits) seen in Fig.~\ref{fig:eq_lkt_seta}. Plotting time lag
vs. reflection fraction from the {\tt eqpair} fits, we see that there is
indeed a correlation between the two, as shown in Fig.~\ref{fig:lagsvsref},
where higher reflection is seen to imply longer time lags of the hard
variability behind the soft variability.  If the reflection fraction is
indeed truly measuring the relative amount of cold material being
illuminated by the corona, then perhaps these data are arguing for low
compactness/soft spectra being associated with a larger coronal size, which
would yield longer time lags due to increased propagation or diffusion path
lengths while simultaneously illuminating a larger region of the cold disc.
We comment upon this further in \S\ref{sec:simple}.

\section{Soft State Observations}\label{sec:soft}

The soft state observations discussed here, P40108\_01 and P40108\_02,
occurred shortly prior to a return to the hard state.  A discussion of
brighter soft state observations of \gx, taken in early 1998 after the
initial transition from the hard to soft state, can be found in Belloni et
al.  \shortcite{belloni:99a}.  Observations P40108\_01 and P40108\_02 are
very characteristic of `classic' BHC soft states (Nowak \nocite{nowak:95a}
1995, and references therein). Both observations exhibit rms variability
$\aproxlt 3$\% between $\approx 10^{-2}$--1\,Hz, and are totally dominated
by Poisson noise residuals at higher frequencies.

\begin{figure*}
\centerline{
\includegraphics[width=0.35\textwidth,angle=270]{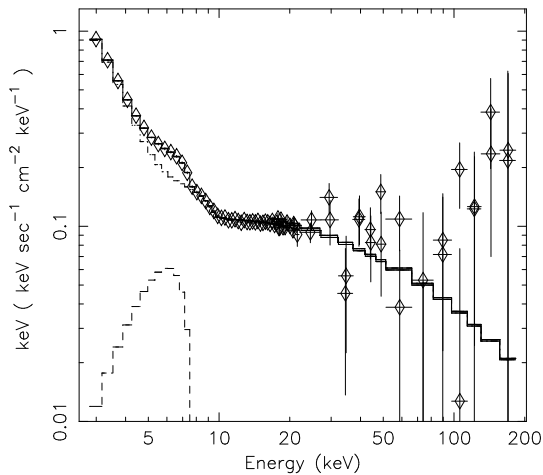}
}
\caption{\small Unfolded energy spectrum of \gx\ (P40108\_01) 
  with the {\tt compps} Comptonization model fitted to the \pca+\hexte\ data.
  \protect{\label{fig:compps_unfold}}}
\end{figure*}

In terms of their spectra, both observations show a strong soft excess at
energies $\aproxlt 8$\,keV, with a $\Gamma \approx 2$ power law at higher
energies; see Fig.~\ref{fig:compps_unfold}.  Such observations are
typically well-described by a multi-temperature disc blackbody with maximum
temperature of $\approx 1$\,keV, plus an additional power law, typically
with a photon index $\aproxgt 2$.  For an example of similar soft state BHC
spectra observed with \rxte, we refer  to our recent work on
LMC~X-1 and LMC~X-3 \cite{nowak:01a,wilms:01a}. For these three soft state
BHC, a broad and strong Fe line seems to be required ($\sim 1$\,keV width,
$\approx 200$--900\,eV equivalent width).  As discussed by Nowak et al.
\shortcite{nowak:01a}, there is some worry that such a line could be an
artifact due to the phenomenological fit components, namely a
multi-temperature disc blackbody and a power law, crossing over one another
at an energy $\sim 6$\,keV.  A broad line could be required to remove an
artificially created ``inflection point''.

As with our LMC~X-1 and LMC~X-3 spectra, to assess this possibility for the
soft spectra of \gx, we turn to the {\tt compps} model, and choose the
``slab geometry'' (i.e., a planar disc and corona, with seed photon
injection occurring at the disc-corona interface at the midplane). As this
model self-consistently calculates the high energy tail from the seed
photon distribution and the coronal parameters, there is less worry about
creating an artificial Fe line.  In addition to the coronal parameters, we
also allow for reflection from an ionized slab, with a relativistically
smeared reflection profile.  (We choose the $\beta=10$ option in the {\tt
  compps} model, and fix the inner and outer disc edge to be $6~GM/c^2$ and
$10^3~GM/c^2$, respectively. The disc Fe abundance is set to 4, and the
disc inclination angle is frozen to $45^\circ$.)  To model the line, we use
the {\tt laor} model in \textsl{XSPEC} \cite{laor:91a}.  Again, we set the
inclination angle to $45^\circ$, and we freeze the inner and outer disc
edge at their maximum values.  We allow the line energy, normalization, and
the disc emissivity index to be free parameters.

Results for these fits are presented in Table~\ref{tab:comp}.  The fits are
very similar to our prior results for LMC~X-1 and LMC~X-3.  Specifically,
the disc inner edge temperatures are $kT_{\rm disc} \approx 540$\,eV, the
fitted coronae are very hot with $kT_{\rm e}\approx130$\,keV, and they are
also optically thin, with $\tau_{\rm es} \approx 0.3$.  In addition, the
fitted Fe lines have energies $\approx 6.8$\,keV (consistent with ionized
iron), are very strong (600 to 900\,eV equivalent widths\footnote{Note that
  we do not quote error bars for the line equivalent widths in
  Table~\ref{tab:comp}, as the error bars are complicated functions of the
  other fit parameters, not just of the line normalization. Generally, the
  equivalent widths deviate from the best fit values by approximately
  $-100$\,eV to $+500$\,eV, depending upon what other parameters are
  allowed to vary.}), and are very broad. Emissivity indices are $>5$,
indicating lines that are \emph{extremely} skewed towards the inner edge of
the disc.  The fitted reflection fractions, in contrast to the line
equivalent widths, are small, with $\Omega/2\pi \aproxlt 0.2$. In this
respect, these observations are similar to P40108\_06 and P40108\_07. The
cold reflector, consistent with the line energies, is ionized.

The unfolded spectrum for observation P40108\_01 is presented in
Fig.~\ref{fig:compps_unfold}.  One question that arises, is the broad,
large equivalent width line an artifact of our assumed model?  For example,
could the higher energy line with a sharp blue edge (due to the emissivity
index being $>5$), in reality be mimicking an unsmeared, neutral reflection
edge at 7.1\,keV?  (The smeared, ionized edge is broader and extends to
higher energies.)  This would be expected if the reflection were due
to a combination of an ionized, relativistic disc and cold, neutral matter
much farther away from the compact object (i.e., a flared outer disc edge,
the companion star, etc.).  We have tried adding a neutral edge at
7.1\,keV, and this does reduce the Fe line equivalent width, but only by
$\approx 300$\,eV.  As for our LMC~X-1 and LMC~X-3 observations
\cite{nowak:01a,wilms:01a}, a fairly strong, broad line is a ubiquitous
feature of all the models that we have attempted to fit to the \gx\ soft
state data.

The other aspect that is to be noted about Fig.~\ref{fig:compps_unfold} is
the possibility of a hardening at energies above $\sim$80\,keV. Due to the
very low hard flux, this hardening is entirely consistent with the
uncertainties of the background measurement (see \S\ref{sec:alternative}).
We note also that we have used the {\tt compps} model to fit non-thermal
Comptonization models, and obtain equally good, purely non-thermal fits.
Thus, these data are unable to distinguish between a thermal and
non-thermal electron distribution for the Comptonizing electrons.

\section{Discussion}\label{sec:discuss}

\subsection{Comparison to Previous Comptonization
  Models}\label{sec:compare}

As discussed in \S\ref{sec:coronal}, the {\tt compps}, {\tt eqpair}, and
{\tt kotelp} models endeavor to be physically motivated, realistic
descriptions of Comptonization in compact object systems.  The results that
we have found here with these models are in many ways comparable to
previous analyses of \gx, as well as the spectrally similar source,
Cyg~X-1.  For example, previous work with versions of the {\tt compps}
model have found $\tau_{\rm es} \sim 1$ and $kT \sim 100$\,keV for Cyg~X-1
\cite{poutanen:97a} and $\tau_{\rm es} \sim 1$ and $kT \sim 50$\,keV for
\gx\ \cite{gierlinski:97a}.  Our own previous analysis of joint
\pca$+$\hexte\ data of Cyg~X-1 using the {\tt kotelp} model found
$\tau_{\rm es} \sim 1$ and $kT \sim 90$\,keV \cite{dove:98a}.  On the other
hand, we previously have found somewhat lower temperature ($kT \sim
20$--40\,keV) and higher optical depth ($\tau_{\rm es} \sim 3$) coronae for
\gx\ \cite{wilms:99aa}.  However, unlike the aforementioned studies which
were broad-band ($\approx 2$-- $>200$\,keV), the work of Wilms et al.
\shortcite{wilms:99aa} fit the low energy (\pca) and high energy (\hexte)
data separately.  Recent work fitting broad-band data of Cyg~X-1 (Maccarone
\& Coppi, in prep.) with the {\tt eqpair} model suggests similar coronal
parameters as discussed above, and interestingly enough suggests at times a
large equivalent width line (200\,eV) with a low reflection fraction
($\Omega/2\pi \sim 0.2$), similar to what we have found for several of our
observations of \gx.  Thus, there is at least general consensus that
``realistic'' Compton coronae models with $\tau_{\rm es} \sim 1$, $kT \sim
100$\,keV, and low reflection fractions, $\Omega/2\pi \aproxlt 0.5$,
provide reasonable descriptions of hard state black holes. (See
\S\ref{sec:soft} for references to recent work concerning soft state black
holes, as well as Frontera et al. \nocite{frontera:01a} 2001, Gierli\'nski et
al.  \nocite{gierlinski:99a} 1999.)

The above studies, however, for the most part considered either single
observations, or a fairly small group of observations.  Studies that have
considered large groups of observations analyzed in a systematic fashion
\cite{wilms:99aa,zdziarski:99a,revnivtsev:99a} have tended to focus on more
phenemonological models (e.g., power-laws with reflection).  In addition,
some of these studies have ignored or lacked high energy data above
$\aproxgt 30$\,keV (e.g., Revnivtsev, Gilfanov, \& Churazov
\nocite{revnivtsev:99a} 2001, Zdziarski, Lubi\'nski, \& Smith
\nocite{zdziarski:99a} 1999).  Within these studies, there has been
disagreement as to the interpretation of the results, even given the same
data (see, for example, Wilms et al.  1999 and Revnivtsev, Gilfanov, \&
Churazov \nocite{revnivtsev:99a} 2001).  For this reason, as well as the
concerns outlined in \S\ref{sec:consider}, in the following sections we
examine our results more carefully.  We begin by comparing our data to
observations of the Crab nebula and pulsar.

\subsection{Crab Ratios}\label{sec:crab}

\begin{figure*}
\centerline{
\includegraphics[width=0.44\textwidth]{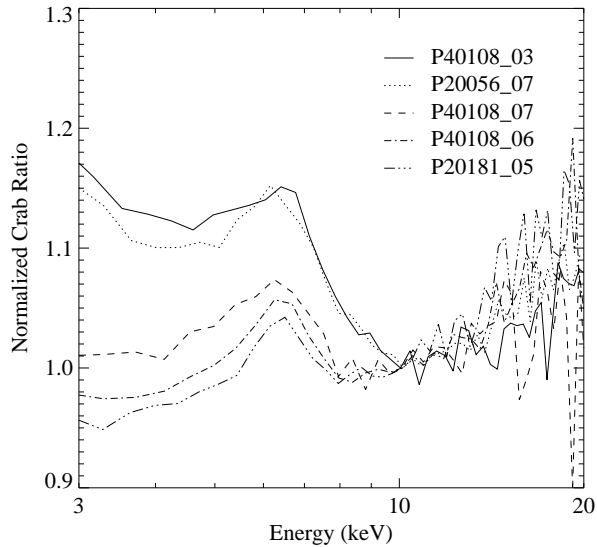}}
\caption{\small The normalized Crab ratio for observations P20056\_07 and
  P20181\_05 with respect to the Crab observation of 1997 September 5, and
  for observations P40108\_03 and P40108\_06 with respect to the Crab
  observation of 1999 February 24. See text for a complete description.
  \protect{\label{fig:crab}}}
\end{figure*}

In Fig.~\ref{fig:crab} we show the normalized ratio for observations
P20056\_07 and P20181\_05 with respect to the Crab nebula and pulsar
observation of 1997 September~05, and for observations P40108\_03 and
P40108\_06 with respect to the Crab nebula plus pulsar observation of 1999
February~24.  Compared to direct fitting of spectral models, such ratio
plots have the advantage that they are relatively insensitive to the
uncertainties in the calibration of the response matrix
\cite{santangelo:98a}, although they are less sensitive to detailed
features than spectral fitting \cite{kreykenbohm:98a}. The Crab ratios were
generated by first subtracting the background from both the Crab and the
\gx\ observations. The Crab spectrum was then multiplied by $E^{\Gamma_{\rm
    Crab}-\Gamma_{\rm 339}}$ where $\Gamma_{\rm Crab}=2.18$ (see Appendix)
and where $\Gamma_{\rm 339}=1.73$ (i.e., the photon index for observation
P20181\_04). All spectra were then normalized to the same flux level at
10\,keV. The Crab ratio was then obtained by dividing the \gx\ pulse height
analyzer (pha) data by the modified Crab spectrum.

Fig.~\ref{fig:crab} shows most of the features that we have also found in
the detailed spectral fitting and thus provides us with additional
confidence that the general behaviour seen was not due to the response
matrix uncertainty.  Higher fluxes clearly yield softer spectra.  Most of
this variation is due to the soft X-rays, while the hard spectrum shape
stays approximately constant. This is completely consistent with the
results we previously obtained by modelling some of these data with a broken
power-law model, where we found that the variation in the power-law index
at softer X-rays was nearly twice as great than that at higher energies
(specifically, see Table~5 of Wilms et al.  \nocite{wilms:99aa} 1999). We
comment on this further in \S\ref{sec:real} below.

The Crab ratios also highlight the broad feature present at $\approx
6.4$\,keV, which in many models is consistent with a broad Fe line. In
fact, fits with realistic Comptonization models always seem to require such
a broad Fe line, as we first noted with fits to \rxte\ data of Cyg~X-1
(Dove et al. \nocite{dove:98a} 1998; who used a very early version of the
\pca\ response matrix).  With every subsequent revision of the \pca\ 
response matrix, realistic Comptonization models of both Cyg~X-1 and \gx\ 
data have continued to require the presence of such a broad Fe line
\cite{wilms:99aa}.  Taking the orthodox interpretation of the feature
present in the Crab ratios as a broad Fe line, these ratios provide us with
more confidence in some of the unusual results obtained in
\S\ref{sec:eqpair}.

First, observation P20181\_05 does indeed have the smallest equivalent
width line, as was determined by the fits.  In addition, the line is very
narrow, and is essentially consistent with the \pca\ resolution.  Also
shown in Fig.~\ref{fig:crab} is the fact that the P40108 observations
(i.e., the observations taken after the return to the hard state in 1999)
indeed do have larger equivalent width, and for the most part broader, Fe
lines.  This is true even into quiescence, as shown by the Crab ratio for
observation P40108\_07.  Note also that despite the fact that observation
P40108\_06 and P20181\_05 have similar flux levels, similar fitted
compactnesses, and similar fitted reflection fractions, the Crab ratio
shows that P40108\_06 has the clearly stronger and broader Fe line (cf.
Revnivtsev, Gilfanov, \& Churazov \nocite{revnivtsev:99a} 2001).  Again,
however, as discussed in \S\ref{sec:eqpair}, the \emph{fitted} physical
line widths, $\sigma$, are complicated by one's assumptions for the line
energy, the assumed reflection properties, and the continuum spectrum.

\begin{figure*}
\centerline{
\includegraphics[width=0.44\textwidth]{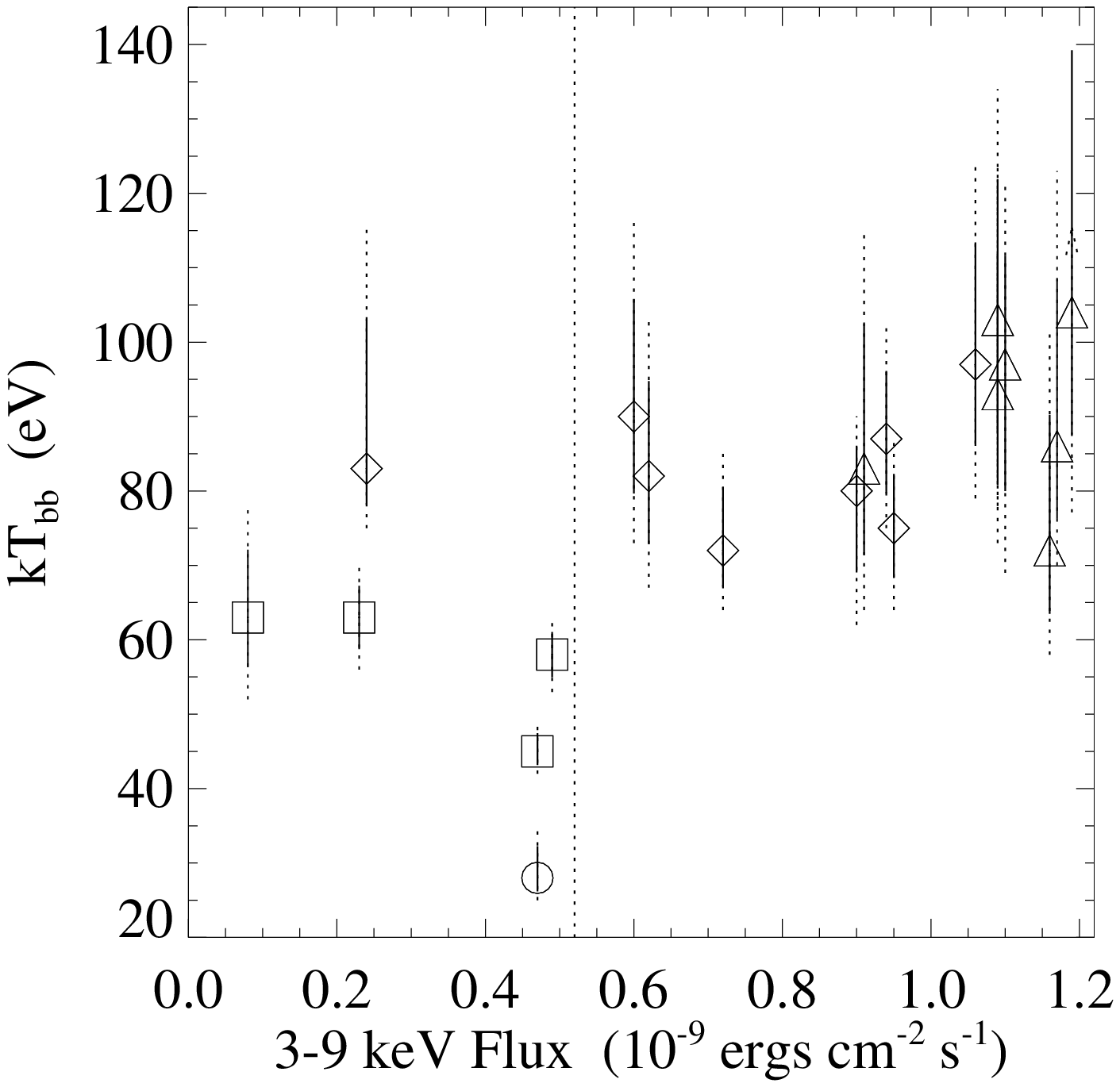}}
\centerline{
\includegraphics[width=0.44\textwidth]{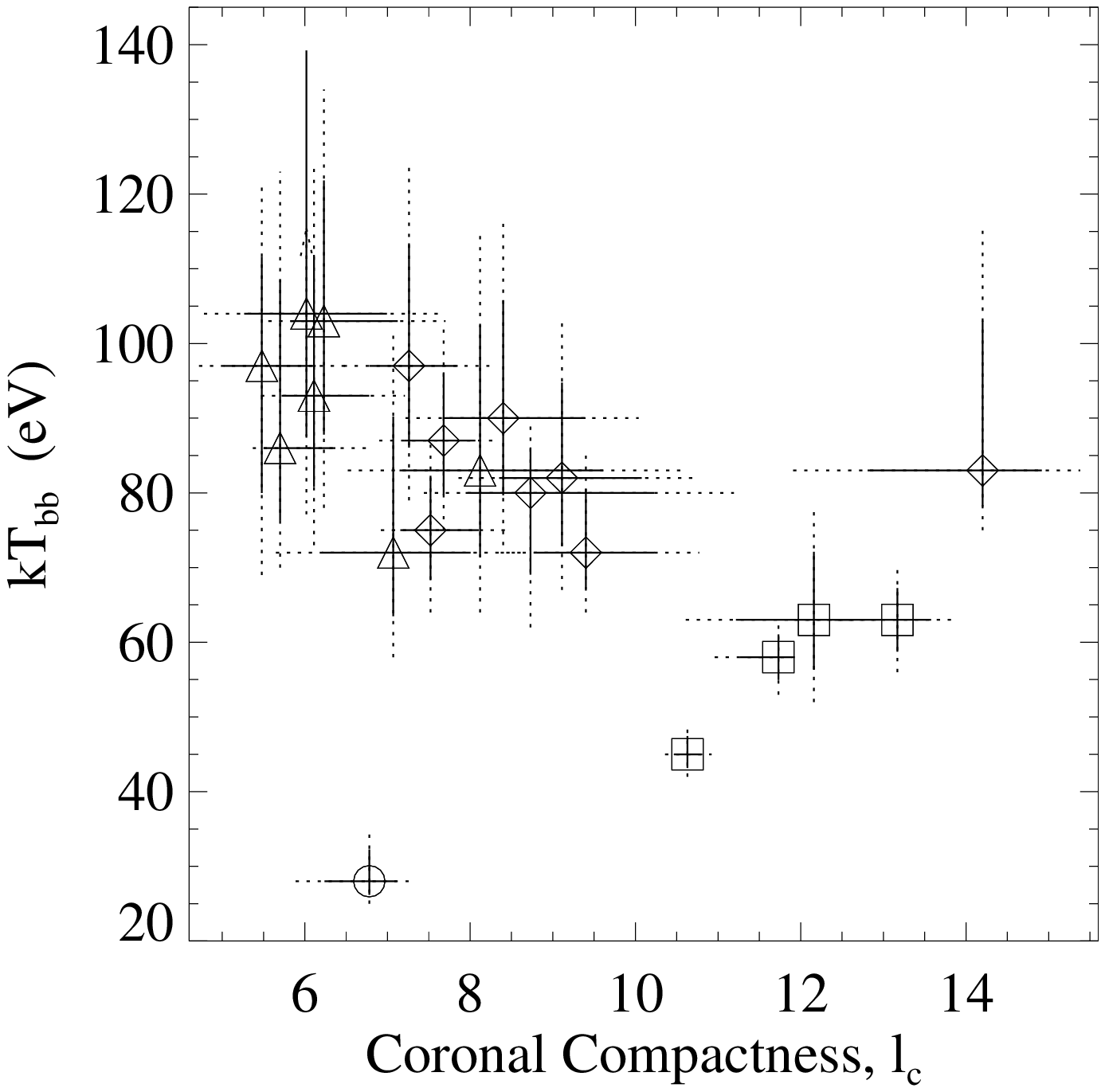}}
\caption{\small {\it Top:} Seed photon temperature vs. 3--9\,keV \pca\
  flux, for the {\tt eqpair} models. {\it Bottom:} Seed photon temperature
  vs. coronal compactness, for the {\tt eqpair} models.}
  \label{fig:temp}
\end{figure*}

The Crab division is too insensitive to reveal some of the finer effects
found by our modelling. For example, one trend seen in the {\tt eqpair}
fits to the data is for the temperature of the seed photons to decrease
with decreasing flux. In the simple sphere+disc coronal model, one expects
the approximate relationship $kT_{\rm bb} \propto \ell_{\rm c}^{-1/2}
F_{\rm bol}^{1/4}$, where $F_{\rm bol}$ is the total flux of the system
(see \S\ref{sec:simple} below).  Thus, utilizing the results of
Table~\ref{tab:log_table} and \ref{tab:kot}, one expects an approximately
factor of 2 change in the seed photon temperature from the brightest to the
faintest observation, which is slightly larger than actually found for the
P20181 and P20056 observations.  Dramatic changes, however, are found for
the first few observations after the return to the hard state in 1999, as
we show in Fig.~\ref{fig:temp}.  The seed photon temperature is seen to be
greatly lowered, and then slowly recovers to the previously observed trend
over the course of the next several months of decline into quiescence.  As
discussed in \S\ref{sec:fits}, such a low seed photon temperature was
required to reproduce a soft excess at energies $\aproxlt 4$\,keV, where we
chose very small systematic errors, thus making the trend extremely
statistically significant.  As seen from Fig.~\ref{fig:crab}, however,
\emph{fractionally} this trend is subtle and small, and therefore we cannot
use the Crab ratios to rule out the possibility of a systematic trend in
the (time-dependent) \pca\ response matrices that we use.

One argument in favor of these features \emph{not} being systematic is that
the transition to the hard state coincides with the turn on of the radio
emission, which initially appears in observation P40108\_03 with a mildly
optically thin radio spectrum and large amplitude (tens of percent)
variations on $\aproxgt 10$\,minute time scales \cite{corbel:00a}.  It is
tempting to associate the unusual soft excess/low fitted seed photon
temperatures with the initial formation of the radio outflow.

\subsection{Is the `$\Gamma$-$\Omega/2\pi$ Correlation' Real?}\label{sec:real}

As discussed above, the Crab ratios clearly show that the \gx\ hard state
spectra can be well-described by a broken power law plus a broad line, with
the low energy (softer) power law varying more than the high energy
(harder) power law \cite{wilms:99aa}.  Using more physical models, this
effect can be found again in the relation between the reflection fraction
and the slope of the underlying power law \cite{zdziarski:98a}.  Since the
shape of the spectrum above 10\,keV changes only slightly while the soft
spectral index shows a strong variation, the comparably constant spectrum
above 10\,keV is reproduced by having a stronger reflection component for
brighter/softer spectra.  This ability of a reflected component to
phenomenologically harden a soft power law has caused some, including
ourselves \cite{wilms:99aa}, to question the reality of the putative
$\Gamma$-$\Omega/2\pi$ correlation.  The Crab ratios, however, remove any
doubt. \emph{Viewed from the purely phenomenological perspective of a
  broken power law model, the `$\Gamma$-$\Omega/2\pi$ correlation' is
  real.}\footnote{Note that the soft state \pca\ spectra presented in
  \S\ref{sec:soft} can also be reasonably well represented by the same
  models used in Table~5 of Wilms et al. \shortcite{wilms:99aa}; i.e., a
  weak disc blackbody with peak temperature $\sim 250$\,eV, a gaussian
  line, and a broken power law.  For those observations, however, the Fe
  line has an equivalent width $\approx 1.5$\,keV, and the broken power law
  indices are $\approx 4.4$, 2.2 for the lower and higher energy power
  laws, respectively.}

Viewed from a physical perspective, the exact nature of these two power law
components is still uncertain.  As measured with the \textsl{Oriented
  Scintillation Spectrometer Experiment} (\textsl{OSSE}) on board the
\textsl{Compton Gamma Ray Observatory} (\textsl{CGRO}), the high energy
spectrum of \gx\ is very well fit by an exponentially cut-off power law,
without any additional reflected component \cite{grove:98a}. \hexte\ 
spectra fitted by themselves yield very similar results \cite{wilms:99aa}.
Reflection models are \emph{not} detecting any characteristic ``spectral
curvature'' in the 30--100\,keV band.  Instead, the reflection models are
primarily sensing the break between the low and high energy power laws,
with a larger break requiring a greater reflection fraction.

Even here, the exact value of the ``reflection fraction'' is subject to
much uncertainty.  As we first discussed for Cyg~X-1 \cite{dove:98a}, since
its $\aproxgt 10$\,keV spectra can be well-described by a spectrum
\emph{without} reflection, the fitted reflection fraction is strongly
dependent upon one's assumptions about the nature of the $\aproxlt 10$\,keV
spectrum.  For Cyg~X-1, we saw that if we allowed a large equivalent width
and broad Fe line, and a higher effective temperature than `usual' soft
excess, no reflection was required (again, using earlier versions of the
\pca\ responses matrices).  On the other hand, the {\tt kotelp} model,
which does have an effective reflection fraction of $\Omega/2\pi \approx
0.3$, also fit those data.  This is similar to the situation here, where
the fits of \gx\ discussed in \S\ref{sec:pca_eqpair}--- which allowed for
high temperature seed photons and large equivalent width, broad lines---
yielded systematically different ``reflection fractions'' than the fits of
\S\ref{sec:eqpair}. The same, physically realistic Comptonization model,
which self-consistently calculates the Comptonized spectrum from the seed
photon spectrum and coronal parameters, yields different, nearly equally as
good, answers for different sets of starting values.  Thus, it is very
difficult to uniquely determine the `best' continuum model and thereby
obtain the most `accurate' value for the reflection fraction (cf.\ di Salvo
et al. \nocite{disalvo:01a} 2001).

Again, however, viewed purely from the phenomenological perspective of a
broken power law, the \emph{relative} trends are giving us information
concerning the degree of the break between the low and high energy power
laws.  As shown in Fig.~\ref{fig:crab}, those observations that show little
or no break in the power law also yield fitted reflection fractions close
to zero.  This is even given the fact that the Fe line, which may or may
not be related to the same component that causes the `reflection', does not
vanish.

Fig.~\ref{fig:crab} also provides an explanation for one of the other
results discussed by Wilms et al. \shortcite{wilms:99aa}.  Specifically,
Wilms et al. \shortcite{wilms:99aa} showed that the P20181 
observations discussed here could be fit by an ionized reflector model with
a weak and narrow Fe line.  The $\Gamma$-$\Omega/2\pi$ correlation was
absent, but a hardness-ionization parameter correlation was instead
detected: softer spectra showed a more highly ionized reflector.  Fitting a
continuum model consisting of a soft excess (modelled as a 250\,eV maximum
temperature disc blackbody), and a power law \emph{fitting predominantly
  the 4--7\,keV data}, the power law break can then be viewed as an edge,
with the position of the edge shifting to higher energies (as would be
appropriate for a more ionized reflector) for softer spectra, as is
apparent in Fig.~\ref{fig:crab}. Such models quite acceptably describe the
data, and in fact can also be shown to describe the P20056 observations as
well.  Again, we see that given the poor spectral resolution of \rxte\ 
coupled with the inherent degeneracies of the spectral models (themselves
consisting of broad, overlapping components), it is difficult to obtain a
unique interpretation of the observed features.

\subsection{A (Too?) Simple Coronal Model}\label{sec:simple}

In \S\ref{sec:timing}, we found that for the {\tt kotelp} model fits the
characteristic PSD frequencies scaled approximately $\propto \ell_{\rm
  c}^{-3/2}$.  Here we provide a simple hypothesis for this behaviour.  The
{\tt kotelp} model fits a relative compactness of the corona as compared to
that of the disc.  Let us assume that the transition radius between the
disc and the corona occurs at a radius of $R_{\rm t}$, while the bulk of
the energy release from the corona is due to falling through the
gravitational potential drop to the marginally stable orbit at $R_{\rm m}$.
The compactness of the corona is proportional to the energy release in the
corona divided by the coronal radius, i.e.  $\propto {R_{\rm m} R_{\rm
    t}}^{-1}$.  The \emph{intrinsic} compactness of the disc, on the other
hand, goes as its gravitational energy release ($\propto R_{\rm t}^{-1}$)
divided by its characteristic radius ($R_{\rm t}$) and thus scales as
$\propto R_{\rm t}^{-2}$.  The \emph{relative} compactness of corona to
disc is therefore $l_{\rm c} \propto R_{\rm t}/R_{\rm m}$.  This is
essentially what we fit with the {\tt kotelp} model.

The fits of \S\ref{sec:kotelp} showed that $l_{\rm c}$ ranged from $\approx
1$ (for the observation after \gx\ returned to the hard state in 1999), to
$\approx 7$ for some of the faintest observations.  In the above simple
model, the flux-hardness anti-correlation then can be ascribed to the
corona-disc transition radius varying from $R_{\rm t} \approx R_{\rm m}$ to
$R_{\rm t} \approx 7~R_{\rm m} \aproxlt 40~GM/c^2$, while the accretion
rate drops by a factor of $\aproxgt 4$ (assuming that the 3--200\,keV flux
is a fair measure of the bolometric flux).

If we take the view point of Psaltis \& Norman \shortcite{psaltis:00a} that
the observed PSD is made up of resonant time scales from the corona-disc
transition region, many of these characteristic frequencies scale as
$R_{\rm t}^{-3/2}$ and therefore should scale as $l_{\rm c}^{-3/2}$ in the
above scenario, consistent with the results shown in
Fig.~\ref{fig:qpovsstuff}. One implicit assumption made here is that the
accretion rate, $\dot M$, scales out of the ratio between the coronal and
disc compactness. That is, we are assuming an efficient accretion disc.  A
further assumption is that the scaling of the frequencies is indeed
$\propto R_{\rm t}^{-3/2}$.  Whereas this is true for the dynamical,
thermal (for fixed $\alpha$), vertical epicyclic, and radial epicyclic (for
$R \gg 8~GM/c^2$) frequencies, this is not true for the Lense-Thirring
precession or viscous damping frequencies.

Although the distance, absolute luminosity, and mass of \gx\ are not
well-known (see Zdziarski et al. \nocite{zdziarski:98a} 1998 for a thorough
discussion of these points), it has $L \aproxlt 5\%~L_{\rm Edd}$ if its
fractional Eddington luminosity is comparable to other hard state GBHC
(Nowak \nocite{nowak:95a} 1995, and references therein).  Thus, in any thin
disc Shakura-Sunyaev model, whether it is gas or radiation
pressure-dominated, one expects the viscous time scale to be several
thousand seconds at $R\sim20~GM/c^2$, even for $\alpha$ relatively large.
This is substantially longer than the low-frequency cutoff time scales
presented in Fig.~\ref{fig:qpovsstuff}. In addition, the ratio between the
viscous time scale and the thermal time scale is approximately $\propto
(L/L_{\rm Edd})^{-2} \aproxgt 400$, which is a larger dynamic range than
that presented in Fig.~\ref{fig:qpovsstuff}.  Let us assume, however, that
the transition region between a corona and disc is \emph{fixed} and is
relatively a large fraction of the transition radius, on the order of $H/R
\sim 0.1$--$0.3$. The dynamic range of frequencies shown in
Fig.~\ref{fig:qpovsstuff} then can be consistent with including the viscous
time scale, and furthermore the viscous frequency would also scale as
$R_{\rm t}^{-3/2}$.

Within this simple model, compactness changes are achieved solely by
varying the coronal radius, and not by changing the underlying geometry.
This is in contrast to the suggestion of Zdziarski, Lubi\'nski, \& Smith
\shortcite{zdziarski:99a}, who hypothesize that the hardness \emph{and}
reflection fraction are regulated by the degree of `overlap' between the
cold disc and quasi-spherical corona (more overlap yields both softer
spectra and greater reflection fraction). We had previously suggested
\cite{dove:98a} that some amount of overlap was required to produce the
broad, large equivalent width line that was required for {\tt kotelp}
models of both Cyg~X-1 and \gx.  We have attempted such simulations with
the {\tt kotelp} code; however, our results indicate that the coronal
region overlapping the disc cools to the inverse Compton temperature,
$kT_{\rm IC} \approx 4$\,keV, and thus fails to yield the required
additional reflection.  \emph{This does not mean such models are ruled out,
  rather that one needs to carefully model additional, substantial heating
  in the overlap region in order to maintain the local coronal
  temperature.}  It has been suggested that such local heating can arise
from thermal conduction between the corona and disc \cite{meyer:94a},
although it has not been rigorously shown that such heating is sufficient.

The above `overlap' model has been suggested as possibly being applicable
to an overall `$\Gamma$-$\Omega/2\pi$ correlation' common to Seyfert~1
galaxies \emph{and} galactic black hole candidates \cite{zdziarski:99a}.
However, as discussed above, the viscous time scale in a thin, low
luminosity accretion disc about a low mass black hole is hundreds to
thousands of seconds.  Thus, each \emph{individual} observation discussed
here is integrated over many viscous time scales, while different
observations can, and likely do, represent vastly different intrinsic
accretion rates through the disc.  This is in contrast to a $10^8\,M_\odot$
black hole in a Seyfert~1 (e.g., NGC~5548; see Chiang et al.
\nocite{chiang:00a} 2000, and references therein), which likely has a
characteristic viscous time scale of \emph{years}.  The week to week
spectral variations of NGC~5548 discussed by Chiang et al.
\shortcite{chiang:00a} therefore correspond to thermal time scales and
faster.  Such thermal time scales are integrated over in a few seconds in
our observations of \gx.  Thus, whereas their are undoubtedly commonalities
between Seyfert~1's and GBHC, it would be surprising if the exact same
models were responsible for the spectral correlations seen in both.

\begin{figure*}
\centerline{
\includegraphics[width=0.44\textwidth]{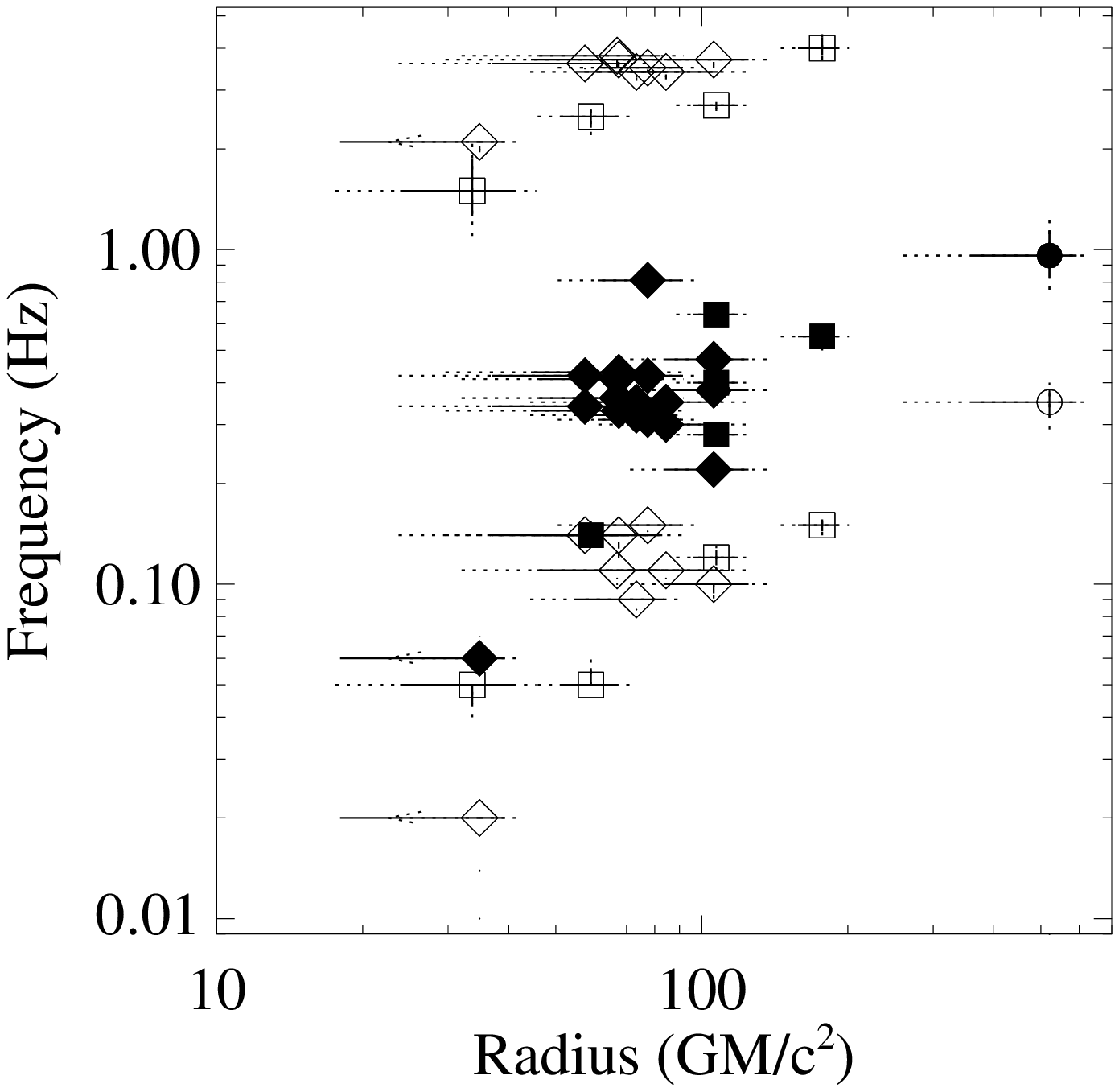}}
\centerline{
\includegraphics[width=0.44\textwidth]{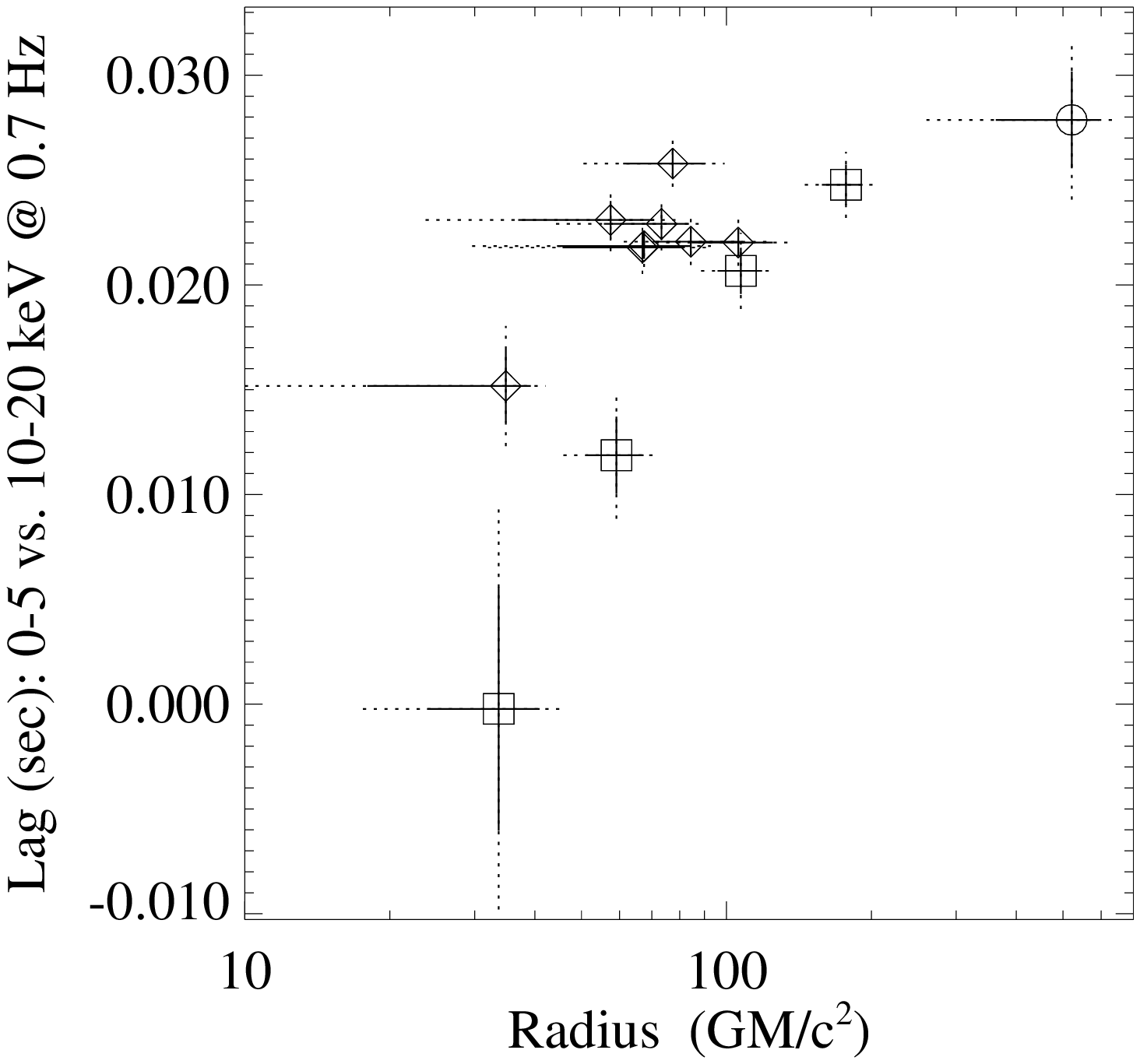}}
\caption{\small   {\it Top:} Characteristic  PSD frequencies vs. ``effective
  radius'' for the seed photon distribution from the {\tt eqpair} fits.
  {\it Bottom:} Time lags between soft and hard variability (average PSD
  frequency 0.7\,Hz) vs. effective radius from the {\tt eqpair} fits. For
  clarity, only the higher signal-to-noise P20181 and P40108 observations
  are shown.  \protect{\label{fig:paolo_r}}}
\end{figure*}

A major failure of the above simple model is that it does not explain the
putative reflection fraction correlation with compactness.  To consider
this further, we turn to the {\tt eqpair} model results, which do allow for
reflection changes.  The {\tt eqpair} model fits agree in rough outline,
but not in detail, with the {\tt kotelp} model fits.  Specifically, both
require coronal temperatures $\aproxgt 150$\,keV (although {\tt eqpair}
fits tend to have optical depths lower by a factor of approximately 2--3),
and both are consistent with overall low reflection fractions of
$\Omega/2\pi \aproxlt 0.5$.  This reflection fraction, however, is clearly
variable in the {\tt eqpair} fits.  In addition, the {\tt eqpair} model
also requires a seed photon temperature variation, which in turn relates to
the ``effective coronal radius'' for these models. Since we have fixed the
seed photon compactness to one, the fitted {\tt eqpair} coronal compactness
scales as $F_{\rm bol}/(\sigma_{\rm SB} T_{\rm bb}^4 R_{\gamma}^2)$, where
$F_{\rm bol}$ is the bolometric flux, $\sigma_{\rm SB}$ is the
Stefan-Boltzman constant, $T_{\rm bb}$ is the temperature of the seed
photons, and $R_{\gamma}$ is the ``effective radius'' for the seed photon
distribution, which for the {\tt eqpair} model is assumed to be on the
order of the coronal radius.  Assuming that the 3--200\,keV flux is a fair
measure of the bolometric flux, and taking a distance of 4\,kpc and a mass
of 4\,${\rm M}_\odot$ for \gx, this effective radius, in units of $GM/c^2$,
is approximately
\begin{equation}
  60  \left ( \frac{d}{4\,{\rm kpc}} \right ) 
   \left ( \frac{4~{\rm M}_\odot}{M} \right ) 
   \left ( \frac{100\,{\rm eV}}{kT_{\gamma}} \right )^2 
   \left ( \frac{F_{\rm bol}}{10^{-9}~\frac{{\rm erg}}{{\rm
           cm^2~s}}} \right )
   \ell_{\rm c}^{-\frac{1}{2}} ~.
\end{equation}
We plot characteristic PSD frequencies vs. this radius in
Fig.~\ref{fig:paolo_r}. 

Note that in the hysteretic/overlap region (the brightest seven P20181
observations), the effective radius and frequencies are roughly constant
and, within the large uncertainties, $\aproxlt 100~{GM/c^2}$.  However, for
the 1999 return to the hard state, the radius trends are exactly
\emph{opposite} of those implied by the {\tt kotelp} fits.  Specifically,
the effective radius is seen to be very large, and then shrinks. This
behaviour, although counter to an association of the PSD \emph{frequencies}
with a coronal radius, is in agreement with the association of the
\emph{time lags} with a characteristic size scale, as we also show in
Fig.~\ref{fig:paolo_r}.  Is this necessarily in contradiction to the {\tt
  kotelp} results?

For the 1999 hard state observations, the large reduced $\chi^2$'s achieved
for most of the {\tt kotelp} models were primarily due to residuals in the
3--4\,keV band (with some contribution from an ``edge-like'' residual near
10\,keV). Otherwise, the {\tt kotelp} models fit the data reasonably well,
especially at high energies.  The low seed photon temperatures found with
the {\tt eqpair} model were largely driven by the 3--4\,keV residuals. We
speculate here that the ``true'' model might be a combination of the
suggested compactness/frequency/time-lag/radii relations above. That is,
perhaps the correct model involves the radii of the \emph{base} of the
corona increasing with decreasing flux/increasing compactness, while the
\emph{vertical} extent of the corona decreases for decreasing
flux/increasing compactness.  The initial transition to the hard state,
exhibited by observation P40108\_03, could be associated with the formation
of a large vertical scale jet, with X-ray emission extending to nearly
1000~$GM/c^2$, while its base has a radius on the order of the marginally
stable orbit.  The characteristic PSD frequencies could then be generated
within the disc, while the characteristic time lags could be generated via
propagation along the length of the jet. An extended jet would also
``view'' a much larger extent of the disc, and thereby have a greater
reflection fraction.  (The slightly low reflection fraction exhibited by
observation P40108\_03 given its long time lag, as seen in
Fig.~\ref{fig:lagsvsref}, might also be attributed to some amount of
beaming away from the disc, as suggested, e.g., by Beloborodov
\nocite{beloborodov:99a} 1999.)  The reflection fraction would naturally
decrease as the jet height decreases and the jet base widens.

Clearly the above suggestion requires detailed theoretical modelling to
verify.  Furthermore, it is subject to great observational uncertainty.
Specifically, are the seed photon temperature trends shown in
Fig.~\ref{fig:temp} real or systematic errors?  However, we note that the
large effective radius implied by the {\tt eqpair} fit to observation
P40108\_03 implies the possibility for this observation, and the subsequent
two observations, of having a large contribution from \emph{synchrotron}
seed photons, as opposed to solely thermal seed photons.  Taking the fitted
{\tt eqpair} coronal temperature and optical depths, and hypothesizing a
magnetic field in equipartition with the thermal energy of the corona, then
the expected flux (at 4\,kpc, assuming a 4\,${\rm M}_\odot$ black hole) due
to synchrotron emission within the corona is on the order of
\begin{equation}
F_{\rm sync}\sim 3\times 10^{-12} ~{\rm erg s^{-1} cm^{-2}} 
     ~ \left ( \frac{R_{\rm c}}{GM/c^2} \right )
       \left ( \frac{kT_{\rm c}}{200\,{\rm keV}} \right )^2 
       \left ( \frac{\tau_{\rm es}}{0.3} \right )^2 ~~.
\end{equation}
Given a typical bolometric flux of $10^{-8}~{\rm erg~s^{-1}~cm^{-2}}$ and
an implied ``seed photon'' flux of $\approx 10^{-9}~{\rm
  erg~s^{-1}~cm^{-2}}$, we see that the synchrotron flux only becomes
comparable to the thermal seed photon flux for an effective coronal radius
$\aproxgt 300\,GM/c^2$, i.e. the implied effective radius of observation
P40108\_03.  Again, this observation, coincident with the initial
transition to the hard state, showed unusually large radio variability on
10 minute time scales, as well as exhibited a quasi-optically thin radio
spectrum \cite{corbel:00a}. Thus, the low fitted seed photon temperature
for the {\tt eqpair} model may have been indicating the need to consider
synchrotron seed photons.

\subsection{Implications for ADAFs}\label{sec:adaf}

An Advection Dominated Accretion Flow was first suggested as a model for
the hard state spectrum of Cyg~X-1 by Ichimaru \shortcite{ichimaru:77a}.
Since that time, there has been much further research and applications to
spectra of GBHC.  One of the key questions in applying ADAF models has
been, where is the transition between the hot, geometrically thick
advective flow and the cool, geometrically thin outer disc located?  Esin,
McClintock, \& Narayan \shortcite{esin:97c} had originally proposed that
the soft to hard state change in GBHC was associated with an increase of
the ``transition radius'' from $\approx 10~GM/c^2$ to $\approx
10^4~GM/c^2$. Esin et al. \shortcite{esin:98a} later revised this range,
for models of Cyg~X-1, to $\approx 60 \rightarrow 400~GM/c^2$.  Recent
models of XTE~J1118+480 require the transition radius to be $\aproxgt
100~GM/c^2$ \cite{esin:01a}.

This aspect of the ADAF model can be tested with these observations.  We
see that for both the {\tt kotelp} and {\tt eqpair} models, with the
exception of the first few observations after the 1999 return to the hard
state, the implied coronal radii are all $\aproxlt 100~GM/c^2$.  There is
no evidence for \emph{large amplitude} variations in this radius as the
observed flux decreases \cite{dimatteo:99a}.  In fact, both the {\tt eqpair}
and {\tt kotelp} models suggest that for the faintest observation,
P40108\_07, which is 15 times fainter than the brightest hard state and 5
times fainter than the initial transition back to the hard state in 1999,
that the effective coronal radius has actually shrunk in comparison to the
previous hard state observations at higher fluxes.

Evidence for a small coronal radius for observation P40108\_07 also comes
from the observed Fe line, which in Fig.~\ref{fig:crab} clearly exhibits a
broad, red tail. Fitting the 3--20\,keV \pca\ data of this observation with
a combination of a power law and a relativistic {\tt diskline} model
\cite{fabian:89a}, we find the following limits on the \emph{inner} radius
of this emission.  Assuming a line at 6.4\,keV, with a $45^\circ$
inclination for the disc, and a line emissivity that is $\propto R^{-3}$,
the best fit inner radius is $14^{+93}_{-8}~GM/c^2$ (90\% confidence
level).  Replacing the power law with a power law plus reflection (i.e.,
Magdziarz \& Zdziarski \nocite{magdziarz:95a} 1995; here with a best fit
value of $\Omega/2\pi = 0.2$), we find an inner disc radius of
$6^{+16}_{-0}~GM/c^2$.  Replacing the reflected power law with an {\tt
  eqpair} model as in \S\ref{sec:eqpair}, the inner disc radius is
$13^{+34}_{-7}~GM/c^2$. Such small implied inner disc radii are strongly
counter to what would be the best evidence for the ``typical'' ADAF model.

Of course, it is possible that the advection dominated region only exists
at radii $\aproxlt 100~GM/c^2$.  The arguments presented above about the
scaling of the PSD time scales with the {\tt kotelp} compactness parameter,
which assumed efficient accretion, are only very weak, indirect evidence
against advection domination.  However, none of the observations here
require advection domination either, and there is no compelling evidence
for any efficiency change from the soft to hard state spectra.  The
observations are all consistent with the spectral changes solely being
determined by accretion rate changes. (Further evidence of this comes from
the fact that the radio flux closely \emph{linearly} tracks the X-ray flux,
all the way into quiescence; Corbel et al. 2001, in preparation.)

\subsection{Alternative Models?}\label{sec:alternative}

\begin{figure*}
\centerline{
\includegraphics[width=0.44\textwidth,angle=270]{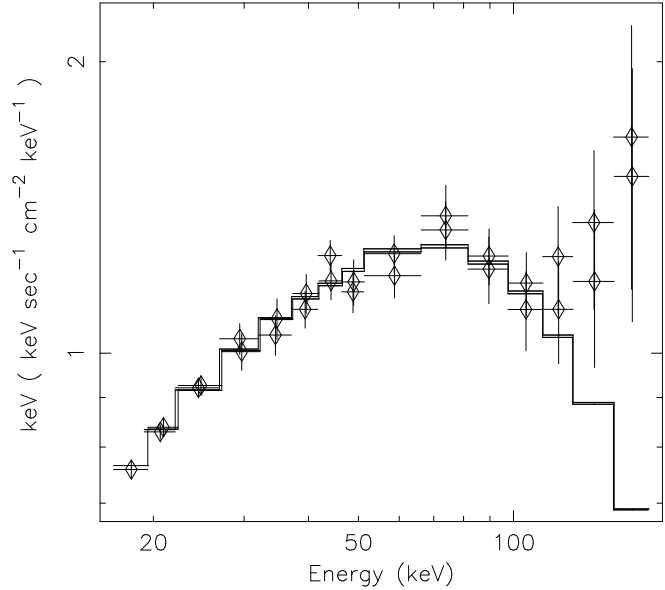}
}
\caption{\small The unfolded \hexte\ spectrum of observation
  P20181\_01. The data have been fit with a {\tt kotelp} model.
  \protect{\label{fig:tail}}}
\end{figure*}

As shown in Fig.~\ref{fig:tail}, some of our observations (most notably
P20181\_01) exhibit a hardening above $\sim$100\,keV in the \hexte\ data.
One possible explanation would be that the hardening is hinting at presence
of a high-energy power-law tail as seen, e.g., in Cyg X-1
\cite{mcconnell:00a}. Power-law tails are generally attributed to the
presence of a non-thermal electron population in the Comptonizing plasma
\cite{bednarek:90a,gierlinski:97a,coppi:99a}.  Alternatively, weak
power-law tails have also been explained in terms of thermal Comptonization
in a Maxwellian plasma with a temperature gradient \cite{skibo:95b}. In
these models, the temperature gradient serves to make the ``effective''
electron distribution non-thermal.  See McConnell et al.
\shortcite{mcconnell:00a} for a discussion of these mechanisms. In galactic
black holes, power-law tails are generally associated with the soft state
or with transitions into the soft state. As these transitions are typically
associated with enhanced radio emission \cite{fender:00b}, one is tempted
to see the presence of the power-law tail as an indication of the
non-thermal electron distribution in a radio outflow \cite{markoff:01a}.

A possible alternative explanation for the power-law tail, however, is that
it is due to instrumental effects: The tail could also be due to an
incomplete background subtraction that might be caused either by the
uncertainty in the estimation of the \hexte\ live-time or by a transient
background source. Such checks are especially important since the \hexte\ 
background contains two strong line complexes below 100\,keV.  In order to
test for the presence of a background source we extracted individual
\hexte\ background spectra for each of the two background dwells of the
\hexte\ clusters and then compared these spectra to each other. In the case
of observation P20181\_01, the difference between the plus and minus
background position for cluster B is $3.6\times 10^{-2}$\,cps, consistent
with statistical fluctuations. On the other hand, for cluster A the
difference between the two background dwells amounts to almost 1\,cps. The
difference spectrum between the two dwells, however, does not resemble the
observed hardening in \gx.  Furthermore, we note that we see the tail in
both \hexte\ clusters which makes it quite improbable that this 1\,cps
difference is the cause of the hard tail. We conclude that the question of
the hard tail in \gx\ is still open for discussion, but that its
association with a problem of the background subtraction is not strongly
suggested. Observations with more sensitive gamma-ray detectors are needed
to determine whether this tail is a real feature.

\section{Summary}\label{sec:summary}

We have presented spectral and timing analysis of 22 individual
observations of the galactic black hole candidate \gx, as observed by the
\textsl{Rossi X-ray Timing Explorer}.  Several of these observations were
coincident with \textsl{Advanced Satellite for Cosmology and Astrophysics},
radio, and/or optical observations.  Our chief results for the X-ray
analyses are as follows.
\begin{itemize}
\item As has been previously observed, the soft state has a strong soft
  excess below $\approx 7$\,keV, a weak hard tail, and very little X-ray
  variability.  Also in agreement with previous observations, the hard
  state, although exhibiting a spectral break at $\approx 10$\,keV, does
  not show a strong soft excess, does have a strong hard energy tail, and
  exhibits large amplitude, rapid X-ray variability.
\item Generally, the hard state exists at lower bolometric luminosities
  than the soft state; however, in terms of 3--9\,keV flux, there appears
  to be a ``hysteresis'' in the state transitions, as first discussed by
  \cite{miyamoto:95a}.  The degree of this hysteresis in terms of the
  bolometric flux is unclear.
\item Brighter flux states tend to show softer overall spectra; however,
  the first few observations after the 1999 return to the hard state are
  unusually soft.  (This is the first ``return'' to a hard state observed
  by \rxte\ in \gx.)  In addition, we have discovered that the trend of
  hardening spectra with decreasing flux is broken for the lowest flux
  observation.
\item Both the soft state and hard state spectra can be reasonably well
  represented by Comptonization models, with parameters similar to previous
  analyses.  However, there is a great deal of degeneracy in these models,
  with different models producing comparable fits. The {\tt kotelp} and
  {\tt eqpair} hard state models both require coronal temperatures
  $\aproxgt 100$\,keV and both are consistent with reflection fractions
  $\Omega/2\pi \aproxlt 0.5$; however, the former requires an approximately
  factor of three greater optical depth.  In addition, there are basic
  differences between these models (e.g., 100\% of the seed photons pass
  through the corona in the {\tt eqpair} model, whereas $\approx 30\%$ of
  the seed photons pass through the corona in the {\tt kotelp} model).
\item The hard state spectra show a break at $\approx 10$\,keV, with the
  spectrum hardening at higher energies.  Modelling this as a broken power
  law, the degree of this break increases with brighter/softer spectra.
  This has been dubbed the `$\Gamma$-$\Omega/2\pi$ correlation' in previous
  analyses with reflection models. Within the context of reflection models,
  in contrast to previous analyses, we find that the ``reflection
  fraction'' is better correlated with 3--9\,keV flux than with the
  spectral hardness.  
\item Most models require the presence of a broad, large equivalent width
  Fe line.  For the 1997 hard state, the equivalent width of this line is
  correlated with the reflection fraction.  For the 1999 hard state, unlike
  for previous observations, the line remains broad and at relatively
  constant equivalent width, even into quiescence.  This is counter to the
  simplest expectations of ADAF models.  Ratios of the \gx\ data to
  observations of the Crab nebula and pulsar show that these results
  represent real effects, and not systematic errors.
\item Timing analysis shows that brighter, softer spectra tend to have
  higher characteristic PSD frequencies than fainter, harder spectra, as
  has been previously observed in both \gx\ and Cyg~X-1.  We have, however,
  included more variability components than previous analyses, and
  furthermore we have correlated our results with coronal parameters,
  rather than a power-law index.  We here have discovered also that the
  variability time lags show the opposite trends from the PSD frequencies.
  Bright, soft spectra tend to show the longest time lag of the hard X-ray
  variability behind the soft X-ray variability.
\item A correlation between time lag and `reflection fraction' was also
  newly discovered. The longest time lag was seen for the first observation
  taken after the 1999 return to the hard state, similar to observations of
  `failed state transitions' in Cyg~X-1 \cite{pottschmidt:00a}.  The 1999
  return of \gx\ to its hard state also exhibited unusual radio properties.
\item Variability properties, both for the characteristic PSD frequencies
  and measured time lags, were most uniform in the ``hysteretic'' range of
  3--9\,keV fluxes (cf. Revnivtsev, Gilfanov, \& Churazov
  \nocite{revnivtsev:99a} 2001).
\item We have hypothesized that a radiatively efficient corona, wherein the
  base of the corona/jet decreases and its height increases, with
  increasing flux/softness, might be able to explain a number of the
  observed spectral-variability correlations.
\item More speculatively (due to uncertainties in instrument backgrounds
  and calibration), there is new evidence within the brightest, softest
  observations for a power law extending beyond the spectral roll-over
  usually associated with thermal Comptonization.  The first few
  observations after the 1999 return to the hard state also exhibited an
  unusual soft excess, which we speculated might be related to the turn on
  of the radio jet.
\end{itemize}
Our own inclination is to believe that the hard tail observed for
observation P20181\_01 and that the soft excesses observed for observations
P40108\_03-\_05 represent real phenomena, and not systematic effects.
Their full implications can only be addressed with improved observations.
For the former, there is some hope that \textsl{Integral} will provide
refined high energy observations.  For the latter, simultaneous
\textsl{XMM-Newton/RXTE} or \textsl{Chandra/RXTE} observations will be
crucial for describing the seed photon regime of the spectra.  This will be
especially important, for example, in breaking the degeneracy between the
600\,eV seed photon/50\,keV corona fits (\S\ref{sec:pca_eqpair}) and the
100\,eV seed photon/150\,keV corona fits (\S\ref{sec:eqpair}) given by the
{\tt eqpair} models.  Furthermore, the ratio of the Comptonized flux to the
seed photon flux (e.g., the ``amplification factor'') has implications for
the coronal geometry.  In this sense, the {\tt eqpair} and {\tt kotelp}
models are very different, and in fact, the optical (seed photon) to X-ray
(Comptonized) flux might be constraining Seyfert 1 geometries in this
manner (Chiang \& Blaes 2001; in preparation).  Finally, none of the
currently available Comptonization models are truly adequate for describing
these data.  Theoretical work, accounting for jet-like geometries and the
presence of the \emph{known} radio emitting outflow need to be more fully
explored.

\smallskip This research has been supported partially by NASA grant
NAG5-3225 (MAN).  We would like to acknowledge useful conversations with C.
Bailyn, J.  Chiang, P. Coppi, S. Corbel, R. Fender, R. Jain, T. Maccarone,
C.  Reynolds, and R. Soria.  M.A.N. would like to thank the Yale Dept. of
Astronomy, where much of this work was performed, for their hospitality.
This research has made use of data obtained through the High Energy
Astrophysics Science Archive Research Center Online Service, provided by
the NASA/Goddard Space Flight Center.

\appendix

\section{Data Analysis Methods}\label{sec:anal}
\subsection{RXTE Data Analysis}\label{sec:rxte_anal}

All \rxte\ results in this paper were obtained using the standard
\rxte\ data analysis software, {\tt ftools} version 5.0 with the
corresponding \pca\ response matrix, and \hexte\ response matrix dated
{\tt 97mar20c}.  Data selection criteria were that the source
elevation was larger than 10$^\circ$ above the earth limb and data
measured within 30\,minutes of passages of the South Atlantic Anomaly
or during times of high particle background (as expressed by the
``electron ratio'' being greater than 0.1) were ignored.  For one
archival observation, P20056\_04, these criteria produced only 45\,sec
of usable data, in contrast to the discussion of this observation
found in Revnivtsev, Gilfanov, \& Churazov \shortcite{revnivtsev:99a}.
Analysis of archival Crab observations indicated that low residual,
uniform results (in terms of measured spectral slope of the Crab from
observation to observation) were best achieved by restricting the
analysis to the first anode layer of the proportional counter units
(PCUs).  We therefore only consider this top layer throughout this
work.  For both spectral and timing analyses, we combined the data
from all active PCUs (typically four or five PCUs were active at any
given time).

For spectral fitting, we limited the energy range of the \pca\ data from 3
to 22\,keV and the energy range of the \hexte\ from 19 to 198\,keV. \pca\ 
channels were grouped so that each had a minimum of 20 counts, while
\hexte\ channels were grouped by: a factor of 3 for channels 16--20, a
factor of 5 for channels 22--51, a factor of 15 for channels 52--126, and a
factor of 24 for channels 127--198.  To take into account the calibration
uncertainty of the \pca\ we applied channel dependent systematic
uncertainties in a manner similar to that described by Wilms et
al.\,\shortcite{wilms:99aa}.  These uncertainties were determined from a
power-law fit to observations of the Crab nebula and pulsar.  As described
on the \rxte\ Guest Observer Facility web pages, the high voltage gain of
the \pca\ detectors was changed in mid-March 1999 (demarking the change
from \pca\ `Gain Epoch 3' to `Gain Epoch 4').  For all the P20181 and
P20056 observations, as well as observations P40108\_01--\_04 (`Epoch 3'),
we applied 0.3\% systematic errors in \pca\ channels 0--10, 1\% systematic
errors in channels 11--18, 0.5\% systematic errors in channels 19--29, 1\%
systematic errors in channels 30--51, and 2\% systematic errors in channels
52--128. For observations P40108\_05--\_07 (`Epoch 4'), these same
systematic errors were applied instead to \pca\ channels 0--8. 9--15,
16--24, 25--43, and 44--128, respectively. Background subtraction of the
\pca\ data was performed using the `SkyVLE' model, as for our previous
studies of \gx\ \cite{wilms:99aa}.

As discussed by Wilms et al. \shortcite{wilms:99aa}, there is a systematic
difference between the \pca\ and the \hexte\ detectors for the best fit
photon index for observations of the Crab nebula and pulsar.  \pca\ fits
yield a Crab photon index of $\Gamma=2.18$, while \hexte\ fits yield
$\Gamma=2.08$.  We have made no attempt to account for any slope
differences; however, we have accounted for the cross-calibration
normalization uncertainties of the \pca\ and the \hexte\ instruments
relative to each other by introducing a multiplicative constant for each
detector in all of our fits.  Typically, if the \pca\ constant was fixed at
1, the multiplicative constants were $\approx 0.9$ for the \hexte\ A
cluster, and a few percent lower for the \hexte\ B cluster.

In all of our spectral fits, we have modelled the neutral hydrogen column
using the cross sections of Morrison \& McCammon \shortcite{morrison:83a}.
Furthermore, unless otherwise noted, we have fixed the value for \gx\ to
$6\times10^{-21}~{\rm cm^{-2}}$.  In the parameter tables below, italicized
parameter values indicate that the parameter was frozen at that value for
that analysis.

For the variability analyses, we follow the methods of Nowak et al.
\shortcite{nowak:99a}, and references therein.  Throughout this work we
have adopted  the PSD normalization of Belloni \& Hasinger
\shortcite{belloni:90a}, where integrating over positive Fourier
frequencies yields the root mean square (rms) variability of the
lightcurve.  We created lightcurves with a time resolution of
$2^{-8}$\,sec; therefore, the Nyquist frequency was 128\,Hz for all
observations.  For the PSD fits discussed above, however, we fit the data
from $3\times10^{-3}$\,Hz to an upper cutoff frequency, which is listed in
Table~\ref{tab:qpo1} and Table~\ref{tab:qpo2} for each observation.
Furthermore, we logarithmically binned each PSD over a range of Fourier
frequencies, $\Delta f/f$, that varied between 0.06--0.15, depending upon
the statistics of each observation.

We performed variability analysis between two energy bands: 0--5\,keV and
10--20\,keV.  These energy bands corresponded to PCA pha channels 0--13 and
27--53 for the P20181 observations and for observations P40108\_03-\_04,
and to PCA pha channels 0--10 and 22--44 for observations P40108\_05-\_07.
The P20056 observations used a data mode with very limited high energy
resolution, therefore the high energy band for those observations
corresponds to 10--100\,keV, i.e., PCA pha channels 26--249.  Since this is
such a broad range, which includes many noise dominated channels, we did
not calculate the time lags for the P20056 observations.

\subsection{ASCA Data Extraction}\label{sec:asca_anal}

We extracted data from the two solid state detectors (\sis 0, \sis 1) and
the two gas detectors (\gis 2, \gis 3) on board \asca\ by using the standard
ftools as described in the ASCA Data Reduction Guide \cite{day:98a}.  We
chose circular extraction regions with radii of $\approx 4$\,arcmin for the
SIS detectors, and $\approx 6$\,arcmin for the \gis\ detectors.  We
excluded approximately the central 1\,arcmin of the \sis\ detectors to
avoid the possibility of photon pile-up.  We used the {\tt sisclean} and
{\tt gisclean} tools (with default values) to remove hot and flickering
pixels. We filtered the data with the strict cleaning criteria outlined by
Brandt et al.  \shortcite{brandt:96a}; however, we took the larger value of
7\,$\mbox{GeV}/c$ for the rigidity.  The background was measured from
rectangular regions on the two edges of the chip farthest from the source
(\sis\ data), or from annuli with inner radii $>6$\,arcmin (\gis\ data).
These data were cleaned and filtered in the same manner as the source
files.

We combined the two \sis\ detectors into a single spectrum, and we combined
the two \gis\ detectors into a single spectrum, properly weighting the
response matrices and backgrounds.  Furthermore, we rebinned the spectral
files as follows.  For the \sis\ spectra, the channels were grouped: by a
factor of 5 for channels 17--246, by a factor of 10 for channels 247--266,
by a factor of 15 for channels 267--311, by a factor of 20 for channels
312--351, and by a factor of 29 for channels 352--380.  For the \gis\ 
spectra, the channels were grouped: by a factor of 10 for channels 18--617,
by a factor of 15 for channels 618--752, by a factor of 30 for channels
753--932, and by a factor of 39 for channels 933--971.  Note that these
observations were taken rather late in the lifetime of \asca, after the
\sis\ detectors in particular had sustained heavy radiation damage.
Although the \sis\ and \gis\ detectors were in good agreement with one
another from 1.8--10\,keV, they were noticeably different from one another
in the 1--1.8\,keV range.  (The \gis\ spectra were systematically softer
than the \sis\ spectra in this energy range.)  Taking this as possibly due
to \sis\ radiation damage, we only considered \sis\ data in the 2--10\,keV
band, whereas we considered \gis\ data in the 1--10\,keV band.

We accounted for the cross-calibration uncertainties of the \sis\ and \gis\ 
instruments relative to each other and relative to \rxte\ by introducing a
multiplicative constant for each detector in all of our fits.  Note that,
relative to observations of the Crab nebula and pulsar, \asca\ fluxes are
systematically lower than \rxte\ fluxes by $\approx 80\%$, which we have
confirmed with our simultaneous \asca/\rxte\ observations.  Throughout this
work we always quote \pca\ fluxes; however, this needs to be borne in mind
when comparing the flux levels discussed here to the \asca\ flux levels
presented in Table~1 of Wilms et al. \shortcite{wilms:99aa}.


\subsection{Fit Parameter Tables}

\begin{table*}
\caption{\small Parameters for models fit to \rxte\ (\pca+\hexte) hard
  state data of \gx. Left: {\tt kotelp}; 72 degrees of freedom (DoF)
  for P20181\_01-P40108\_04; 66 DoF for P40108\_05-\_07. Right: {\tt
  eqpair}; 70 DoF for P20181\_01-P40108\_04; 64 DoF for
  P40108\_05-\_07. \label{tab:kot}}

{\small
      \begin{center}

\begin{tabular}{lllllrllllllr}
\hline
\tabspace
Obs. & ${\ell_c}$ & $\tau_{\rm es}$ & $\sigma_{\rm line}$ & EW &
  $\chi^2$ &  ${\ell_c}$ & $\tau_{\rm es}$ & $kT_{bb}$ &
  $\Omega/2\pi$ & $\sigma_{\rm line}$ & EW & $\chi^2$ \\
\tabspace
&&& (keV) & (eV)  & & & ($10^{-2}$) & (eV) & ($10^{-2}$) & (keV) & (eV) & \\
\tabspace
\hline
%
\tabspace
P20181\_01 
   & \errtwo{3.3}{0.4}{0.3} & \errtwo{1.2}{0.2}{0.2}
   & \errtwo{0.8}{0.1}{0.1} & \errtwo{197}{28}{28}
   & 105 
   & \errtwo{7.3}{1.0}{0.8} & \errtwo{54}{10}{10}
   & \errtwo{97}{27}{18} & \errtwo{39}{6}{6}
   & \errtwo{1.0}{0.2}{0.2} & \errtwo{174}{41}{41}
   & 40 \\
\tabspace
{\it P20181\_01} 
   & \errtwo{3.5}{0.2}{0.2} & \errtwo{1.1}{0.1}{0.1}
   & \errtwo{0.73}{0.05}{0.05} & \errtwo{175}{10}{10}
   & 215 
   & \errtwo{6.9}{1.5}{0.2} & \errtwo{60}{2}{11}
   & \errtwo{106}{6}{26} & \errtwo{39}{4}{6}
   & \errtwo{0.85}{0.08}{0.04} & \errtwo{156}{24}{13}
   & 81 \\
\tabspace
P20181\_02 
   & \errtwo{3.3}{0.4}{0.3} & \errtwo{1.3}{0.2}{0.2}
   & \errtwo{0.8}{0.1}{0.1} & \errtwo{192}{23}{23}
   & 177 
   & \errtwo{7.7}{0.6}{0.8} & \errtwo{49}{7}{7}
   & \errtwo{87}{15}{12} & \errtwo{43}{7}{5}
   & \errtwo{1.0}{0.2}{0.2} & \errtwo{174}{42}{33}
   & 43 \\
\tabspace
P20181\_03 
   & \errtwo{3.6}{0.4}{0.4} & \errtwo{1.2}{0.2}{0.2}
   & \errtwo{0.8}{0.1}{0.1} & \errtwo{180}{23}{22}
   & 144
   & \errtwo{8.7}{2.5}{1.3} & \errtwo{42}{10}{41}
   & \errtwo{80}{10}{18} & \errtwo{37}{7}{4}
   & \errtwo{0.9}{0.2}{0.2} & \errtwo{160}{36}{41}
   & 57 \\
\tabspace
P20056\_01
   & \errtwo{3.5}{0.4}{0.5} & \errtwo{1.1}{0.3}{0.2}
   & \errtwo{0.8}{0.2}{0.2} & \errtwo{175}{30}{29}
   & 75 
   & \errtwo{8.1}{2.5}{1.6} & \errtwo{48}{15}{25}
   & \errtwo{83}{32}{19} & \errtwo{33}{9}{10}
   & \errtwo{1.0}{0.2}{0.3} & \errtwo{185}{38}{40}
   & 52 \\
\tabspace
P20056\_02
   & \errtwo{2.8}{0.3}{0.3} & \errtwo{1.2}{0.3}{0.2}
   & \errtwo{0.8}{0.1}{0.1} & \errtwo{194}{33}{33}
   & 103 
   & \errtwo{6.2}{1.5}{0.7} & \errtwo{55}{7}{12}
   & \errtwo{103}{31}{25} & \errtwo{44}{8}{6}
   & \errtwo{1.0}{0.2}{0.3} & \errtwo{174}{42}{47}
   & 52 \\
\tabspace
P20056\_03
   & \errtwo{2.5}{0.3}{0.2} & \errtwo{1.5}{0.1}{0.3}
   & \errtwo{0.8}{0.1}{0.2} & \errtwo{204}{31}{34}
   & 110 
   & \errtwo{6.1}{1.1}{0.6} & \errtwo{54}{19}{11}
   & \errtwo{93}{31}{20} & \errtwo{54}{9}{10}
   & \errtwo{1.3}{0.2}{0.2} & \errtwo{227}{38}{46}
   & 62 \\
\tabspace
P20056\_05
   & \errtwo{2.0}{0.3}{0.1} & \errtwo{1.5}{0.1}{0.4}
   & \errtwo{0.9}{0.1}{0.1} & \errtwo{215}{35}{37}
   & 96 
   & \errtwo{7.1}{1.5}{1.4} & \errtwo{39}{17}{14}
   & \errtwo{72}{30}{14} & \errtwo{37}{11}{8}
   & \errtwo{1.1}{0.2}{0.2} & \errtwo{237}{40}{44}
   & 47 \\
\tabspace
P20056\_06
   & \errtwo{2.6}{0.3}{0.3} & \errtwo{1.2}{0.3}{0.2}
   & \errtwo{0.8}{0.1}{0.1} & \errtwo{207}{42}{38}
   & 112 
   & \errtwo{6.0}{1.6}{1.2} & \errtwo{51}{15}{12}
   & \errtwo{104}{58}{27} & \errtwo{47}{12}{11}
   & \errtwo{1.0}{0.2}{0.2} & \errtwo{184}{49}{51}
   & 49 \\
\tabspace
P20056\_07
   & \errtwo{1.9}{0.1}{0.1} & \errtwo{1.5}{0.1}{0.1}
   & \errtwo{0.8}{0.1}{0.1} & \errtwo{206}{30}{28}
   & 105 
   & \errtwo{5.7}{1.1}{0.3} & \errtwo{50}{23}{11}
   & \errtwo{86}{37}{16} & \errtwo{47}{13}{8}
   & \errtwo{0.9}{0.2}{0.2} & \errtwo{212}{33}{39}
   & 51 \\
\tabspace
P20056\_08
   & \errtwo{1.9}{0.1}{0.1} & \errtwo{1.5}{0.1}{0.2}
   & \errtwo{0.8}{0.1}{0.1} & \errtwo{212}{33}{31}
   & 108 
   & \errtwo{5.5}{1.1}{0.8} & \errtwo{54}{22}{8}
   & \errtwo{97}{24}{28} & \errtwo{47}{12}{9}
   & \errtwo{1.0}{0.2}{0.2} & \errtwo{206}{38}{40}
   & 50 \\
\tabspace
P20181\_04 
   & \errtwo{4.4}{0.2}{0.3} & \errtwo{1.1}{0.2}{0.1}
   & \errtwo{0.7}{0.1}{0.1} & \errtwo{148}{25}{32}
   & 109 
   & \errtwo{8.4}{1.6}{1.2} & \errtwo{51}{11}{12}
   & \errtwo{90}{26}{17} & \errtwo{38}{7}{7}
   & \errtwo{0.9}{0.2}{0.3} & \errtwo{138}{43}{39}
   & 45 \\
\tabspace
P20181\_05 
   & \errtwo{7.0}{0.1}{0.3} & \errtwo{1.0}{0.1}{0.1}
   & \errtwo{0.6}{0.2}{0.2} & \errtwo{107}{28}{26}
   & 70 
   & \errtwo{14.2}{1.2}{2.3} & \errtwo{25}{32}{24}
   & \errtwo{83}{33}{8} & \errtwo{18}{7}{8}
   & \errtwo{0.3}{0.3}{0.3} & \errtwo{80}{34}{27}
   & 52 \\
\tabspace
P20181\_06 
   & \errtwo{3.9}{0.3}{0.4} & \errtwo{1.3}{0.2}{0.2}
   & \errtwo{0.7}{0.1}{0.1} & \errtwo{157}{23}{23}
   & 153 
   & \errtwo{9.4}{1.4}{1.0} & \errtwo{40}{9}{21}
   & \errtwo{72}{14}{8} & \errtwo{39}{5}{7}
   & \errtwo{0.9}{0.2}{0.2} & \errtwo{162}{25}{33}
   & 52 \\
\tabspace
P20181\_07 
   & \errtwo{2.8}{0.2}{0.1} & \errtwo{1.5}{0.1}{0.2}
   & \errtwo{0.8}{0.1}{0.1} & \errtwo{177}{22}{22}
   & 175 
   & \errtwo{7.5}{1.0}{0.6} & \errtwo{44}{7}{7}
   & \errtwo{75}{12}{11} & \errtwo{45}{6}{8}
   & \errtwo{1.0}{0.2}{0.2} & \errtwo{189}{25}{33}
   & 66 \\
\tabspace
P20181\_08 
   & \errtwo{4.4}{0.3}{0.2} & \errtwo{1.1}{0.2}{0.1}
   & \errtwo{0.7}{0.1}{0.1} & \errtwo{148}{24}{23}
   & 129
   & \errtwo{9.1}{1.6}{1.2} & \errtwo{46}{11}{12}
   & \errtwo{82}{21}{15} & \errtwo{37}{6}{7}
   & \errtwo{0.8}{0.3}{0.2} & \errtwo{143}{34}{33}
   & 45 \\
\tabspace
P40108\_03 
   & \errtwo{1.1}{0.1}{0.2} & \errtwo{1.6}{0.4}{0.1}
   & \errtwo{0.4}{0.1}{0.1} & \errtwo{123}{26}{23}
   & 149 
   & \errtwo{6.8}{0.5}{0.9} & \errtwo{11}{11}{10}
   & \errtwo{28}{7}{3} & \errtwo{29}{7}{9}
   & \errtwo{0.8}{0.1}{0.1} & \errtwo{174}{26}{25}
   & 49 \\
\tabspace
P40108\_04 
   & \errtwo{2.8}{0.1}{0.1} & \errtwo{1.5}{0.1}{0.2}
   & \errtwo{0.4}{0.1}{0.2} & \errtwo{116}{22}{20}
   & 120 
   & \errtwo{10.6}{0.3}{0.3} & \errtwo{1}{27}{0}
   & \errtwo{45}{4}{3} & \errtwo{33}{5}{6}
   & \errtwo{0.8}{0.2}{0.2} & \errtwo{155}{25}{23}
   & 43 \\
\tabspace
P40108\_05 
   & \errtwo{3.9}{0.2}{0.3} & \errtwo{1.5}{0.0}{0.1}
   & \errtwo{0.5}{0.2}{0.2} & \errtwo{111}{24}{23}
   & 130 
   & \errtwo{11.7}{0.3}{0.8} & \errtwo{1}{23}{0}
   & \errtwo{58}{5}{5} & \errtwo{37}{6}{6}
   & \errtwo{0.8}{0.2}{0.3} & \errtwo{134}{34}{16}
   & 57 \\
\tabspace
P40108\_06 
   & \errtwo{5.1}{0.4}{0.7} & \errtwo{1.4}{0.3}{0.2}
   & \errtwo{0.4}{0.2}{0.2} & \errtwo{124}{27}{27}
   & 56 
   & \errtwo{13.2}{0.7}{1.3} & \errtwo{1}{34}{0}
   & \errtwo{63}{7}{7} & \errtwo{15}{4}{5}
   & \errtwo{0.6}{0.2}{0.2} & \errtwo{158}{40}{38}
   & 54 \\
\tabspace
P40108\_07 
   & \errtwo{3.7}{0.6}{0.8} & \errtwo{0.9}{0.5}{0.2}
   & \errtwo{0.6}{0.3}{0.4} & \errtwo{128}{60}{63}
   & 48 
   & \errtwo{12.2}{0.8}{1.6} & \errtwo{10}{33}{9}
   & \errtwo{63}{15}{11} & \errtwo{7}{18}{7}
   & \errtwo{0.6}{0.3}{0.3} & \errtwo{149}{57}{57}
   & 46 \\
\tabspace
%
\hline
\end{tabular}
      \end{center}
}
\end{table*}



\begin{table*}
\caption{\small Parameters for near face on {\tt kotelp} plus dust
  scattering halo models fit to hard state data of \gx.\label{tab:asca}}    

{\small
      \begin{center}

\begin{tabular}{lcccccccrl}
\hline
\tabspace
Obs. & ${\rm N_H}$ & $f_{\rm halo}$  & $S_{\rm halo}$ & $\ell_{c}$ &
$\tau_{es}$ & $\sigma_{\rm line}$ & EW & $\chi^2$/DoF & Instruments \\  
\tabspace
& ($10^{21}~{\rm cm}^{-2}$) & & & & & (keV) & (eV) \\
\tabspace
\hline
%
\tabspace
P40108\_04
   & {\it 6.0} & & 
   & \errtwo{3.3}{0.4}{0.2} & \errtwo{2.5}{0.1}{0.1}
   & \errtwo{0.6}{0.1}{0.1} & \errtwo{169}{28}{21}
   & 94/72 & \pca, \hexte \\
\tabspace
   & \errtwo{5.1}{0.2}{0.2} 
   & \errtwo{0.40}{0.04}{0.04} & \errtwo{1.8}{0.1}{0.1}
   & \errtwo{4.0}{0.3}{0.3} & \errtwo{3.6}{0.2}{0.3}
   & \errtwo{0.7}{0.2}{0.2} & \errtwo{237}{41}{49}
   & 118/125 & \gis, \sis, \hexte \\
\tabspace
   & \errtwo{5.3}{0.2}{0.3} 
   & \errtwo{0.62}{0.08}{0.06} & \errtwo{1.8}{0.1}{0.1}
   & \errtwo{3.4}{0.3}{0.3} & \errtwo{2.6}{0.1}{0.1}
   & \errtwo{0.7}{0.1}{0.1} & \errtwo{216}{20}{21}
   & 220/172 & \gis, \sis, \pca, \hexte \\
\tabspace
P40108\_05
   & {\it 6.0} & & 
   & \errtwo{4.6}{0.6}{0.3} & \errtwo{2.4}{0.2}{0.3}
   & \errtwo{0.7}{0.1}{0.2} & \errtwo{158}{28}{25}
   & 105/66 & \pca, \hexte \\
\tabspace
   & \errtwo{5.5}{0.1}{0.1} 
   & \errtwo{0.49}{0.07}{0.06} & \errtwo{1.7}{0.1}{0.1}
   & \errtwo{5.9}{0.5}{0.3} & \errtwo{3.0}{0.1}{0.1}
   & \errtwo{0.7}{0.3}{0.2} & \errtwo{264}{15}{71}
   & 155/125 & \gis, \sis, \hexte \\
\tabspace
   & \errtwo{6.1}{0.1}{0.1} 
   & \errtwo{0.81}{0.03}{0.04} & \errtwo{1.7}{0.0}{0.0}
   & \errtwo{4.8}{0.1}{0.1} & \errtwo{2.5}{0.1}{0.1}
   & \errtwo{0.7}{0.1}{0.1} & \errtwo{212}{19}{20}
   & 284/166 & \gis, \sis, \pca, \hexte \\
\tabspace
%
\hline
\end{tabular}
      \end{center}
}
\end{table*}



\begin{table*}
\caption{\small Parameters for the {\tt compps} 
  model fit to
  \rxte\ (\pca+\hexte) soft state data of \gx.\label{tab:comp}}    

{\small
      \begin{center}

\begin{tabular}{llcccccccr}
\hline
\tabspace
Obs. & ${kT_{\rm e}}$ & $\tau_{\rm es}$ & $kT_{dbb}$ &
  $\Omega/2\pi$ & $\Xi$ & ${\rm E_{line}}$ & $\sigma_{\rm line}$ & EW &
  $\chi^2$/DoF \\ 
\tabspace
& (keV) & & (keV) & ($10^{-2}$) & ($10^4$) & (keV) & & (keV) \\
\tabspace
\hline
%
\tabspace
P40108\_01 
   & \errtwo{123}{16}{6} & \errtwo{0.30}{0.02}{0.05}
   & \errtwo{0.54}{0.01}{0.01} & \errtwo{19}{2}{2}
   & \errtwo{1.1}{0.2}{0.6} & \errtwo{6.8}{0.3}{0.2}
   & \errtwo{5.4}{0.7}{0.4} & {0.6}
   & 59/67 \\
\tabspace
P40108\_02 
   & \errtwo{136}{8}{5} & \errtwo{0.28}{0.02}{0.01}
   & \errtwo{0.54}{0.01}{0.01} & \errtwo{20}{2}{2}
   & \errtwo{0.8}{0.4}{0.3} & \errtwo{6.8}{0.2}{0.2}
   & \errtwo{5.2}{0.5}{0.4} & {0.9}
   & 68/67 \\
\tabspace
%
\hline
\end{tabular}
      \end{center}
}
\end{table*}



\begin{table*}
\caption{\small Fits to the 10--20\,keV PSD of \gx\ (1997 hard
  state).\label{tab:qpo1}}    

{\small
      \begin{center}

\begin{tabular}{lclllllr}
\hline
\tabspace
Obs. & $f_{\rm max}$ (Hz) & zfc-Lor$_1$ & QPO$_1$ & QPO$_2$ & QPO$_3$ & 
zfc-Lor$_2$ & $\chi^2$/DoF \\
\tabspace
\hline
%
\tabspace
P20181\_01 & 20
   & \errtwo{A=0.13}{0.01}{0.01} & \errtwo{R=0.06}{0.01}{0.01}
   & \errtwo{R=0.15}{0.01}{0.00} && \errtwa{A=6.7}{0.2}{0.4} 
   & 133/97 \\
\tabspace
  && \errtwo{f=0.14}{0.01}{0.02} & \errtwo{f=0.33}{0.00}{0.00}
   & \errtwo{f=0.43}{0.00}{0.03} && \errtwo{f=3.7}{0.1}{0.2} \\
\tabspace
 &&& \errtwo{Q=15}{12}{4} & \errtwo{Q=1.9}{0.1}{0.2} \\
%
\tabspace
P20181\_02 & 20
   & \errtwo{A=0.15}{0.01}{0.01} & \errtwo{R=0.06}{0.00}{0.01}
   & \errtwo{R=0.11}{0.01}{0.00} & \errtwo{R=0.05}{0.01}{0.01}
   & \errtwa{A=7.0}{0.3}{0.2} 
   & 113/94 \\
\tabspace
  && \errtwo{f=0.15}{0.01}{0.01} & \errtwo{f=0.31}{0.00}{0.01}
   & \errtwo{f=0.42}{0.02}{0.01} & \errtwo{f=0.81}{0.06}{0.05}
   & \errtwo{f=3.5}{0.1}{0.1} \\
\tabspace
 &&& \errtwo{Q=17}{8}{4} & \errtwo{Q=3.5}{0.3}{0.3}
   & \errtwo{Q=3.8}{1.3}{1.0} \\
%
\tabspace
P20181\_03 & 15
   & \errtwo{A=0.15}{0.01}{0.01} & \errtwo{R=0.04}{0.01}{0.01}
   & \errtwo{R=0.17}{0.01}{0.00} && \errtwa{A=7.1}{0.2}{0.2} 
   & 74/56 \\
\tabspace
  && \errtwo{f=0.11}{0.01}{0.01} & \errtwo{f=0.30}{0.00}{0.00}
   & \errtwo{f=0.35}{0.01}{0.01} && \errtwo{f=3.4}{0.2}{0.1} \\
\tabspace
 &&& \errtwo{Q=21}{5}{5} & \errtwo{Q=1.6}{0.1}{0.1} \\
%
\tabspace
P20056\_01 & 15
   & \errtwo{A=0.09}{0.01}{0.02} & \errtwo{R=0.07}{0.01}{0.02}
   & \errtwo{R=0.10}{0.01}{0.03} && \errtwa{A=4.4}{0.8}{0.3} 
   & 37/52 \\
\tabspace
  && \errtwo{f=0.13}{0.02}{0.02} & \errtwo{f=0.37}{0.02}{0.01}
   & \errtwo{f=0.49}{0.05}{0.05} && \errtwo{f=3.4}{0.3}{0.5} \\
\tabspace
 &&& \errtwo{Q=8.1}{2.2}{2.7} & \errtwo{Q=1.9}{0.5}{0.5} \\
%
\tabspace
P20056\_02 & 15
   & \errtwo{A=0.06}{0.01}{0.01} & \errtwo{R=0.15}{0.01}{0.01}
   &&& \errtwa{A=3.5}{0.4}{0.3} & 67/73 \\
\tabspace
  && \errtwo{f=0.14}{0.06}{0.03} & \errtwo{f=0.40}{0.05}{0.05}
   &&& \errtwo{f=4.0}{0.5}{0.4} \\
\tabspace
 &&& \errtwo{Q=1.5}{0.6}{0.3} \\
%
\tabspace
P20056\_03 & 15
   & \errtwo{A=0.07}{0.01}{0.02} & \errtwo{R=0.16}{0.01}{0.01}
   &&& \errtwa{A=3.5}{0.6}{0.3}  & 54/41 \\
\tabspace
  && \errtwo{f=0.10}{0.03}{0.02} & \errtwo{f=0.40}{0.03}{0.03}
   &&& \errtwo{f=4.2}{0.4}{0.4} \\
\tabspace
 &&& \errtwo{Q=1.5}{0.4}{0.2} \\
%
\tabspace
P20056\_05 & 10
   & \errtwo{A=0.04}{0.01}{0.00} & \errtwo{R=0.13}{0.01}{0.01}
   &&& \errtwa{A=2.9}{0.3}{0.2}  & 54/38 \\
\tabspace
  && \errtwo{f=0.16}{0.07}{0.03} & \errtwo{f=0.54}{0.02}{0.04}
   &&& \errtwo{f=4.5}{0.6}{0.8} \\
\tabspace
 &&& \errtwo{Q=2.1}{0.6}{0.5} \\
%
\tabspace
P20056\_07 & 15
   & \errtwo{A=0.05}{0.01}{0.01} & \errtwo{R=0.04}{0.01}{0.01}
   & \errtwo{R=0.11}{0.01}{0.00} && \errtwa{A=2.5}{0.3}{0.2}
   & 58/70 \\
\tabspace
  && \errtwo{f=0.17}{0.02}{0.04} & \errtwo{f=0.57}{0.02}{0.02}
   & \errtwo{f=0.65}{0.03}{0.05} && \errtwo{f=4.7}{0.6}{0.5} \\
\tabspace
 &&& \errtwo{Q=18}{13}{6} & \errtwo{Q=2.0}{0.4}{0.3} \\
%
\tabspace
P20056\_08 & 10
   & \errtwo{A=0.04}{0.01}{0.01} & \errtwo{R=0.12}{0.01}{0.01}
   &&& \errtwa{A=2.1}{0.3}{0.5}
   & 26/29 \\
\tabspace
  && \errtwo{f=0.19}{0.05}{0.02} & \errtwo{f=0.67}{0.04}{0.04}
 &&& \errtwo{f=5.1}{1.6}{1.1} \\
\tabspace
 &&& \errtwo{Q=2.0}{0.5}{0.4} \\
%
\tabspace
P20181\_04 & 15
   & \errtwo{A=0.15}{0.00}{0.01} & \errtwo{R=0.05}{0.00}{0.01}
   & \errtwo{R=0.14}{0.01}{0.00} && \errtwa{A=6.1}{0.2}{0.2} 
   & 78/56 \\
\tabspace
  && \errtwo{f=0.14}{0.01}{0.01} & \errtwo{f=0.34}{0.01}{0.01}
   & \errtwo{f=0.42}{0.01}{0.02} && \errtwo{f=3.6}{0.2}{0.2} \\
\tabspace
 &&& \errtwo{Q=18}{5}{5} & \errtwo{Q=1.7}{0.2}{0.1} \\
%
\tabspace
P20181\_05 & 8
   & \errtwo{A=0.55}{0.13}{0.11} & \errtwo{R=0.26}{0.01}{0.02}
   &&& \errtwa{A=13.5}{1.4}{1.2} 
   & 50/36 \\
\tabspace
  && \errtwo{f=0.02}{0.01}{0.01} & \errtwo{f=0.06}{0.01}{0.00}
   &&& \errtwo{f=2.1}{0.2}{0.1} \\
\tabspace
 &&& \errtwo{Q=0.6}{0.1}{0.0} \\
%
\tabspace
P20181\_06 & 30
   & \errtwo{A=0.18}{0.01}{0.01} & \errtwo{R=0.05}{0.02}{0.01}
   & \errtwo{R=0.17}{0.01}{0.01} && \errtwa{A=7.3}{0.5}{0.5} 
   & 82/64 \\
\tabspace
  && \errtwo{f=0.09}{0.01}{0.01} & \errtwo{f=0.32}{0.01}{0.01}
   & \errtwo{f=0.35}{0.01}{0.02} && \errtwo{f=3.4}{0.1}{0.2} \\
\tabspace
 &&& \errtwo{Q=13}{9}{5} & \errtwo{Q=1.5}{0.2}{0.1} \\
\tabspace
%
%
\tabspace
P20181\_07 & 30
   & \errtwo{A=0.14}{0.00}{0.01} & \errtwo{R=0.05}{0.00}{0.01}
   & \errtwo{R=0.08}{0.00}{0.01} & \errtwo{R=0.15}{0.01}{0.00}
   & \errtwa{A=6.3}{0.2}{0.2} 
   & 97/61 \\
\tabspace
  && \errtwo{f=0.10}{0.01}{0.00} & \errtwo{f=0.22}{0.01}{0.00}
   & \errtwo{f=0.38}{0.01}{0.00} & \errtwo{f=0.47}{0.01}{0.01}
   & \errtwo{f=3.7}{0.1}{0.2} \\
\tabspace
 &&& \errtwo{Q=8.6}{3.3}{2.6} & \errtwo{Q=13}{2}{2}
   & \errtwo{Q=1.8}{0.1}{0.1} \\
%
\tabspace
P20181\_08 & 20
   & \errtwo{A=0.17}{0.01}{0.01} & \errtwo{R=0.07}{0.01}{0.01}
   & \errtwo{R=0.15}{0.01}{0.01} && \errtwa{A=5.9}{0.1}{0.1} 
   & 64/59 \\
\tabspace
  && \errtwo{f=0.11}{0.01}{0.01} & \errtwo{f=0.36}{0.02}{0.01}
   & \errtwo{f=0.41}{0.03}{0.01} && \errtwo{f=3.8}{0.2}{0.3} \\
\tabspace
 &&& \errtwo{Q=4.9}{1.6}{1.2} & \errtwo{Q=1.3}{0.4}{0.4} \\
\tabspace
%
\hline
\end{tabular}
      \end{center}
}
\end{table*}




\begin{table*}
\caption{\small Fits to the 10--20\,keV PSD of \gx\ (1999 hard state).\label{tab:qpo2}}   

{\small
      \begin{center}

\begin{tabular}{lclllllr}
\hline
\tabspace
Obs. & $f_{\rm max}$ (Hz) & zfc-Lor$_1$ & QPO$_1$ & QPO$_2$ & QPO$_3$ & 
zfc-Lor$_2$ & $\chi^2$/DoF \\
\tabspace
\hline
%
\tabspace
P40108\_03 & 8
   & \errtwo{A=0.02}{0.00}{0.00} & \errtwo{R=0.16}{0.01}{0.01}
   &&&& 53/38 \\
\tabspace
  && \errtwo{f=0.35}{0.05}{0.06} & \errtwo{f=0.96}{0.30}{0.20} \\
\tabspace
 &&& \errtwo{Q=0.7}{0.4}{0.2} \\
%
\tabspace
P40108\_04 & 15
   & \errtwo{A=0.11}{0.00}{0.01} & \errtwo{R=0.14}{0.01}{0.00}
   &&& \errtwa{A=3.9}{0.6}{0.2} 
   & 110/95 \\
\tabspace
  && \errtwo{f=0.15}{0.01}{0.01} & \errtwo{f=0.55}{0.02}{0.05}
   &&& \errtwo{f=4.0}{0.5}{0.3} \\
\tabspace
 &&& \errtwo{Q=1.8}{0.1}{0.2} \\
\tabspace
%
%
\tabspace
P40108\_05 & 25
   & \errtwo{A=0.20}{0.01}{0.01} & \errtwo{R=0.07}{0.00}{0.01}
   & \errtwo{R=0.09}{0.00}{0.01} & \errtwo{R=0.05}{0.00}{0.01}
   & \errtwa{A=9.6}{0.4}{0.6} 
   & 56/54 \\
\tabspace
  && \errtwo{f=0.12}{0.01}{0.01} & \errtwo{f=0.28}{0.00}{0.01}
   & \errtwo{f=0.40}{0.01}{0.00} & \errtwo{f=0.64}{0.02}{0.02}
   & \errtwo{f=2.7}{0.1}{0.1} \\
\tabspace
 &&& \errtwo{Q=11}{5}{3} & \errtwo{Q=6.1}{1.4}{1.0}
   & \errtwo{Q=11}{9}{3} \\
%
\tabspace
P40108\_06 & 8
   & \errtwo{A=0.42}{0.04}{0.03} & \errtwo{R=0.18}{0.01}{0.02}
   &&& \errtwa{A=10.7}{0.9}{0.9} 
   & 53/36 \\
\tabspace
  && \errtwo{f=0.05}{0.01}{0.00} & \errtwo{f=0.14}{0.02}{0.01}
   &&& \errtwo{f=2.5}{0.2}{0.3} \\
\tabspace
 &&& \errtwo{Q=1.0}{0.2}{0.1} \\
%
\tabspace
P40108\_07 & 5
   & \errtwo{A=0.43}{0.06}{0.06} & 
   &&& \errtwa{A=8.4}{1.9}{2.0}
   & 20/19 \\
\tabspace
  && \errtwo{f=0.05}{0.00}{0.01} & 
 &&& \errtwo{f=1.5}{0.6}{0.4} \\
%
\tabspace
\hline
\end{tabular}
      \end{center}
}
\end{table*}


\end{document}